\pdfoutput=1
\documentclass[%
 reprint,
 amsmath,amssymb,aps,
]{revtex4-2}

\usepackage{graphicx}
\usepackage{dcolumn}
\usepackage{bm}
\usepackage{subcaption}
\usepackage{caption}
\usepackage{wrapfig}
\usepackage{float}
\usepackage{siunitx}
\usepackage{mathrsfs}
\usepackage{mathtools}
\usepackage{cancel}
\usepackage[utf8]{inputenc}
\usepackage[T1]{fontenc}
\usepackage{mathptmx}
\usepackage{etoolbox}
\usepackage{eqnarray}
\usepackage[export]{adjustbox}
\usepackage[normalem]{ulem}
\usepackage[abs]{overpic}
\usepackage[svgnames, table]{xcolor}
\usepackage{array, booktabs, multirow}

\makeatletter
\def\@email#1#2{%
 \endgroup
 \patchcmd{\titleblock@produce}
  {\frontmatter@RRAPformat}
  {\frontmatter@RRAPformat{\produce@RRAP{*#1\href{mailto:#2}{#2}}}\frontmatter@RRAPformat}
  {}{}
}%
\makeatother
\begin{document}


\title{Surface-directed and bulk spinodal decomposition compete to decide the morphology of bimetallic nanoparticles}
\author{P. Pankaj}
\author{Saswata Bhattacharyya*}%

\author{Subhradeep Chatterjee*}
\email{saswata@msme.iith.ac.in, subhradeep@msme.iith.ac.in}
\affiliation{%
{Department of Materials Science and Metallurgical Engineering},\\
            {Indian Institute of Technology}, {Hyderabad}, {502284}, {Telangana}, {India}
}%

\date{\today}

\begin{abstract}
An embedded-domain phase-field formalism is used for studying phase transformation pathways in 
bimetallic nanoparticles (BNPs). Competition of bulk and surface-directed spinodal decomposition 
processes and their interplay with capillarity are identified as the main determinants of BNP 
morphology. The former is characterized by an effective bulk driving force $\Delta\tilde{f}$ which 
increases with decreasing temperature, while the latter manifests itself through a balance of 
interfacial energies captured by the contact angle $\theta$. The simulated morphologies, namely, 
core-shell, Janus and inverse core-shell, cluster into distinct regions of the 
$\Delta\tilde{f}$-$\theta$ space. Variation of $\theta$ with $\Delta\tilde{f}$ in the Ag-Cu alloy 
system is computed as a function of temperature using a CALPHAD approach in which surface energies 
are estimated from a modified Butler equation. This $\theta-\Delta\tilde{f}$ trajectory for Ag-Cu, 
when superimposed on the morphology map, enables the prediction of different morphological 
transitions as a function of temperature. Therefore, the study establishes a unique thermodynamic 
framework coupled with phase-field simulations for predicting and tailoring nanoparticle morphology 
through a variation of processing temperature.
\end{abstract}

\maketitle

%

\section*{Introduction}

Properties of bimetallic nanoparticles (BNPs) used in diverse fields such as catalysis, photonics, 
spintronics and biomedical-sensing ~\cite{ferrando2016structure,Cortie2011,Medina-Cruz2020} depend
crucially on their morphology. Core-shell (CS) and Janus are the most commonly reported 
morphologies: CS morphology consists of an outer shell of the lower surface energy component 
surrounding a core of the other component, while Janus is characterized by the two components 
forming two opposite faces of a particle with their common interface extending to the surface. An 
inverse core-shell (ICS) morphology, where the higher surface energy component forms the shell, has 
also been observed in a few cases~\cite{Tsuhi_ics, nakamura_ics}.

A fundamental understanding of morphological development in BNPs is crucial for tailoring their 
properties. First principles~\cite{Ferrando2015} and atomistic~\cite{Chandross_2014}
simulations, as well as classical thermodynamics~\cite{Yuan2008} have been employed to find the
lowest energy morphology of BNPs. However, these approaches generally do not address kinetic aspects
of morphological development. Since many commonly used BNP systems like Ag-Cu, Ag-Ni, Au-Co, Co-Cu,
etc. exhibit solid-state immiscibility, spinodal decomposition (SD) presents a likely kinetic 
pathway for morphology development in BNPs~\cite{Radnoczi2017}. 

Phase-field models have been employed very successfully for studying
microstructure development in bulk immiscible systems systems using the Fourier spectral
method~\cite{LQChen1999}. For finite systems with complex geometries, embedding the system of
interest in a larger computational domain has proven to be an effective strategy
for dealing with the non-periodicity of the domain of interest~\cite{BuenoOrovioSIAM2006,
LiCommMathSci2009,Yu2012,Poulsen2016}. In addition to Dirichlet, Neumann and Robin
boundary conditions, Yu et al. developed a methodology~\cite{Yu2012} to impose
a contact angle boundary condition for studying phase transformations in contact with an
external surface. However, if one wishes to capture the \emph{in situ} development of 
three-phase contacts (or their absence) on the surface, a different approach involving
a modification of the system's free energy can be adopted.

Recently, we presented such an embedded-domain phase-field model (EPFM)~\cite{pankaj} 
using which correct contact angles could be naturally recovered without imposing
them directly through boundary conditions. This model was was used to understand how
contact angle and particle size influenced the morphology development in BNPs. However,
the role of temperature, which is an important parameter from a processing perspective,
on BNP morphology was not addressed. Temperature, or equivalently, undercooling below a
critical temperature, can exert a strong influence on the phase separation process by
altering the driving force for bulk SD. In addition, in confined systems such as BNPs,
surface directed SD (SDSD) presents another mode of microstructural evolution
which  may also be influenced by temperature through its effect on surface
and interfacial energies. Interplay between these two alternative mechanisms and their 
interactions with capillarity give rise to different BNP morphologies. In this paper, we
use EPFM to systematically investigate this process, and by carrying out further 
thermodynamic computations, identify the relevant physical parameter groups that can be
used to understand and predict morphology selection in BNPs as a function of temperature.

\section*{Model Formulation}

\begin{figure}[hbtp]
    \centering
    \includegraphics[scale=0.25]{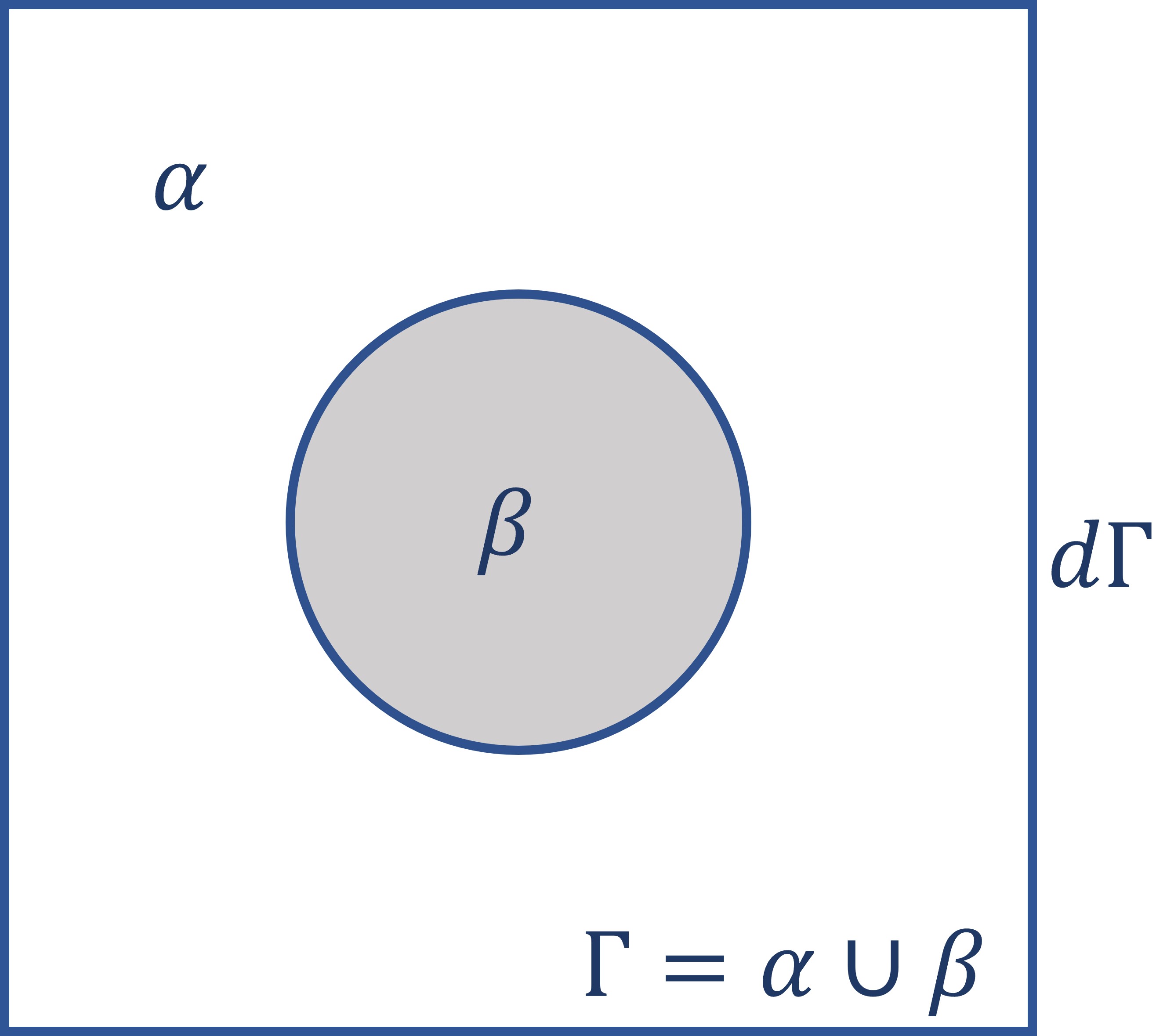}
    \caption{Computational domain $\Gamma$ consisting of an isolated nanoparticle $\beta$
    (physical domain of interest) in  an inert matrix $\alpha$. Periodic boundary conditions
    are imposed on the domain boundary $d\Gamma$.}
    \label{fig:fig0}
\end{figure}

In the EPFM formalism~\cite{pankaj} illustrated schematically in Fig.~\ref{fig:fig0}, a BNP system 
consisting of atomic species A and B is modeled by placing an isolated particle ($\beta$) in an 
inert matrix ($\alpha$). A scaled composition field $c$ describes phase separation of an initially
homogeneous $\beta$ to a mixture of $\beta_1$ and $\beta_2$ phases with equilibrium compositions 
$c_{\beta_1}^e$ and $c_{\beta_2}^e$, respectively. We use 
$c=(X-X_{\beta_1}^{e})/(X_{\beta_2}^{e}-X_{\beta_1}^{e})$ for the scaling where $X$ 
denotes the mole fraction of B and $X_{\beta_i}$ is the equilibrium composition of phase 
$\beta_i$ at a given temperature. An auxiliary non-conserved phase-field variable $\phi$ is
used to distinguish the particle from the matrix; we use a stationary and radially symmetric
\textsf{tanh} profile for $\phi$ that varies smoothly across the particle surface from 0
in the matrix side to 1 inside the particle~\cite{pankaj}. Total free energy of the system is
then expressed as:

\begin{equation}
    \mathcal{F} = \frac{1}{V_m} \int_{\Gamma} \Big(f(c,\phi)
                  + \kappa_\phi|\nabla\phi|^2 + \kappa_c|\nabla c|^2 \Big) d\Gamma,
                  \label{eq:totalFcal}
\end{equation}
where $V_m$ is the molar volume, $f$ is bulk free energy density, $\kappa_c$ and $\kappa_{\phi}$ are
gradient energy coefficients associated with $c$ and $\phi$, respectively, and $\Gamma$ is the volume
of the whole computational domain. 

Defining an interpolating function $h(\phi)=\phi^3(10 - 15\phi + 6\phi^2)$ that 
connects $\alpha$ and $\beta$ (with $\phi$ being 0 and 1 inside $\alpha$ and $\beta$,
respectively), the bulk free energy density $f$ is given as
$f(c,\phi)=h(\phi)f^{\beta} + (1 - h(\phi))f^{\alpha} + \omega(c) g(\phi)$. 
Here $\omega(c)g(\phi) = \omega_0(1 - \chi c)\phi^2(1-\phi)^2$ describes the free energy barrier 
between $\alpha$ and $\beta$, $\omega_0$ sets the barrier height and $\chi$ controls solute 
segregation at the particle-matrix interface~\cite{pankaj}. We use the following forms of free energy
for $\alpha$ and $\beta$:
\begin{align}
 f^{\alpha}&=f_0^m(c-c_{\alpha}^e)^2,\nonumber\\
 f^{\beta}&=f_0^p (c-c_{\beta_1}^e)^2(c-c_{\beta_2}^e)^2.
 \label{eq:bulk-free-energy-dens}
\end{align}
The scaled equilibrium compositions $c_\alpha^e, c_{\beta_1}^e, c_{\beta_2}^e$ are chosen to be 0.5,
0 and 1, respectively, and $f_0^p$ and $f_0^m$ are temperature-dependent constants. 
Evolution of composition field within the particle is described by the Cahn-Hilliard equation:
\begin{equation}
    \frac{\partial{c}}{\partial{t}}=\nabla\cdot{M(\phi)}\nabla\frac{\delta\mathcal{F}}{\delta{c}}.
    \label{eqn:CH}
\end{equation}

We constrain the matrix to remain inert with respect to solute diffusion by choosing 
$M(\phi)=M_ch(\phi)$, with $M_c$ being the effective atomic mobility of solute in the
particle. Eq.~\eqref{eqn:CH} is solved numerically using a semi-implicit Fourier spectral 
method~\cite{cogswell2011, cogs-thesis, Zhu1999}. 
We non-dimensionalize all parameters in the equations using characteristic length $L_c$,
time $\tau_c$ and energy $E_c$. Details of the non-dimensionalization procedure are provided
Supplementary Information A; further details of the model and its numerical implementation
can be found elsewhere~\cite{pankaj}.

Energies of $\alpha$-$\beta_1$, $\alpha$-$\beta_2$ and $\beta_1$-$\beta_2$ interfaces,
designated as $\sigma_1, \sigma_2$ and $\sigma_{12}$, are obtained from the equilibrium
composition profiles $c^e(x)$ across the respective interfaces:
\begin{align}
    \sigma&=\frac{1}{V_m}\int\Big[{f}(c^e(x),\phi(x))
    +\kappa_\phi|\nabla\phi|^2+\kappa_c|\nabla{c^e(x)}|^2\nonumber\\
    &-\Big\{
    (1-c^e(x)) \mu_A^{e} + c^e(x)\mu_B^{e}
    \Big\}
    \Big]dx,
    \label{eq:interfacialenergies}
\end{align}
where $\mu_i^{e}\;(i=A,\;B)$ denotes the equilibrium chemical potential of component
$i$ in any one the coexisting phases across the interface.

The surface and interfacial energies define the contact angle $\theta$ at the triple junction
between the phases as:
\begin{equation}
    \cos\theta=\frac{\sigma_1-\sigma_2}{\sigma_{12}}.
    \label{eq-contact}
\end{equation}

Phase transformations within the particle need not always yield triple junctions at the surface.
Cahn, in his classic paper on ``Critical Point Wetting'', described a condition that precluded the
formation of a triple junction. He defined this \emph{spontaneous} wetting condition to be
$\sigma_1 - \sigma_2 \geq \sigma_{12}$; the equality condition corresponds to $\theta=\ang{0}$,
while $\theta$ become undefined for the inequality condition. In the latter case, the phase
with the low surface energy develops a continuous layer on the surface.  

Non-zero $\chi$ leads to preferential solute segregation to one of the surfaces by creating
imbalance between the `surface' energies $\sigma_1$ and $\sigma_2$ -- its key role on BNP
morphology has already been explored in detail~\cite{pankaj}. Here, we keep $\chi$ and particle
size fixed, and investigate the competition of bulk SD and SDSD in deciding the BNP morphology
(CS, ICS and Janus) by varying $f_0^p$, $\omega_0$ and $\kappa_c$. Note that these variable also
control $\theta$. Since each of these parameter combinations gives rise to one of these three 
morphologies, we performed simulations over a large set of parameters and classified the results
in terms of driving force and contact angle.

\section*{Results and Discussion}

\subsection{Simulation of BNP morphologies}

\begin{figure}[htbp]
\centering%
\begin{subfigure}{0.085\textwidth}
\centering%
    \includegraphics[scale=0.03,trim={39cm 7cm 39cm 7cm},clip]{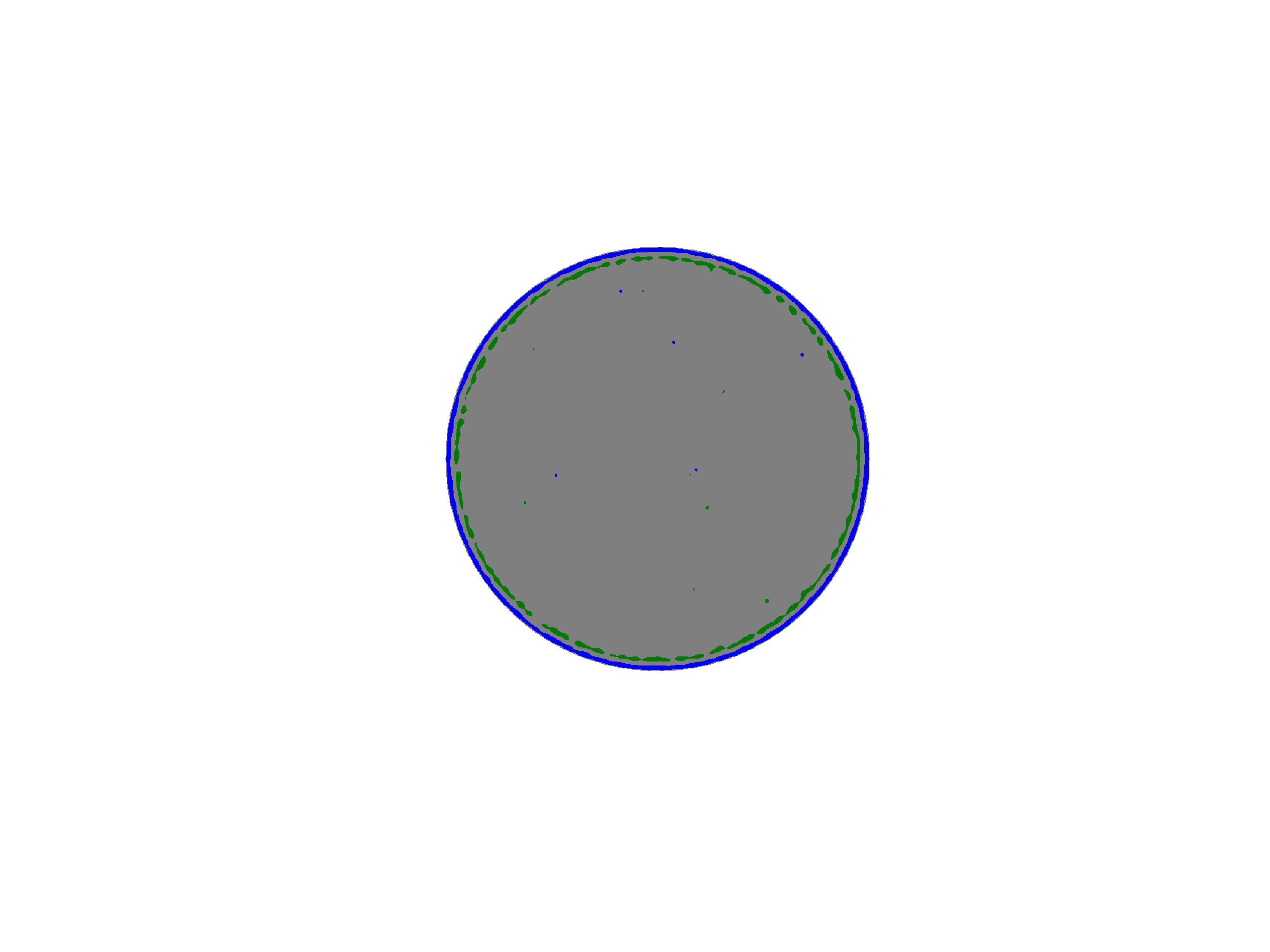}%
   \caption{$t=0.2$}
    \label{fig:fig1a}
\end{subfigure}%
\begin{subfigure}{0.085\textwidth}
\centering%
    \includegraphics[scale=0.03,trim={39cm 7cm 38cm 7cm},clip]{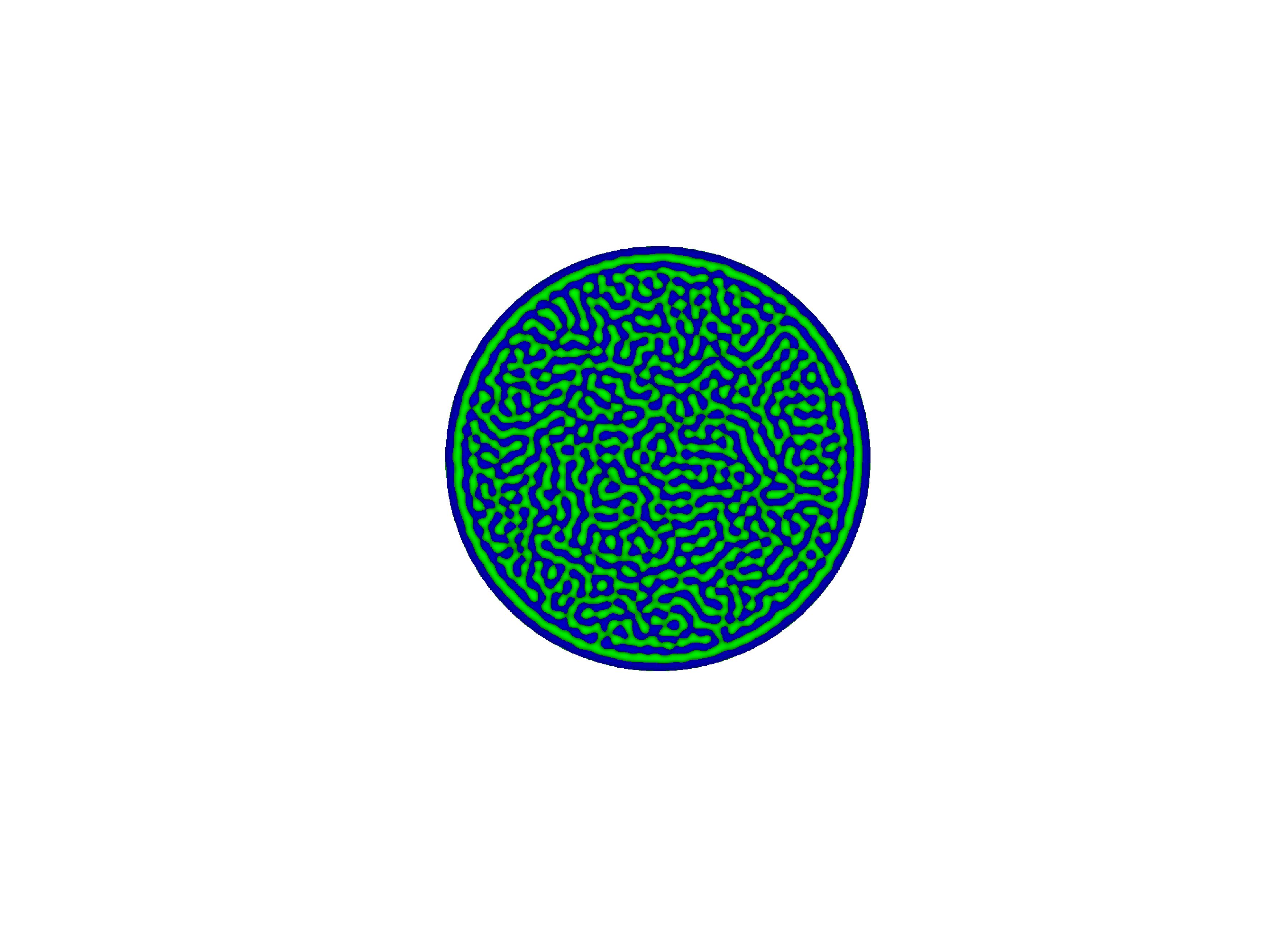}%
   \caption{$t=1$}
    \label{fig:fig1b}
\end{subfigure}%
\begin{subfigure}{0.085\textwidth}
\centering%
    \includegraphics[scale=0.03,trim={39cm 7cm 38cm 7cm},clip]{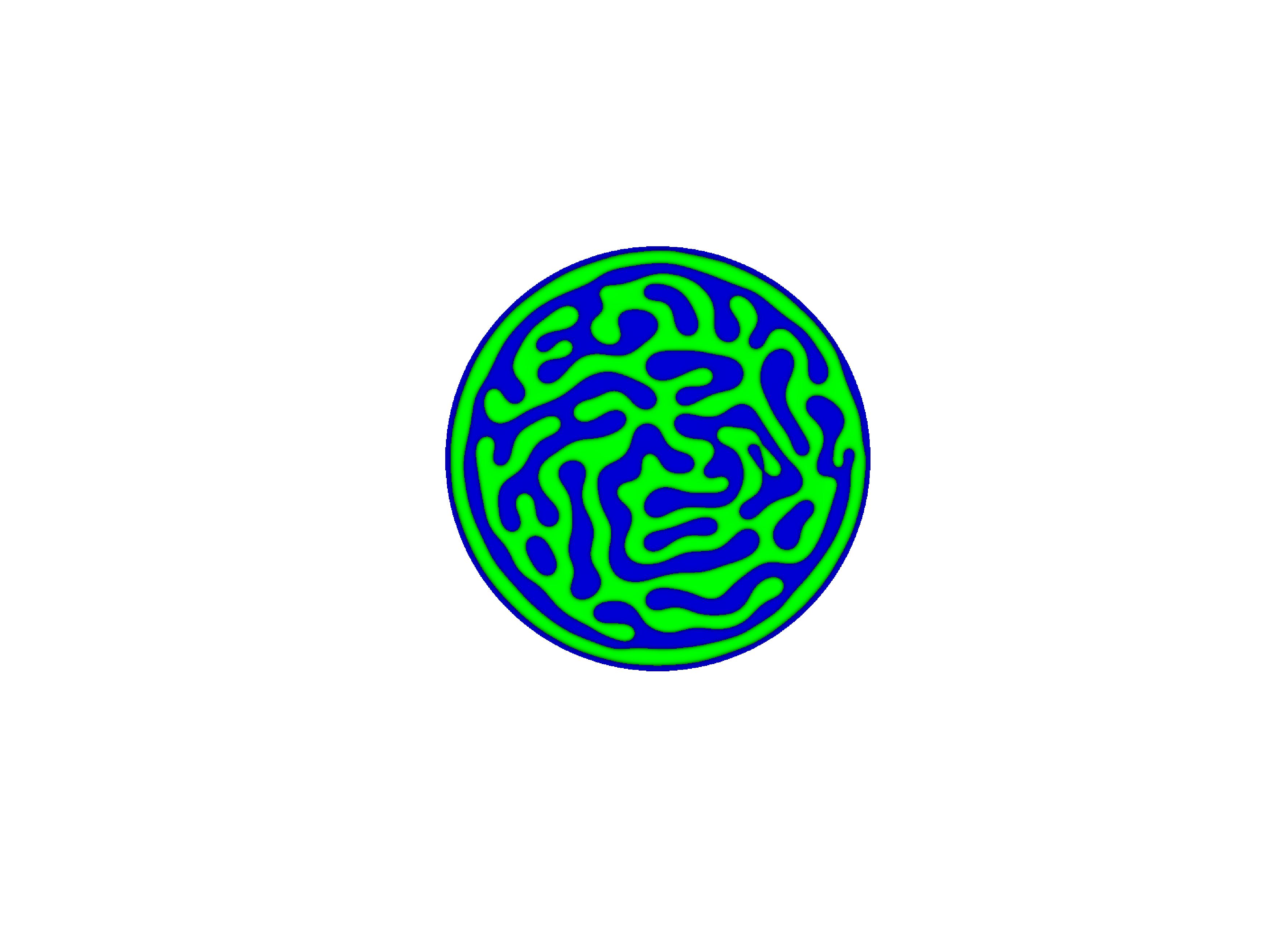}%
   \caption{$t=20$}
    \label{fig:fig1c}
\end{subfigure}%
\begin{subfigure}{0.085\textwidth}
\centering%
    \includegraphics[scale=0.03,trim={39cm 8cm 39cm 8cm},clip]{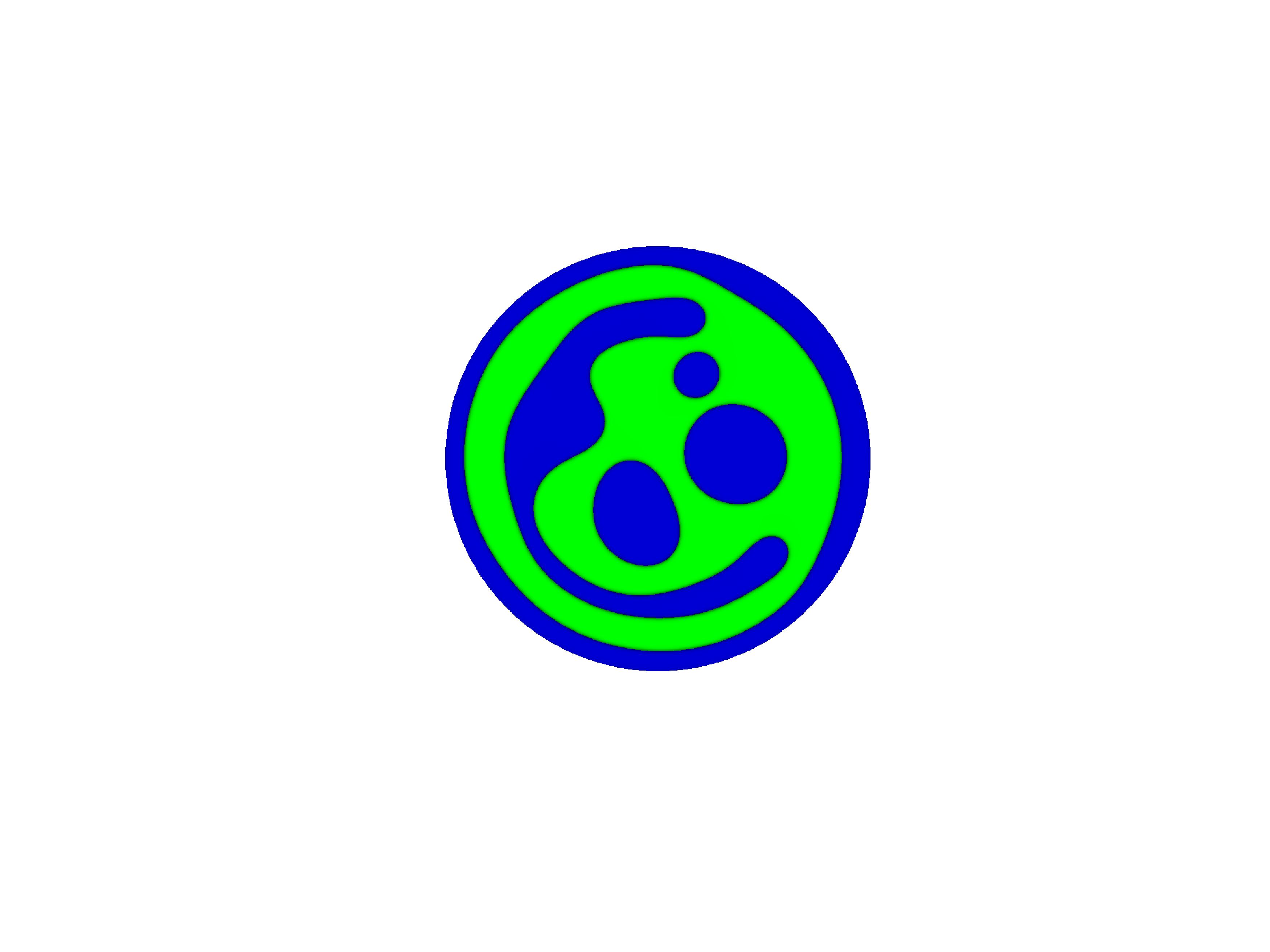}%
   \caption{$t=600$}
    \label{fig:fig1d}
\end{subfigure}%
\begin{subfigure}{0.085\textwidth}
\centering%
    \includegraphics[scale=0.03,trim={39cm 8cm 39cm 8cm},clip]{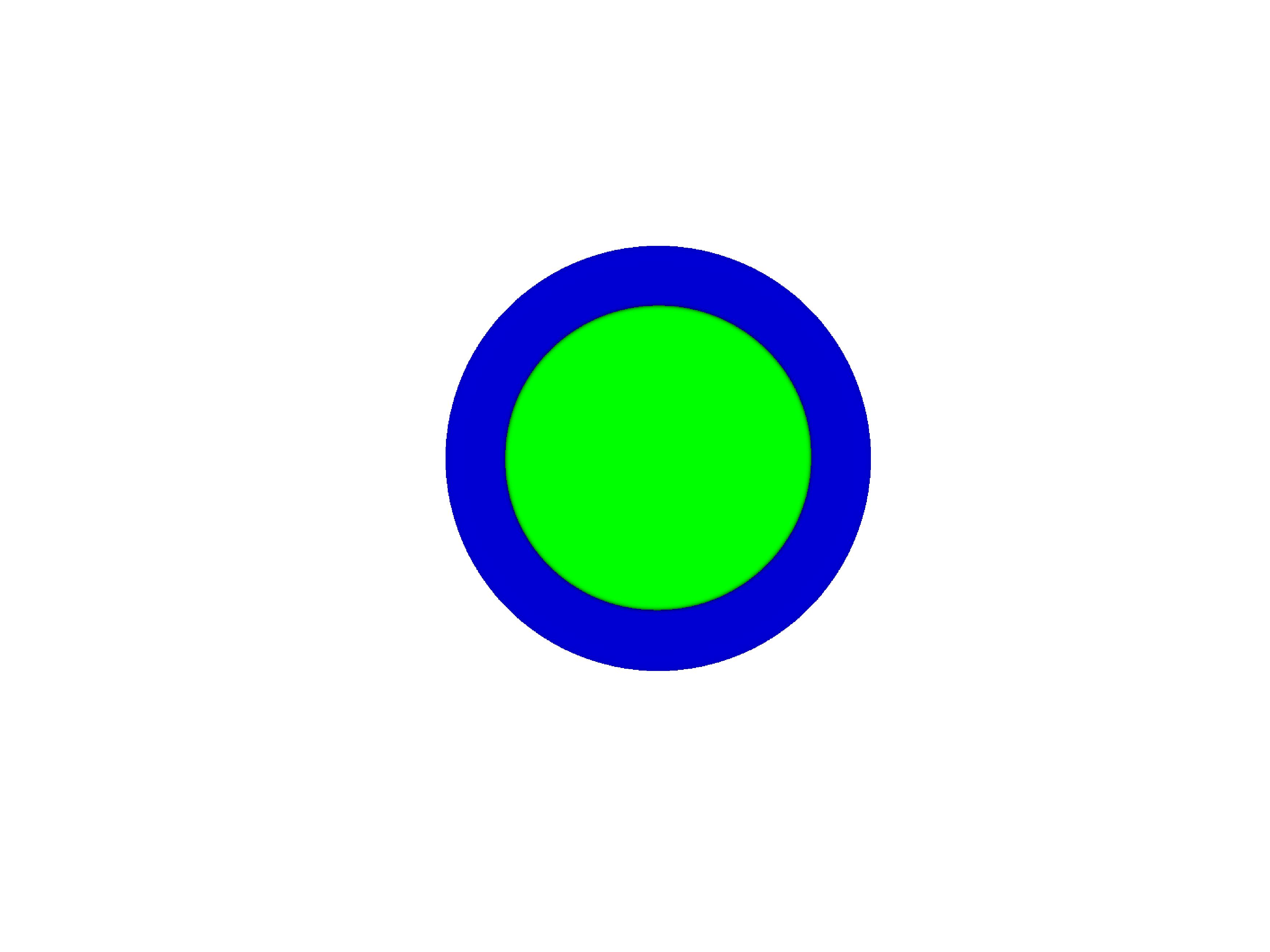}%
   \caption{$t=3500$}
    \label{fig:fig1e}
\end{subfigure}%

\begin{subfigure}{0.085\textwidth}
\centering%
    \includegraphics[scale=0.03,trim={39cm 7cm 39cm 7cm},clip]{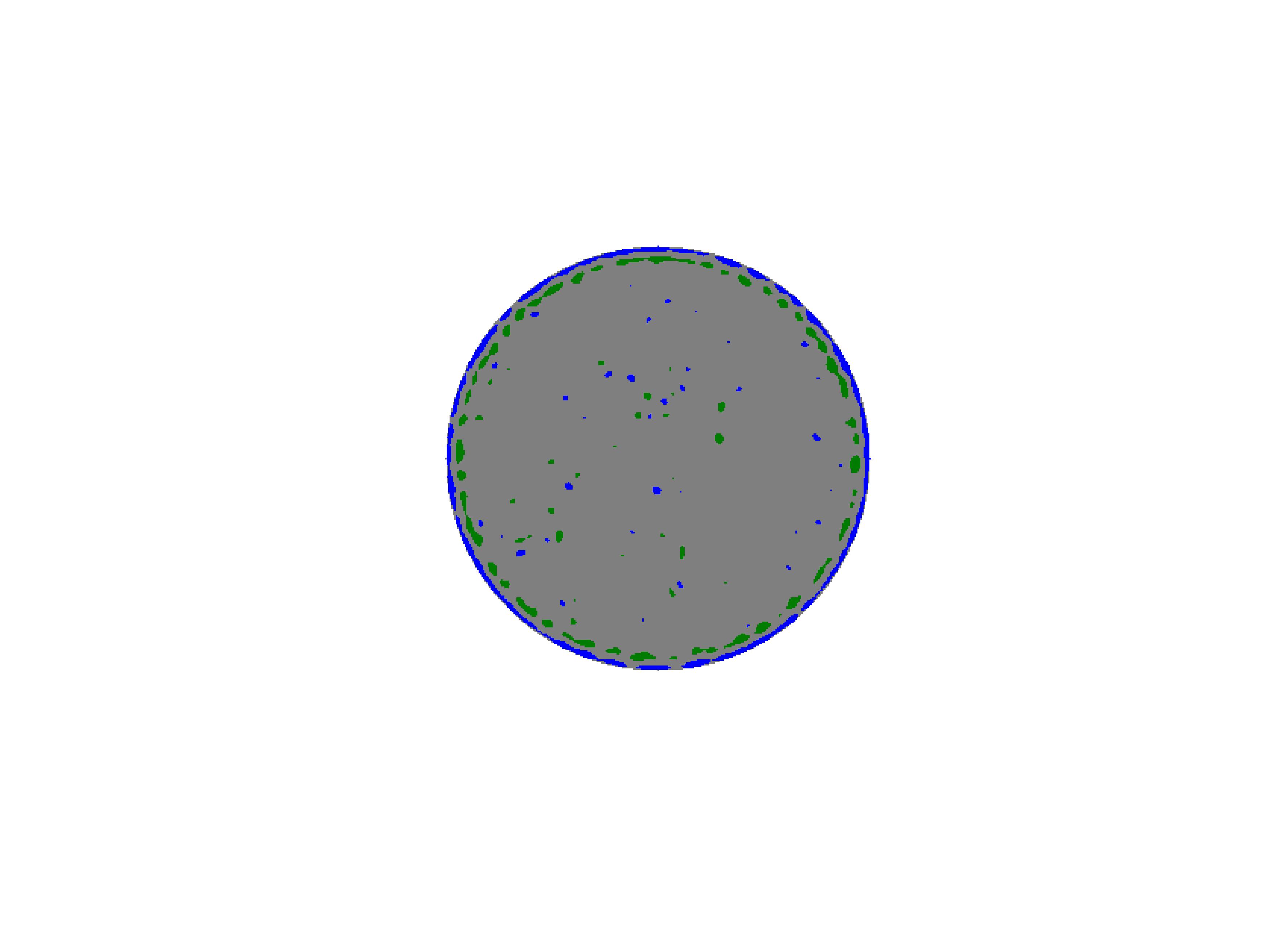}%
   \caption{$t=0.4$}
    \label{fig:fig1f}
\end{subfigure}%
\begin{subfigure}{0.085\textwidth}
\centering%
    \includegraphics[scale=0.03,trim={39cm 7cm 39cm 7cm},clip]{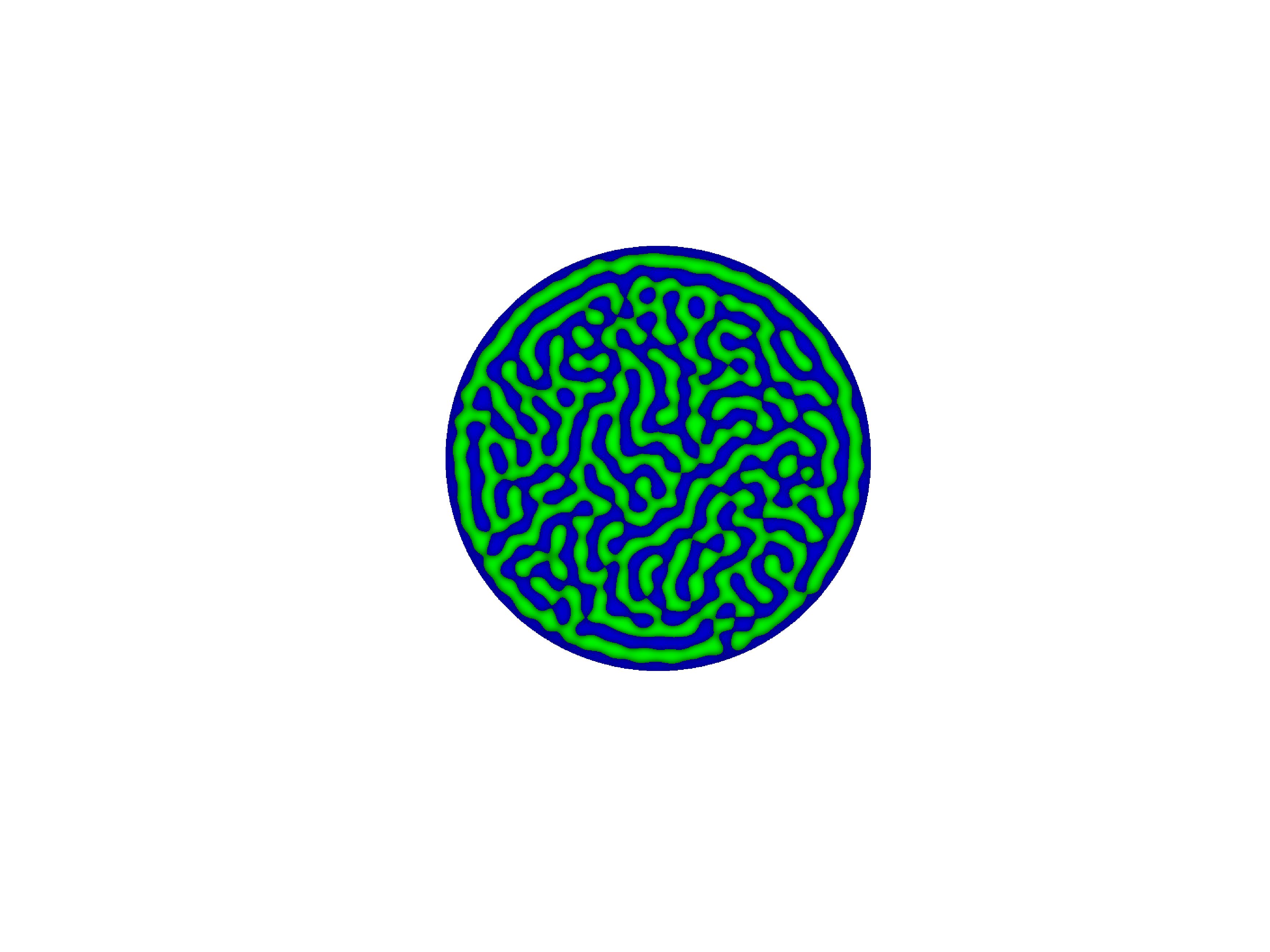}%
   \caption{$t=4$}
    \label{fig:fig1g}
\end{subfigure}%
\begin{subfigure}{0.085\textwidth}
\centering%
    \includegraphics[scale=0.03,trim={39cm 7cm 39cm 7cm},clip]{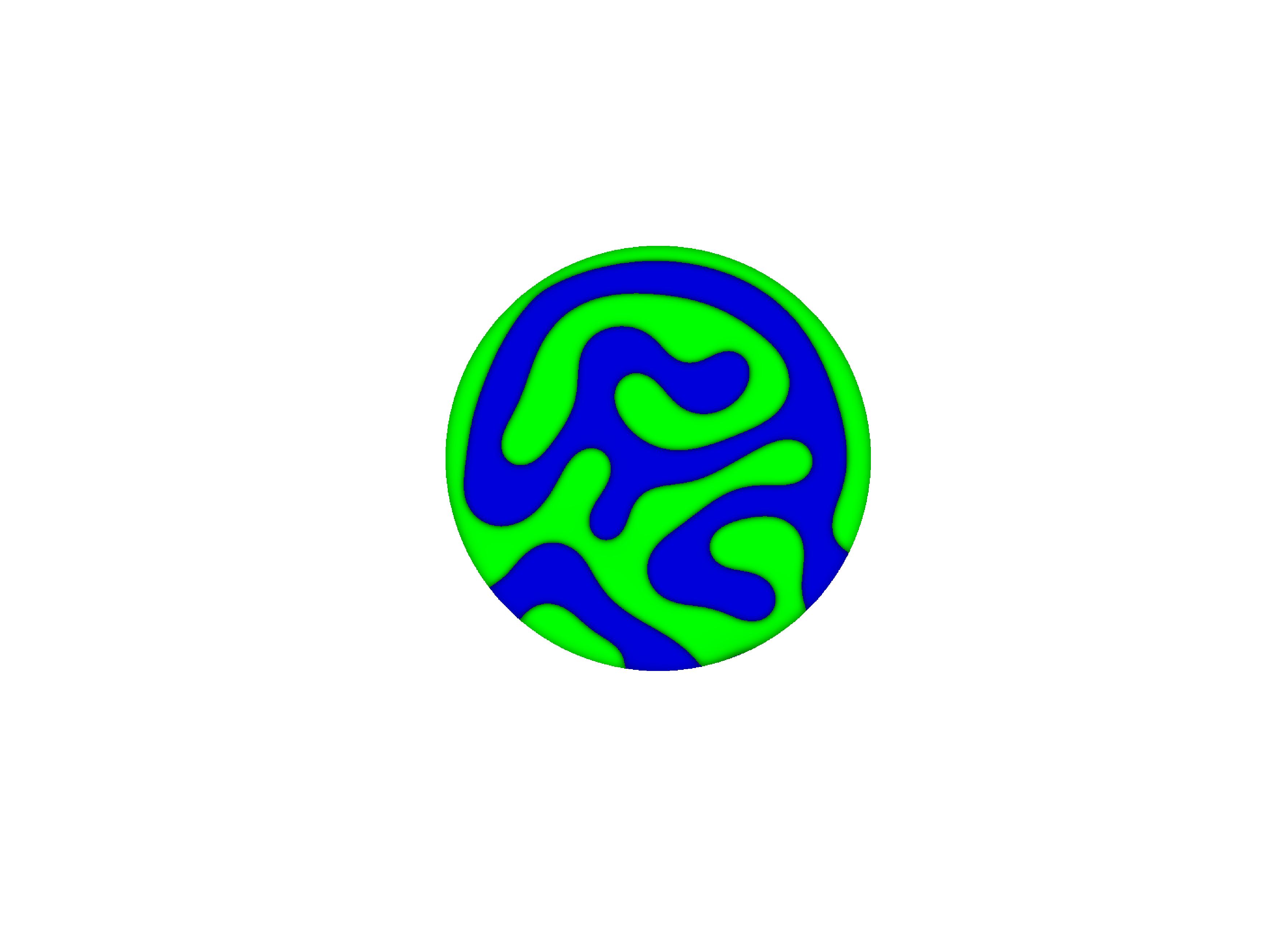}%
   \caption{$t=200$}
    \label{fig:fig1h}
\end{subfigure}%
\begin{subfigure}{0.085\textwidth}
\centering%
    \includegraphics[scale=0.03,trim={39cm 7cm 39cm 7cm},clip]{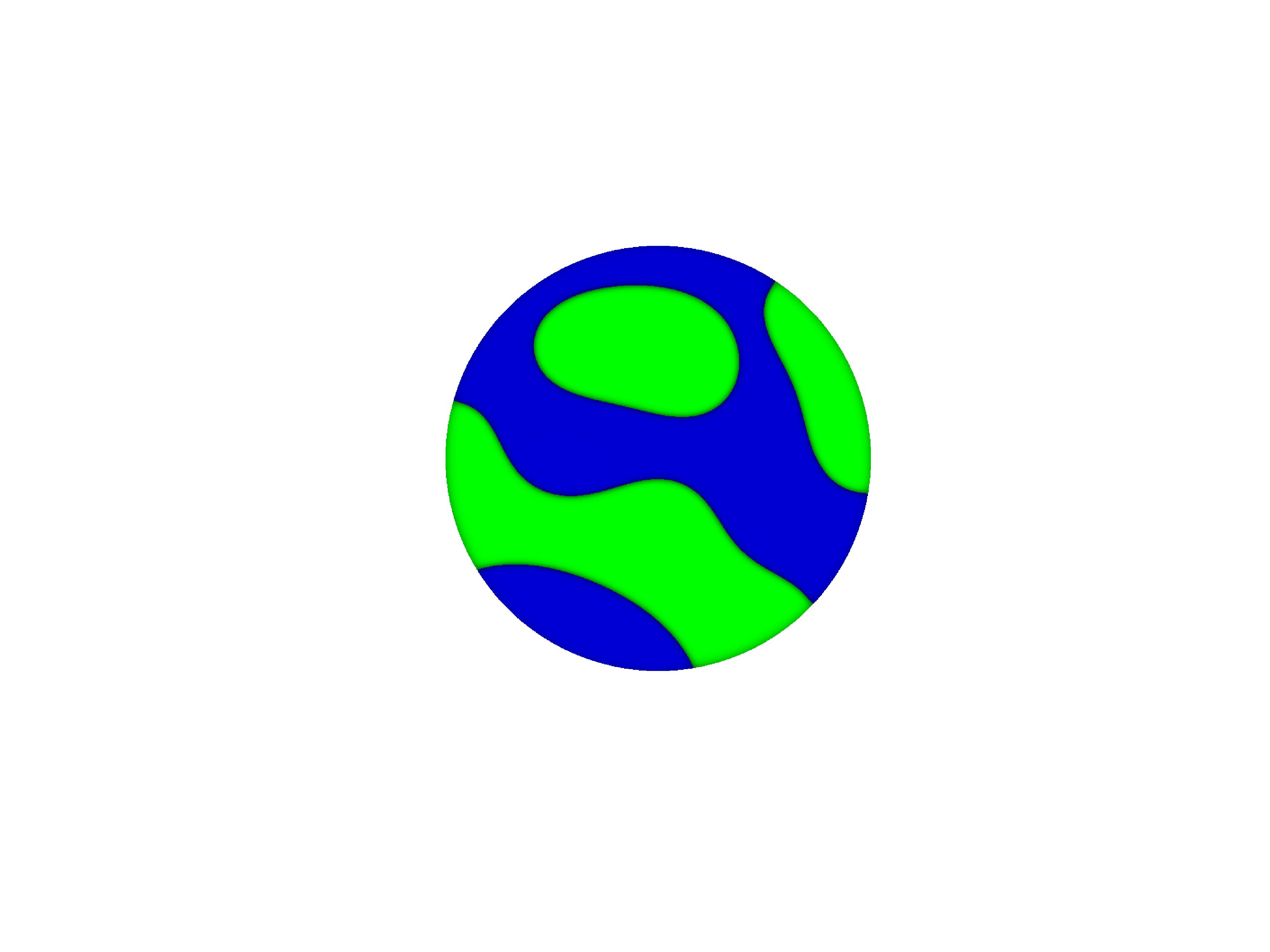}%
    \caption{$t=2000$}
    \label{fig:fig1i}
\end{subfigure}%
\begin{subfigure}{0.085\textwidth}
\centering%
    \includegraphics[scale=0.03,trim={39cm 7cm 39cm 7cm},clip]{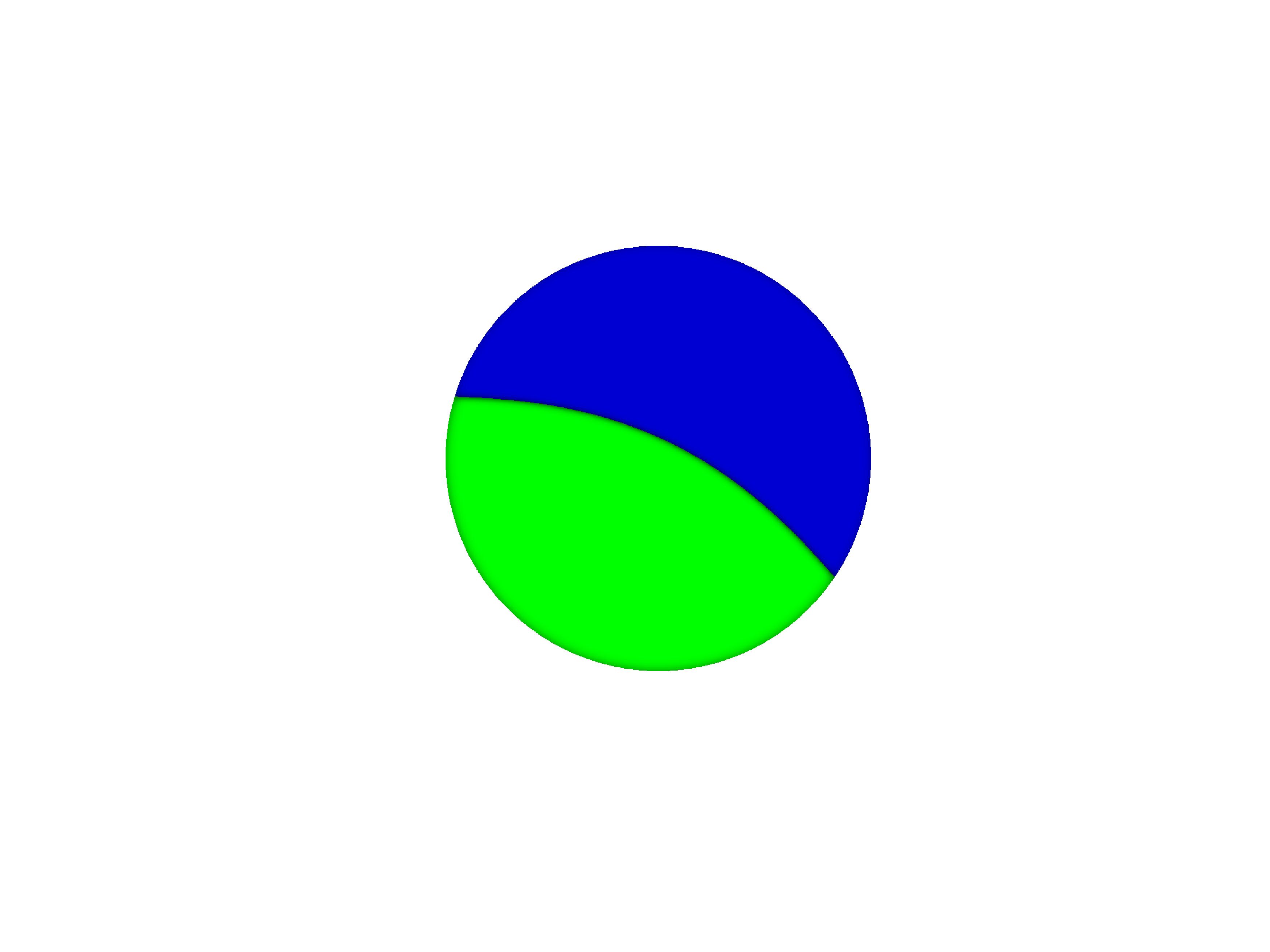}%
   \caption{$t=$3.9E4}
    \label{fig:fig1j}
\end{subfigure}%
\caption{Time snapshots of morphological evolution of stable CS (top row) and Janus (bottom row)
configurations. Non-dimensional times are indicated below each snapshot.}
\label{fig:fig1}
\end{figure}

\begin{figure}[thbp]
\centering%
\begin{subfigure}{0.085\textwidth}
\centering%
    \includegraphics[scale=0.03,trim={39cm 7cm 39cm 7cm},clip]{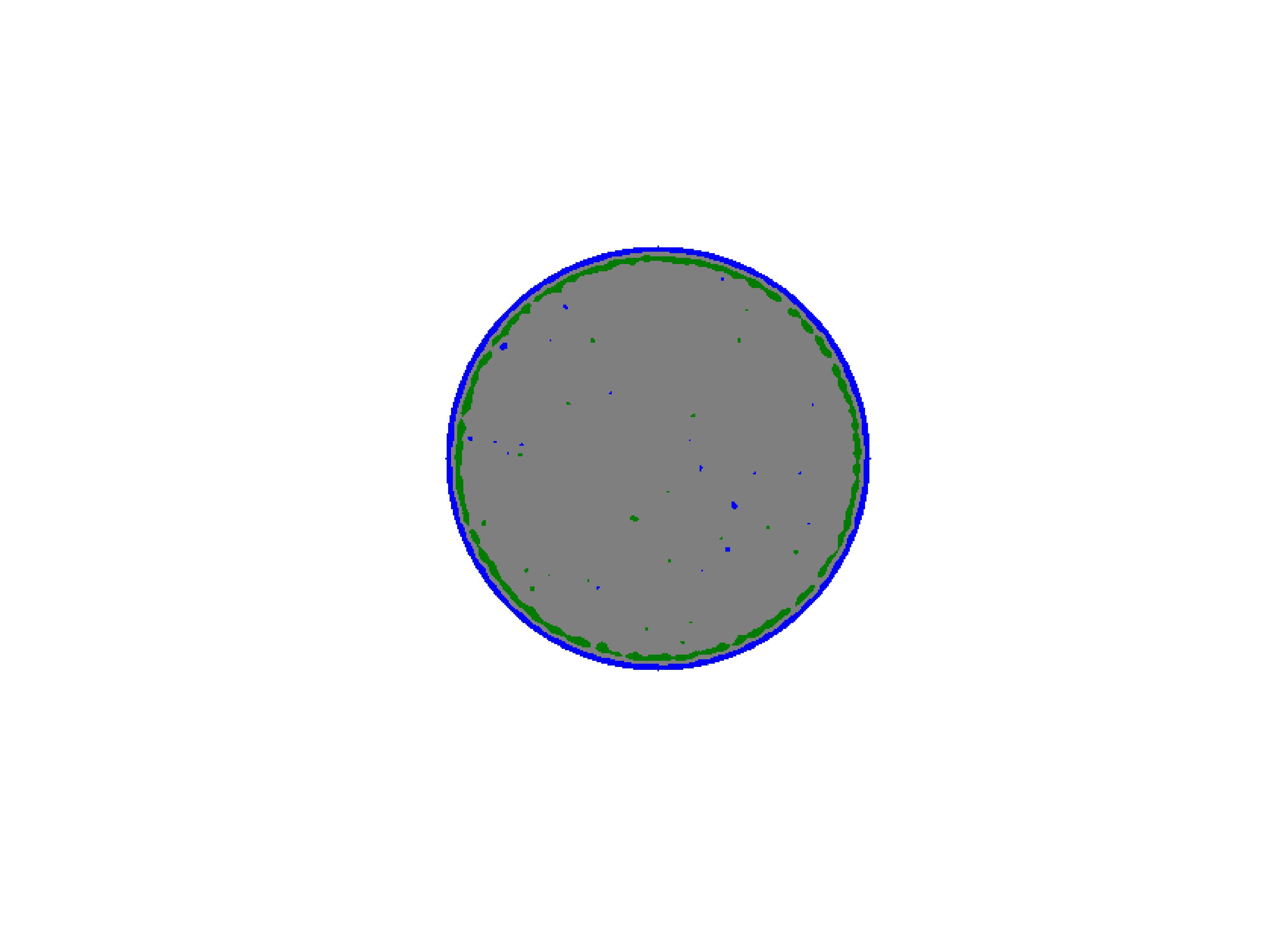}%
   \caption{$t=0.4$}
    \label{fig:fig2a}
\end{subfigure}%
\begin{subfigure}{0.085\textwidth}
\centering%
    \includegraphics[scale=0.03,trim={39cm 7cm 39cm 7cm},clip]{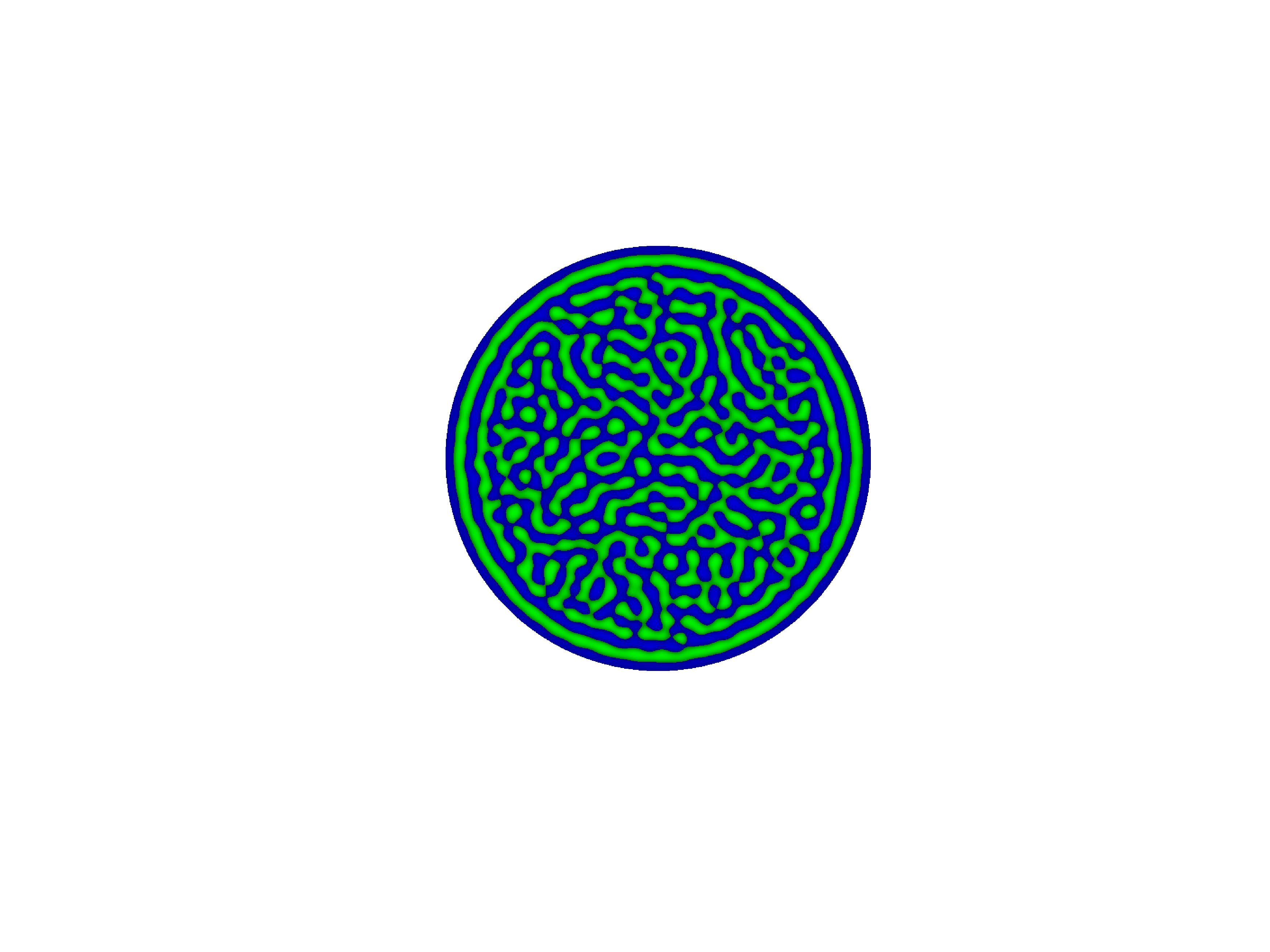}%
   \caption{$t=4$}
    \label{fig:fig2b}
\end{subfigure}%
\begin{subfigure}{0.085\textwidth}
\centering%
    \includegraphics[scale=0.03,trim={39cm 7cm 39cm 7cm},clip]{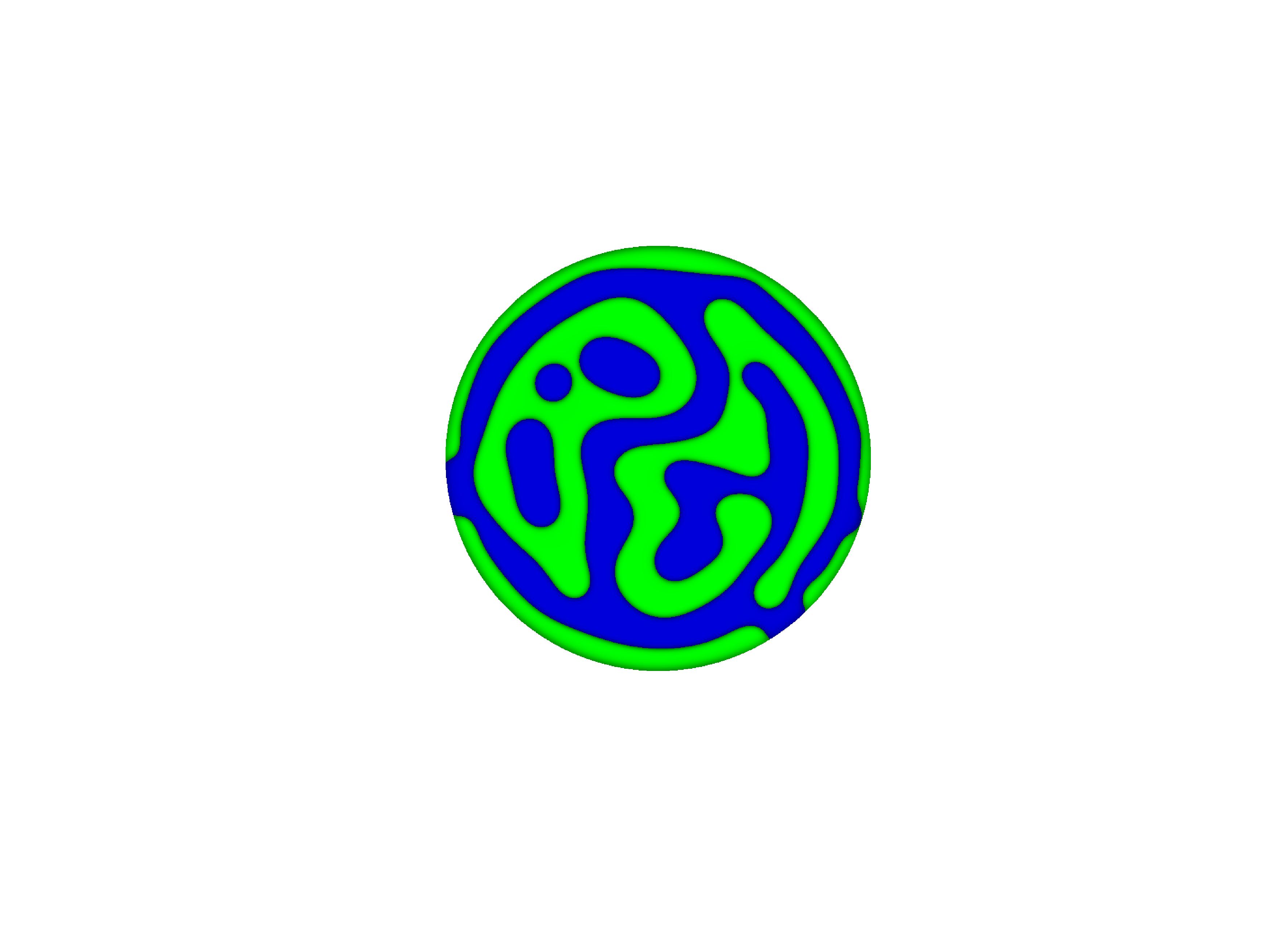}%
   \caption{$t=200$}
    \label{fig:fig2c}
\end{subfigure}%
\begin{subfigure}{0.085\textwidth}
\centering%
    \includegraphics[scale=0.03,trim={39cm 7cm 39cm 7cm},clip]{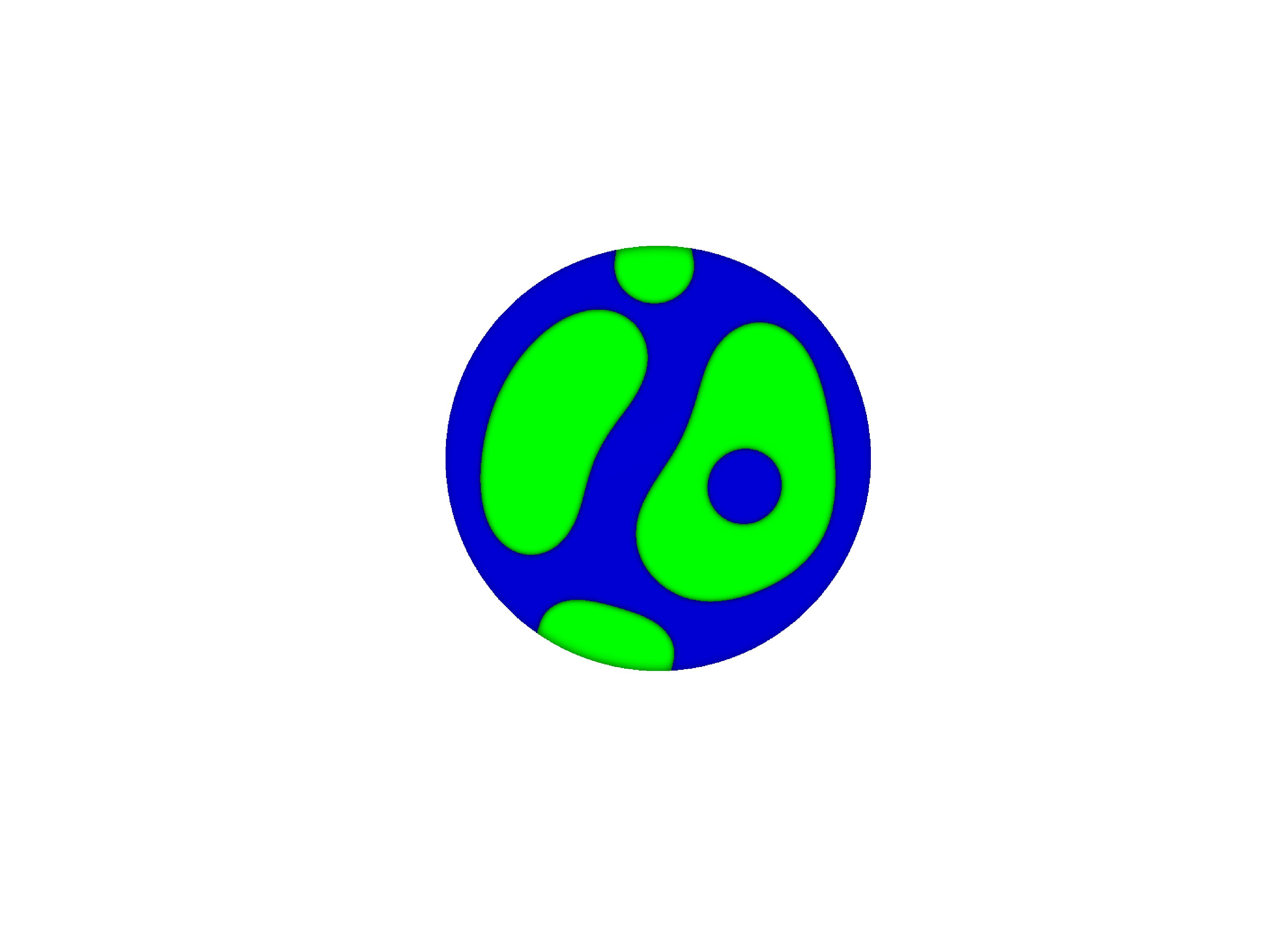}%
   \caption{$t=2000$}
    \label{fig:fig2d}
\end{subfigure}%
\begin{subfigure}{0.085\textwidth}
\centering%
    \includegraphics[scale=0.03,trim={39cm 7cm 39cm 7cm},clip]{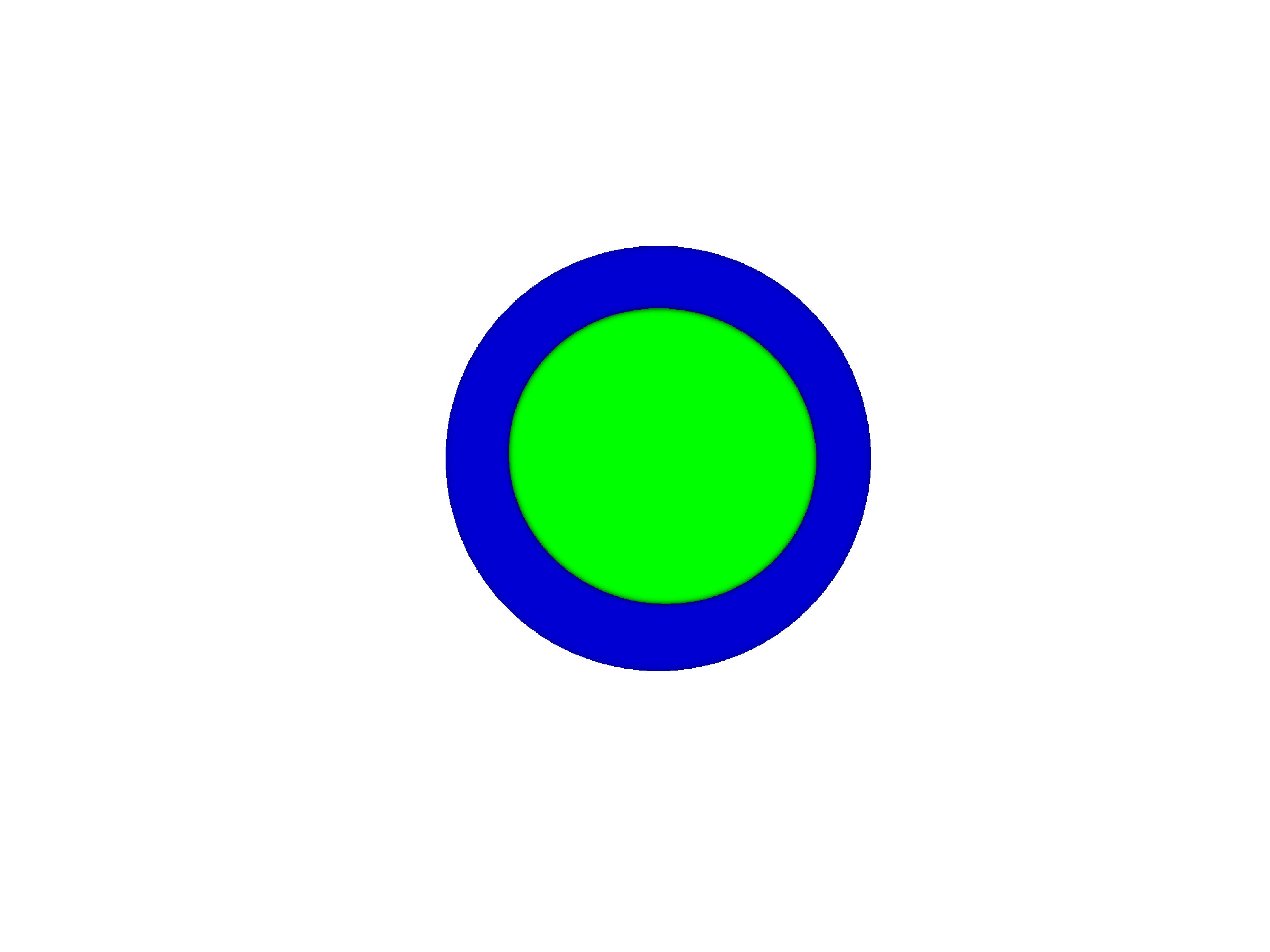}%
   \caption{$t=6000$}
    \label{fig:fig2e}
\end{subfigure}%

\begin{subfigure}{0.085\textwidth}
\centering%
    \includegraphics[scale=0.03,trim={39cm 7cm 39cm 7cm},clip]{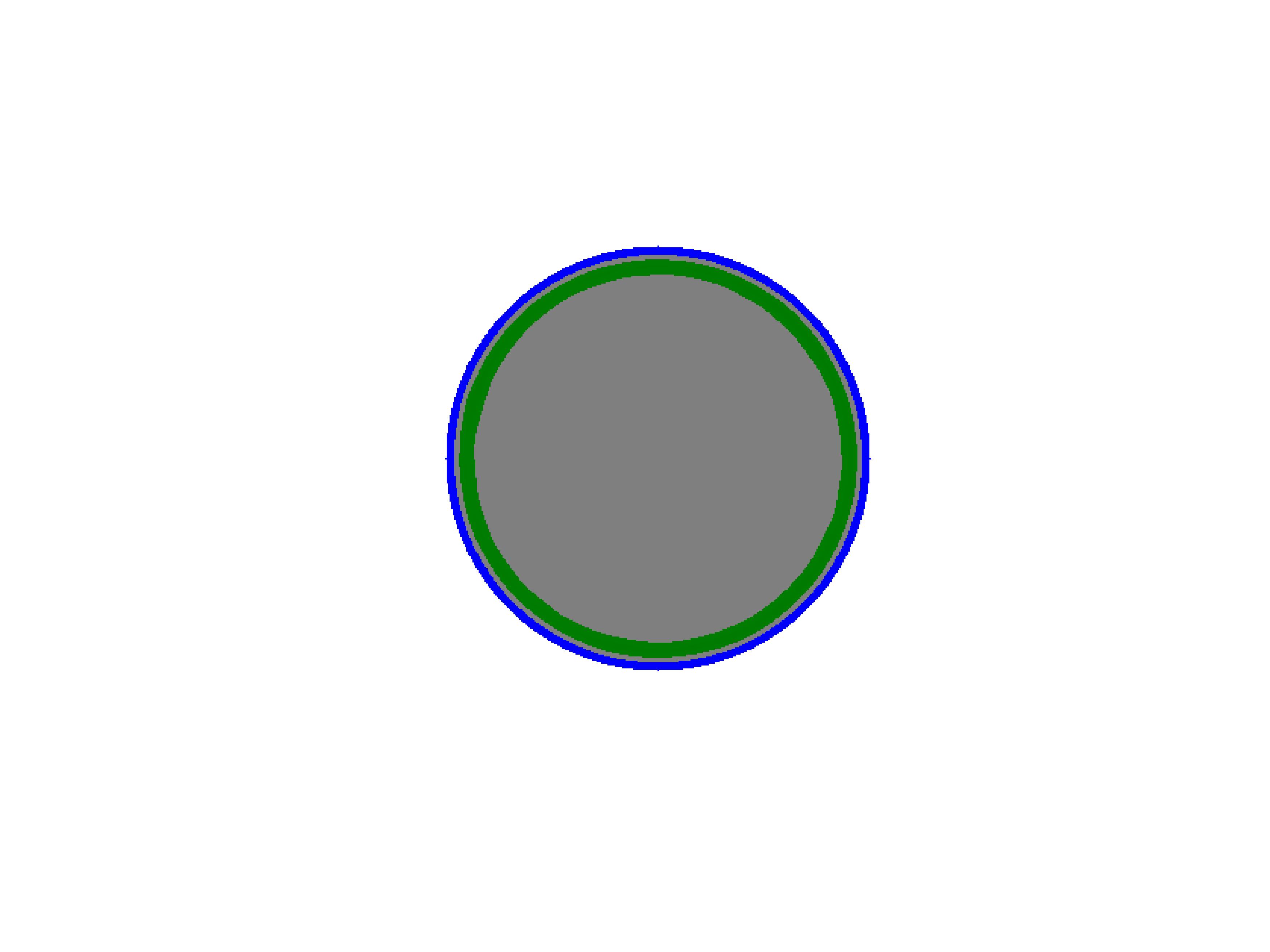}%
   \caption{$t=2$}
    \label{fig:fig2f}
\end{subfigure}%
\begin{subfigure}{0.085\textwidth}
\centering%
    \includegraphics[scale=0.03,trim={39cm 7cm 39cm 7cm},clip]{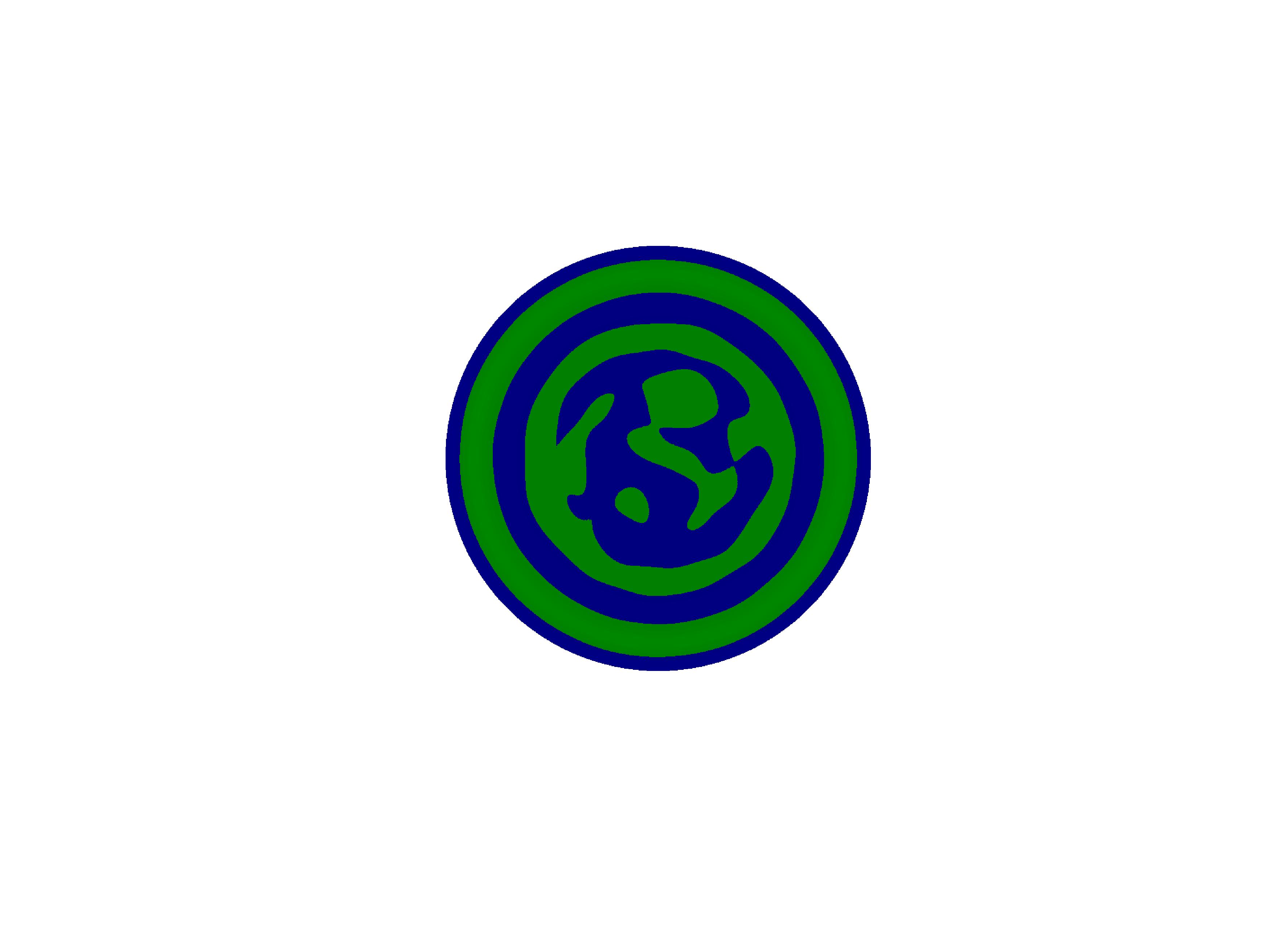}
   \caption{$t=12$}
    \label{fig:fig2g}
\end{subfigure}%
\begin{subfigure}{0.085\textwidth}
\centering%
    \includegraphics[scale=0.03,trim={39cm 7cm 39cm 7cm},clip]{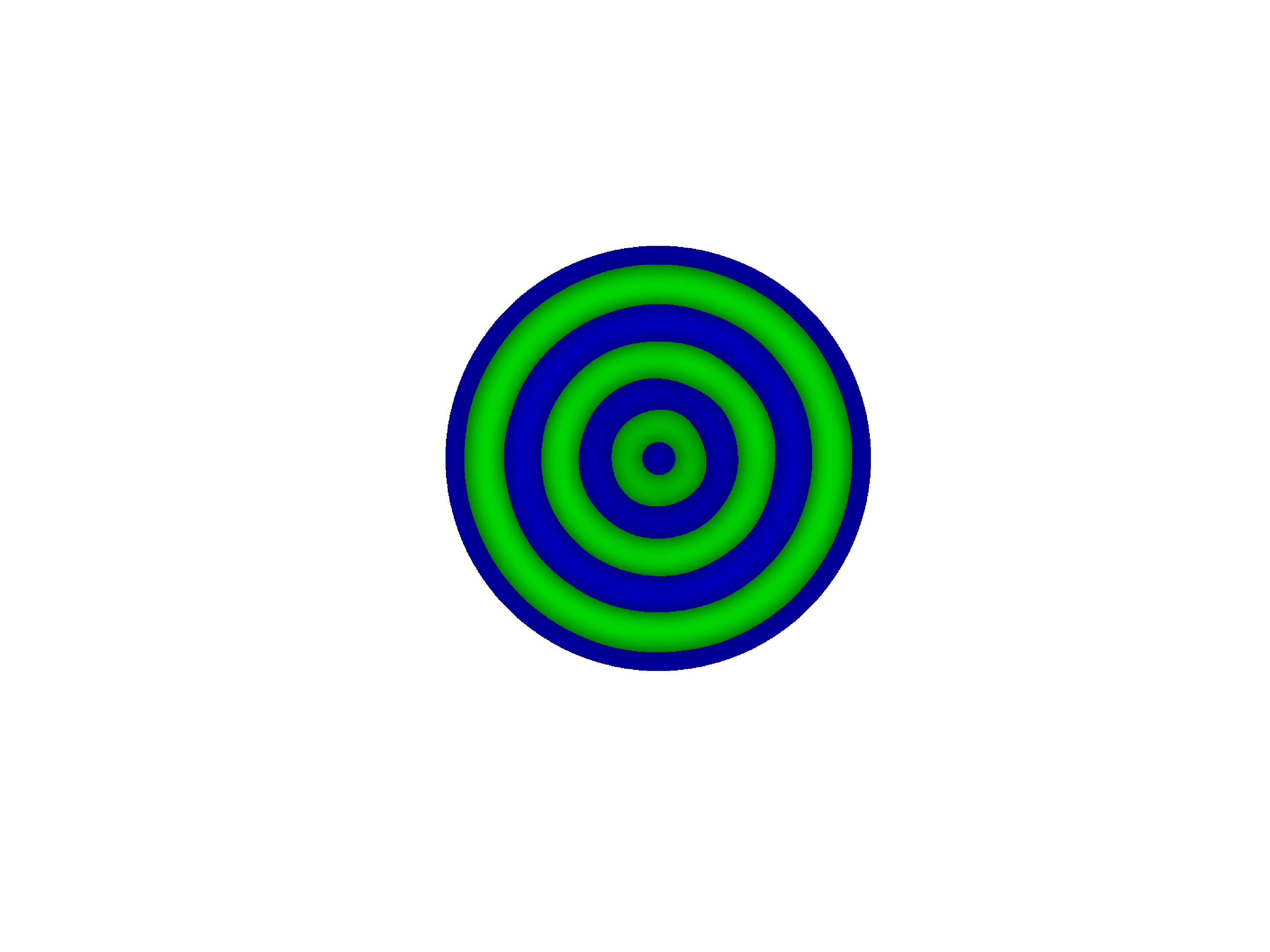}
   \caption{$t=80$}
    \label{fig:fig2h}
\end{subfigure}%
\begin{subfigure}{0.085\textwidth}
\centering%
    \includegraphics[scale=0.03,trim={39cm 7cm 39cm 7cm},clip]{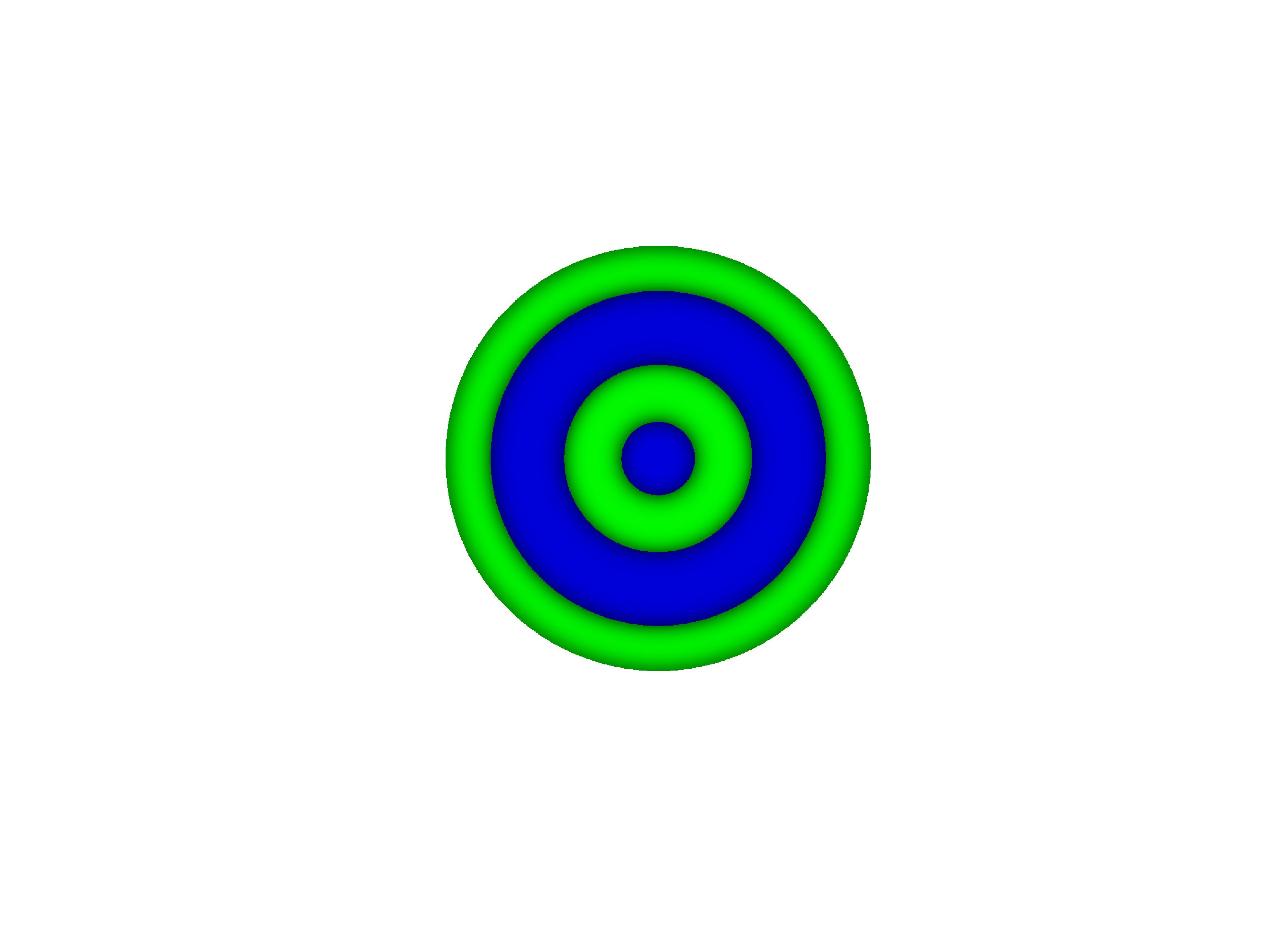}
   \caption{$t=600$}
    \label{fig:fig2i}
\end{subfigure}%
\begin{subfigure}{0.085\textwidth}
\centering%
    \includegraphics[scale=0.03,trim={39cm 7cm 39cm 7cm},clip]{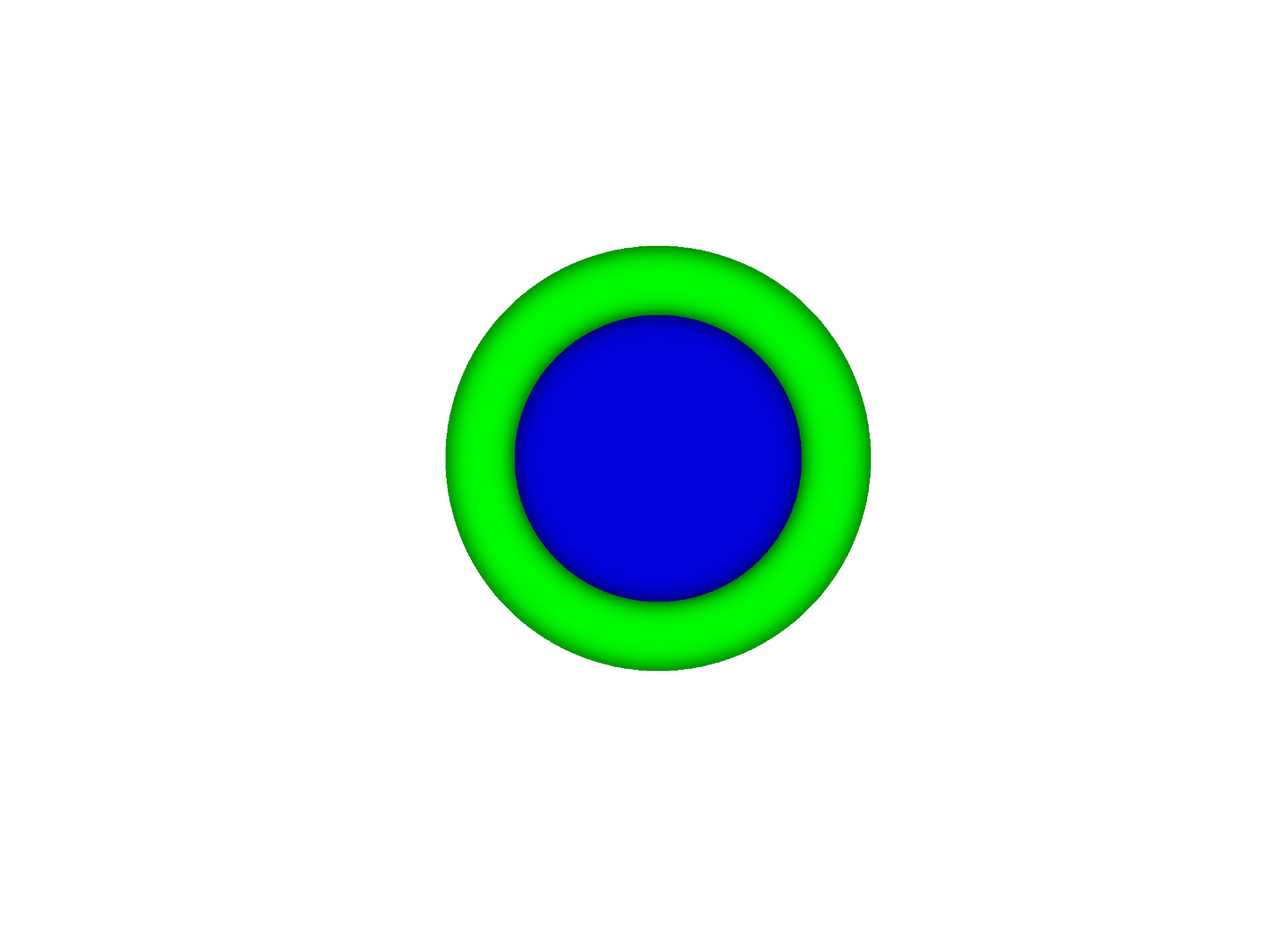}
   \caption{$t=19000$}
    \label{fig:fig2j}
\end{subfigure}%

\begin{subfigure}{0.085\textwidth}
\centering%
    \includegraphics[scale=0.03,trim={39cm 7cm 39cm 7cm},clip]{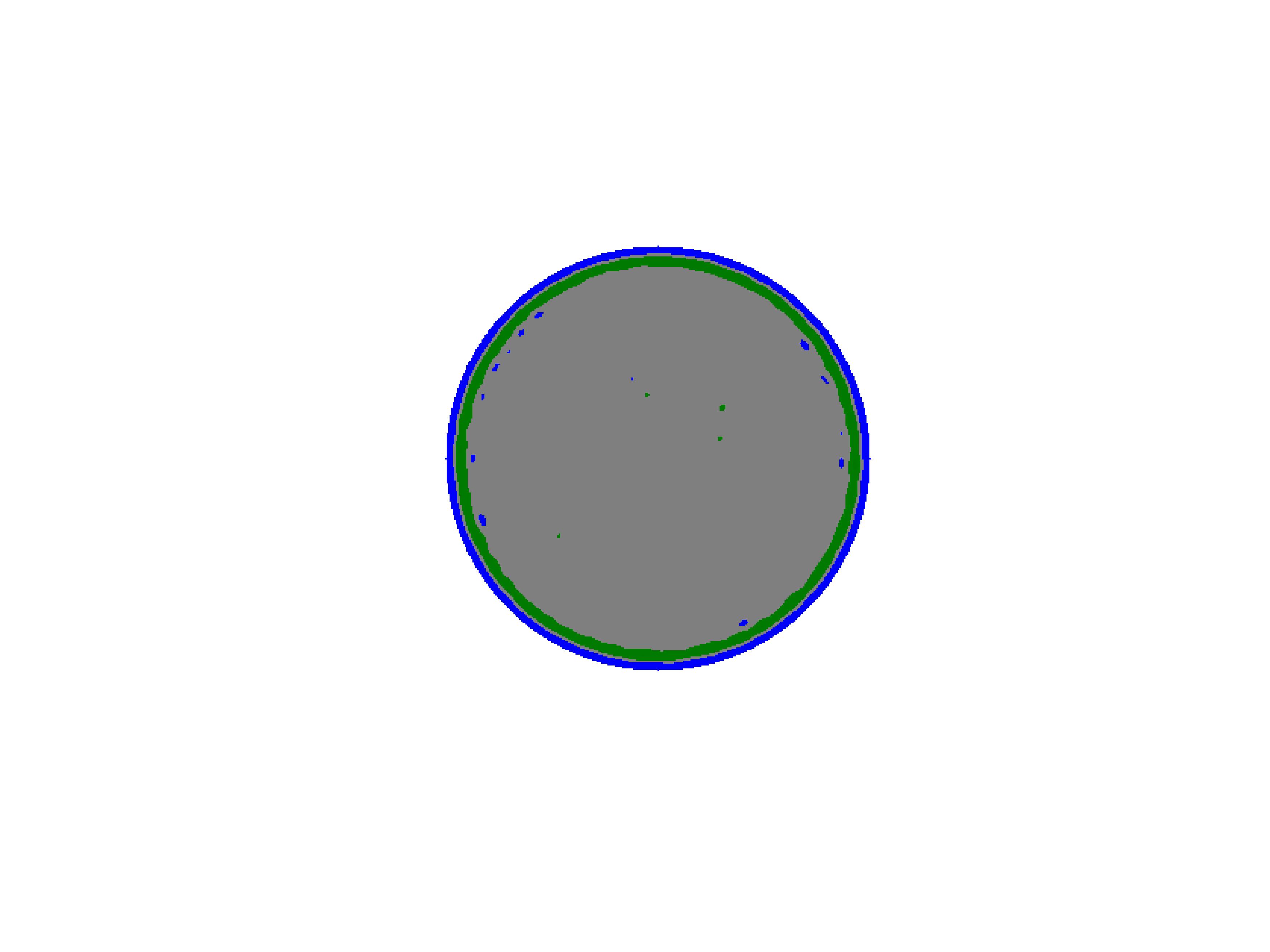}%
   \caption{$t=1$}
    \label{fig:fig2k}
\end{subfigure}%
\begin{subfigure}{0.085\textwidth}
\centering%
    \includegraphics[scale=0.03,trim={39cm 7cm 39cm 7cm},clip]{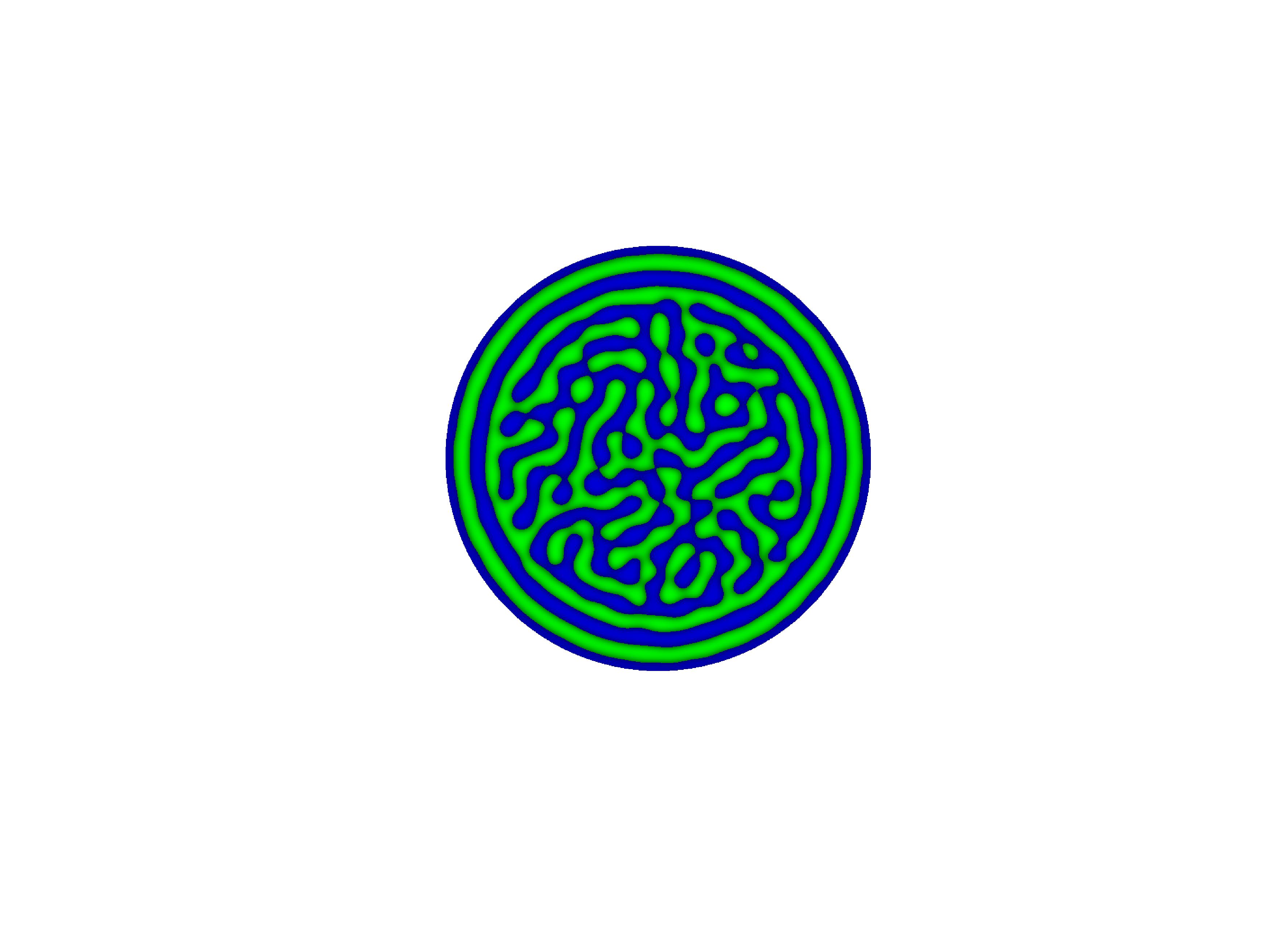}%
   \caption{$t=10$}
    \label{fig:fig2l}
\end{subfigure}%
\begin{subfigure}{0.085\textwidth}
\centering%
    \includegraphics[scale=0.03,trim={39cm 7cm 39cm 7cm},clip]{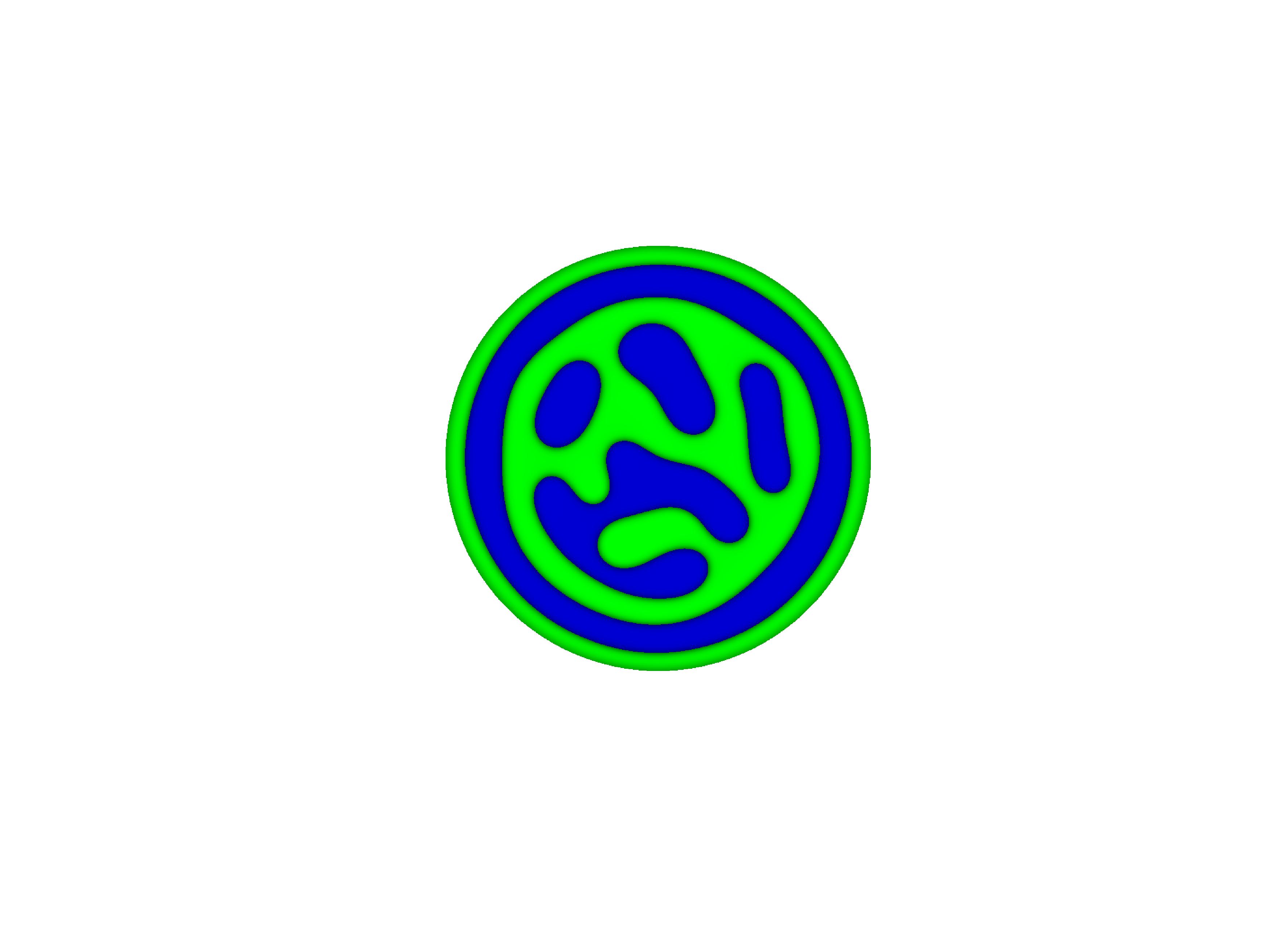}%
   \caption{$t=200$}
    \label{fig:fig2m}
\end{subfigure}%
\begin{subfigure}{0.085\textwidth}
\centering%
    \includegraphics[scale=0.03,trim={39cm 7cm 39cm 7cm},clip]{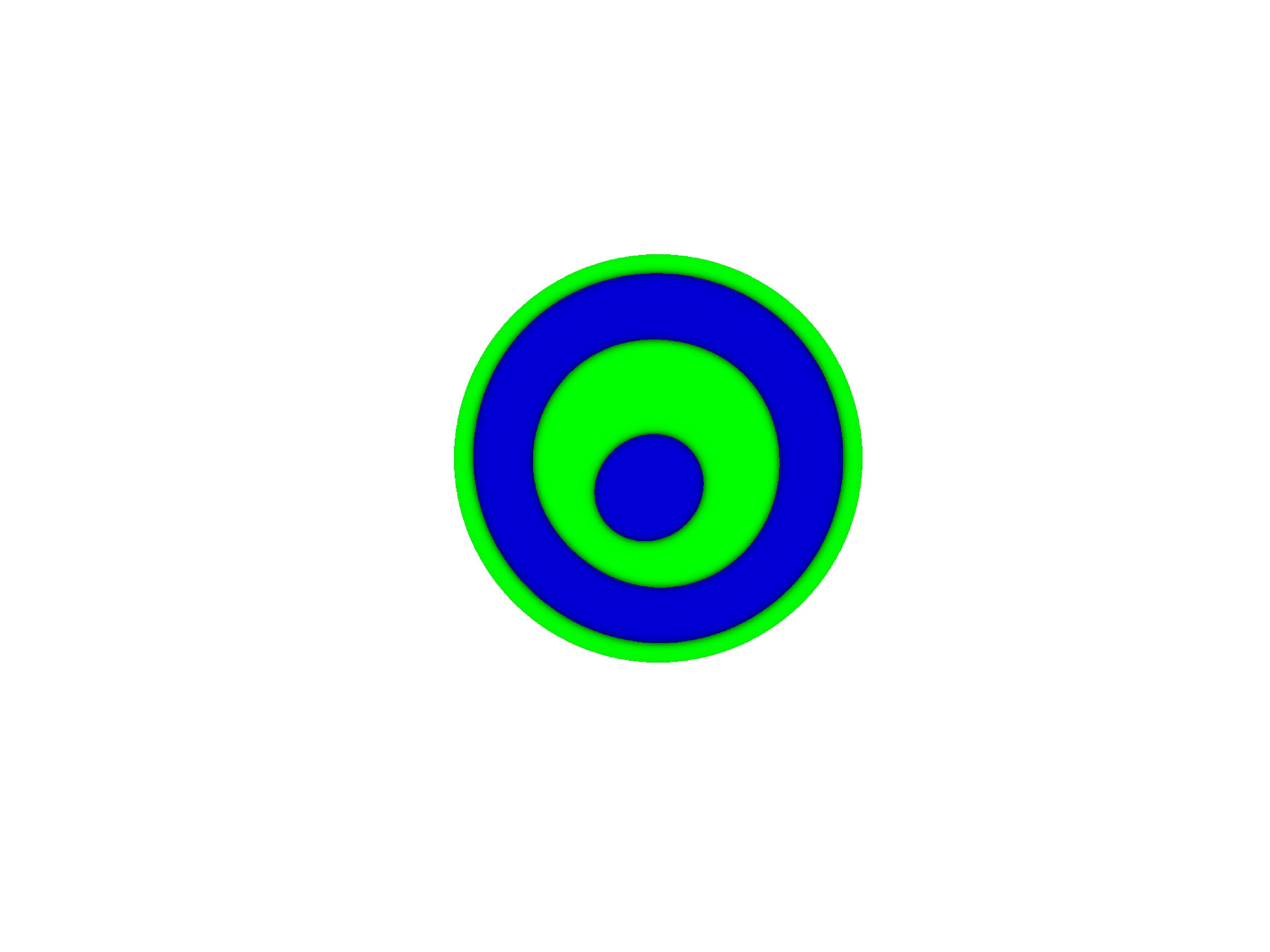}%
   \caption{$t=2000$}
    \label{fig:fig2n}
\end{subfigure}%
\begin{subfigure}{0.085\textwidth}
\centering%
    \includegraphics[scale=0.03,trim={39cm 7cm 39cm 7cm},clip]{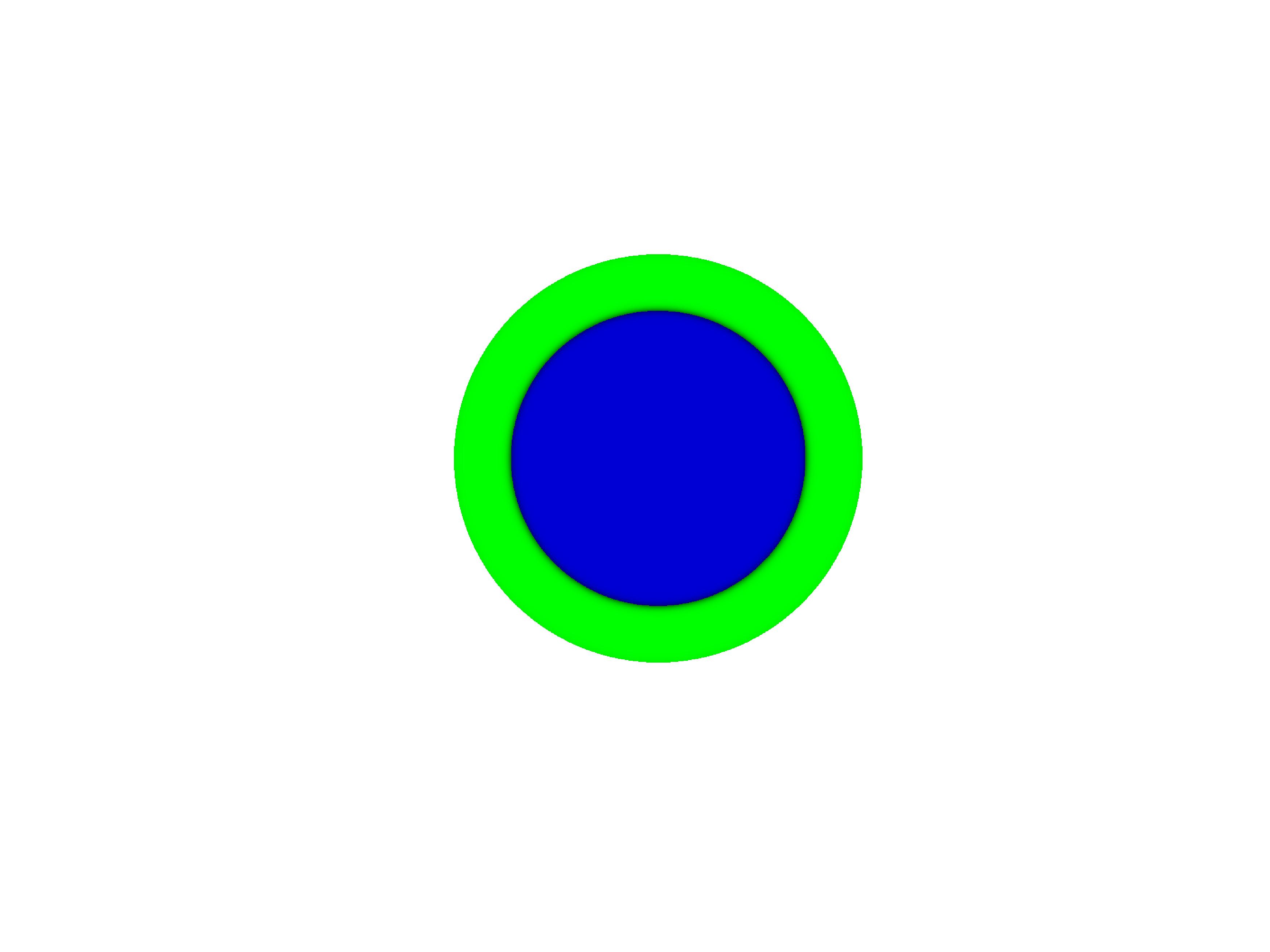}%
   \caption{$t=55000$}
    \label{fig:fig2o}
\end{subfigure}%
\caption{Time snapshots of metastable CS (top row) and ICS (middle and bottom rows) evolution.
Non-dimensional times are indicated below for each snapshot.}
\label{fig:fig2}
\end{figure}

All simulations begin with an initially homogeneous, axisymmetric $\beta$ particle with composition
$c=0.5$ quenched inside the miscibility gap. A small initial noise ($\pm{1}\%$) mimicking thermal
fluctuations is applied to trigger bulk SD.
Figs.~\ref{fig:fig1} and~\ref{fig:fig2} present time snapshots of evolution of stable
and metastable morphologies corresponding to representative parameter sets listed in
Table~\ref{tab:parameters}. In these figures, green color represents the solute-poor $\beta_1$ phase,
blue the solute-rich $\beta_2$ phase, and grey the undecomposed $\beta$ with $c=0.5$. 

\begin{table}[htbp]
 \centering
 \small\caption{Variation of surface and interfacial energies and contact angle $\theta$ with
 model parameters. Figure numbers refer to snapshots of morphological evolution for the
 corresponding set of parameters.}
 \label{tab:parameters}
 \begin{tabular*}{0.5\textwidth}{@{\extracolsep{\fill}}*{9}l}
 \hline
  Set& $f_0^p$ & $\kappa_c$ & $\omega_0$ & $\sigma_1$ & $\sigma_2$ & $\sigma_{12}$ & $\theta$ & Figure no.\\
 \hline
   
   1 & 8 & 1 & 12 & 4.19 & 3.09 & 0.94 & wetting & \ref{fig:fig1a}-\ref{fig:fig1e}\\
   2 & 6 & 2 & 3.75 & 1.77 & 1.44 & 1.15 & \ang{74} & \ref{fig:fig1f}-\ref{fig:fig1j}  \\
   3 & 4 & 1 & 6 & 2.21 & 1.76 & 0.67 & \ang{48}  & \ref{fig:fig2a}-\ref{fig:fig2e}\\
   4 & 2 & 8 & 6 & 2.24 & 1.86 & 1.33 & \ang{73.5} & \ref{fig:fig2f}-\ref{fig:fig2j}\\
   5 & 4 & 2 & 5 & 1.98 & 1.60 & 0.94 & \ang{66} & \ref{fig:fig2k}-\ref{fig:fig2o}\\
 \hline

 \end{tabular*}
\end{table}

In all cases, SDSD precedes bulk-SD and forms alternate solute-rich and solute-lean rings which
grow inward. When bulk SD starts in the interior of the particle, it creates intertwined
compositionally modulated domains that interact with the rings growing from the surface to the
center. Depending on the chosen parameter set, domain coarsening proceeds along
different pathways to produce the final BNP configuration. A Janus structure is formed when
coarsening disrupts the continuity of ring-like structures at the surface, and CS or ICS
results otherwise.

Figs.~\ref{fig:fig1}(a-e) is an example of a typical evolution pattern leading to a stable CS morphology
(set 1 of Table 1). It occurs irrespective of the driving force if the spontaneous wetting condition
is satisfied (\emph{i.e.}, $\sigma_1 - \sigma_2 \geq \sigma_{12}$). For large $\theta$,
if bulk driving force is sufficiently high to break the outermost layer, we obtain stable Janus
morphology. A typical sequence of its formation is presented in Figs.~\ref{fig:fig1}(f-j) which 
corresponds to parameter set 2 of Table~\ref{tab:parameters}.

Parameter sets 3-5 of Table 1 correspond to intermediate values of driving force 
(which scales with $f_0^p$) and $\theta$ which result in metastable configurations
shown in Figure~\ref{fig:fig2}. For example, Figs.~\ref{fig:fig2}(a-e) shows how a
CS morphology can develop for small non-zero $\theta$'s if the driving force for
bulk spinodal is also small. Thus, the spontaneous wetting condition is only a
sufficient condition -- stable CS forms when it is satisfied, but a metastable CS
can develop even when it is not.

On the other hand, one obtains a metastable ICS morphology when the driving force is
small but $\theta$ is large, as demonstrated by the typical evolution patterns in
Figs.~\ref{fig:fig2}(f-j) and Figs.~\ref{fig:fig2}(k-o). When surface effects dominate over the
bulk, an onion-like ring structure develops initially (Figs.~\ref{fig:fig2a}-\ref{fig:fig2b}),
which coarsens subsequently (Figs.~\ref{fig:fig2}(a-e)) to form a metastable ICS 
morphology. It is also possible that we get metastable ICS/CS even in the presence of bulk
spinodal. As shown in Figs.~\ref{fig:fig2}(f-o), the bulk driving force in this case
is not sufficiently high enough to achieve a stable Janus configuration, and domain coarsening
leads to an ICS configuration. Between the two evolutionary paths to ICS, the latter is 
observed when the driving force is higher.

\begin{figure}[thbp]
\centering%
\begin{subfigure}{0.085\textwidth}
\centering%
    \includegraphics[scale=0.03,trim={39cm 7cm 39cm 7cm},clip]{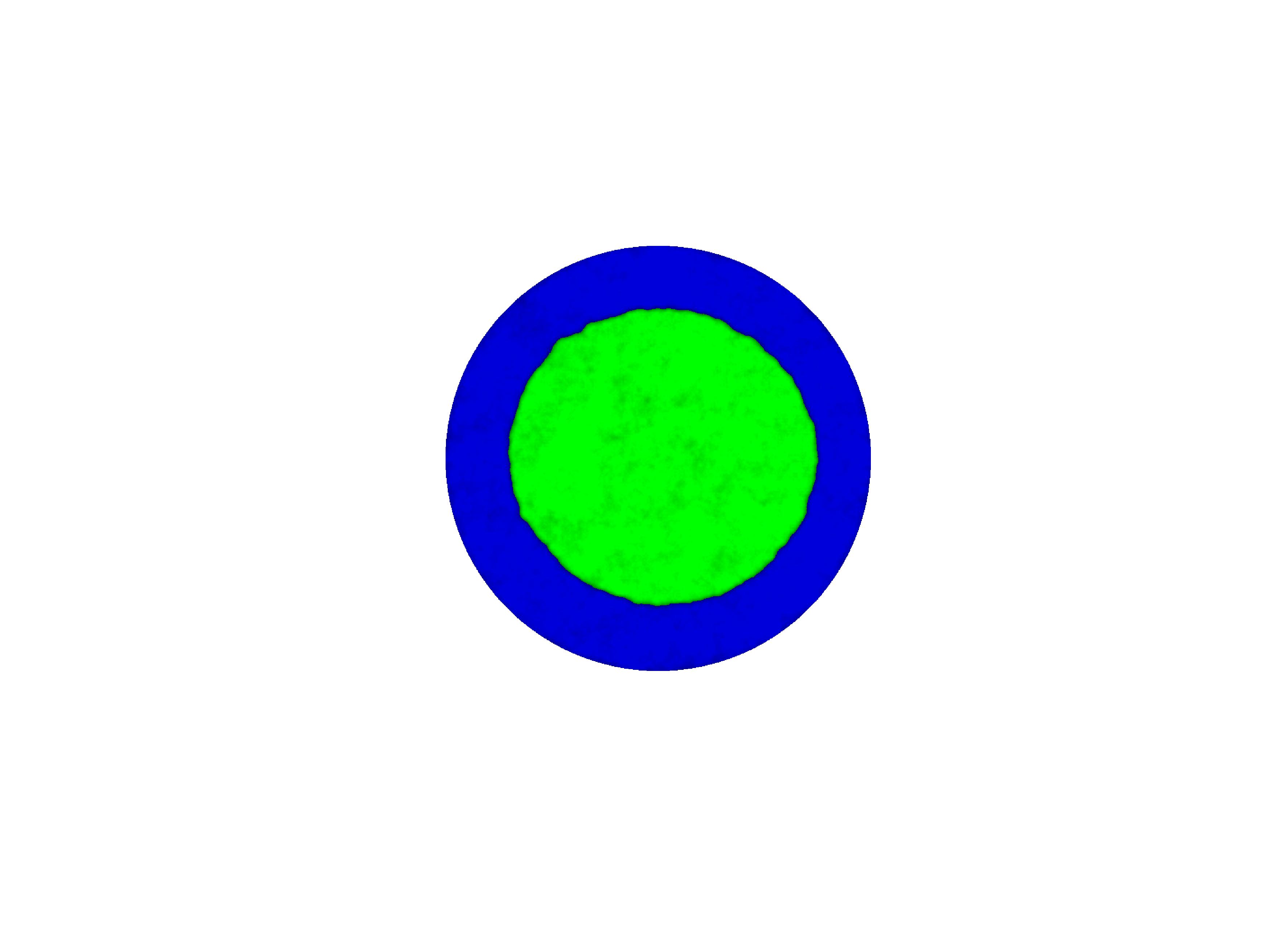}%
   \caption{$t=6000$}
    \label{fig:3a}
\end{subfigure}%
\begin{subfigure}{0.085\textwidth}
\centering%
    \includegraphics[scale=0.03,trim={39cm 7cm 39cm 7cm},clip]{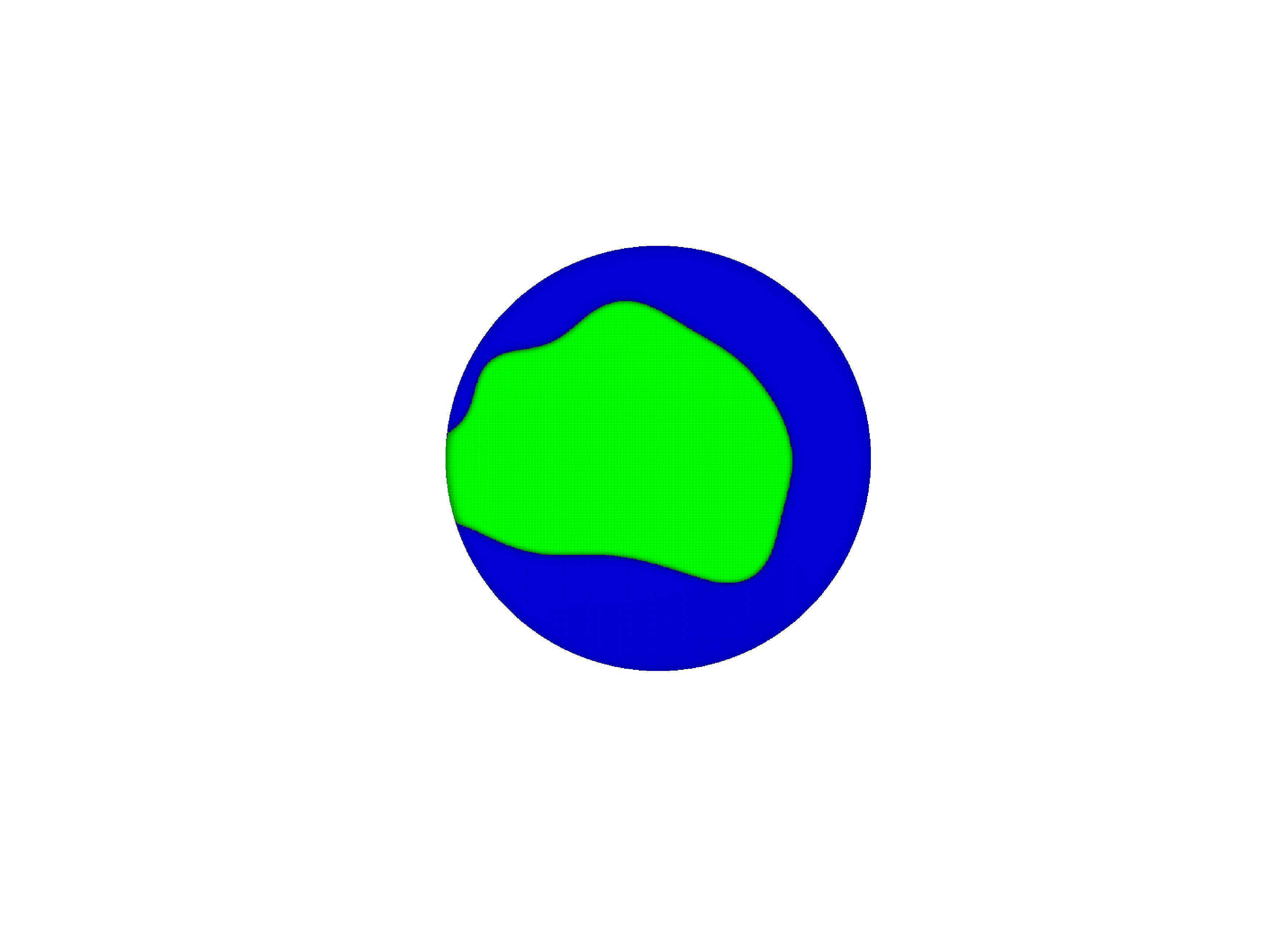}%
   \caption{$t=6300$}
    \label{fig:3b}
\end{subfigure}%
\begin{subfigure}{0.085\textwidth}
\centering%
    \includegraphics[scale=0.03,trim={39cm 7cm 39cm 7cm},clip]{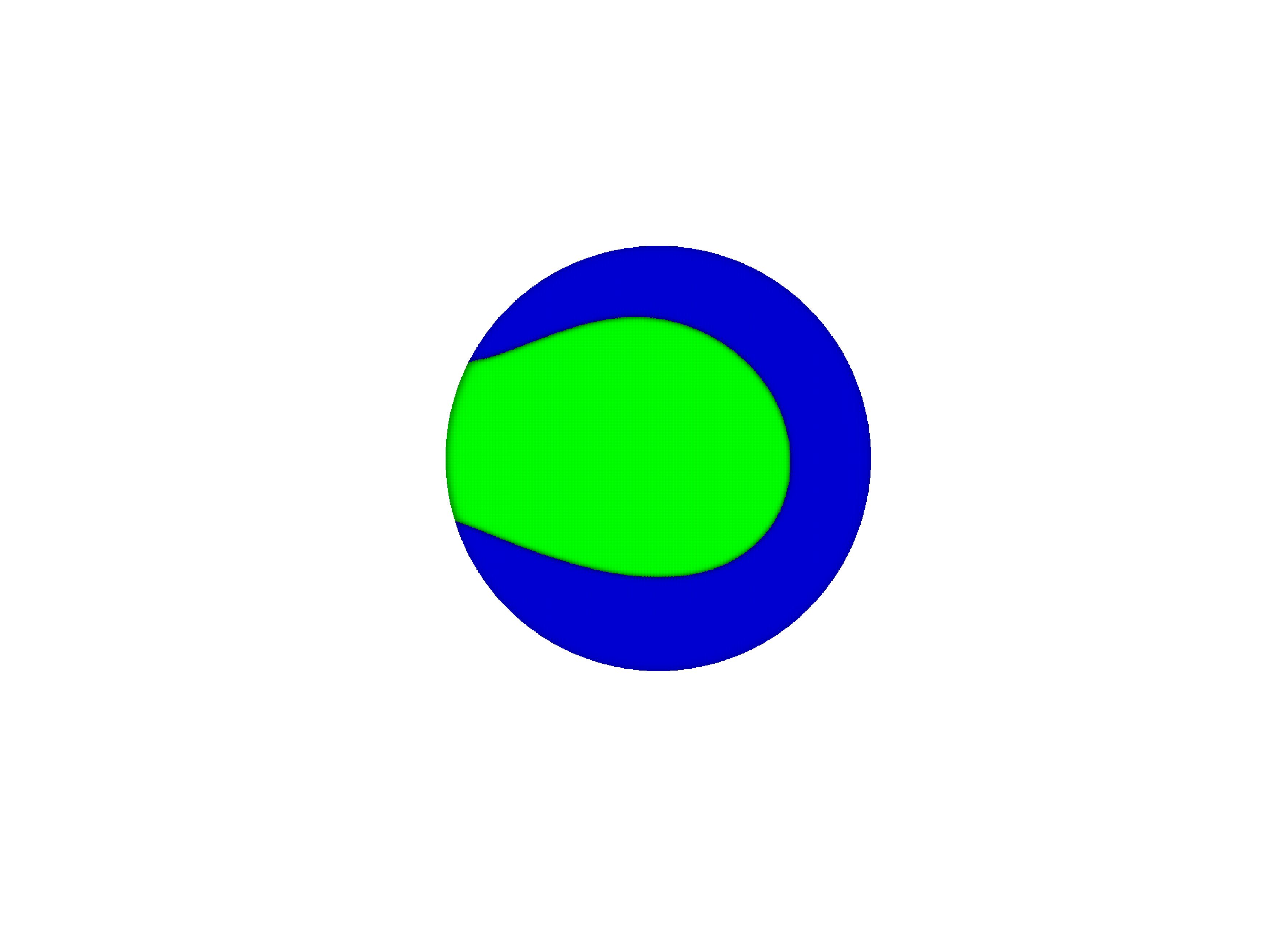}%
   \caption{$t=7000$}
    \label{fig:3c}
\end{subfigure}%
\begin{subfigure}{0.085\textwidth}
\centering%
    \includegraphics[scale=0.03,trim={39cm 7cm 39cm 7cm},clip]{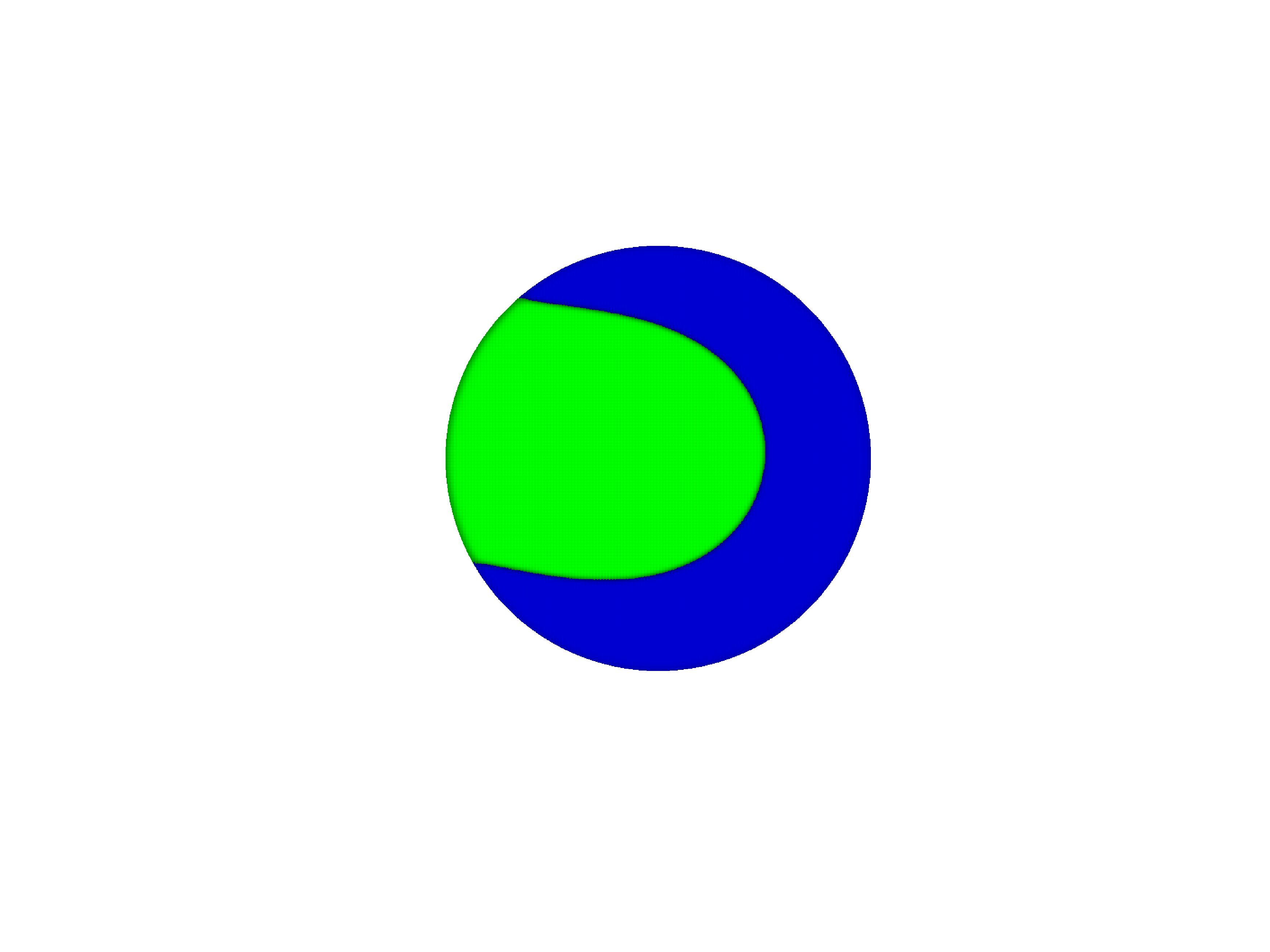}%
   \caption{$t=10000$}
    \label{fig:3d}
\end{subfigure}%
\begin{subfigure}{0.085\textwidth}
\centering%
    \includegraphics[scale=0.03,trim={39cm 7cm 39cm 7cm},clip]{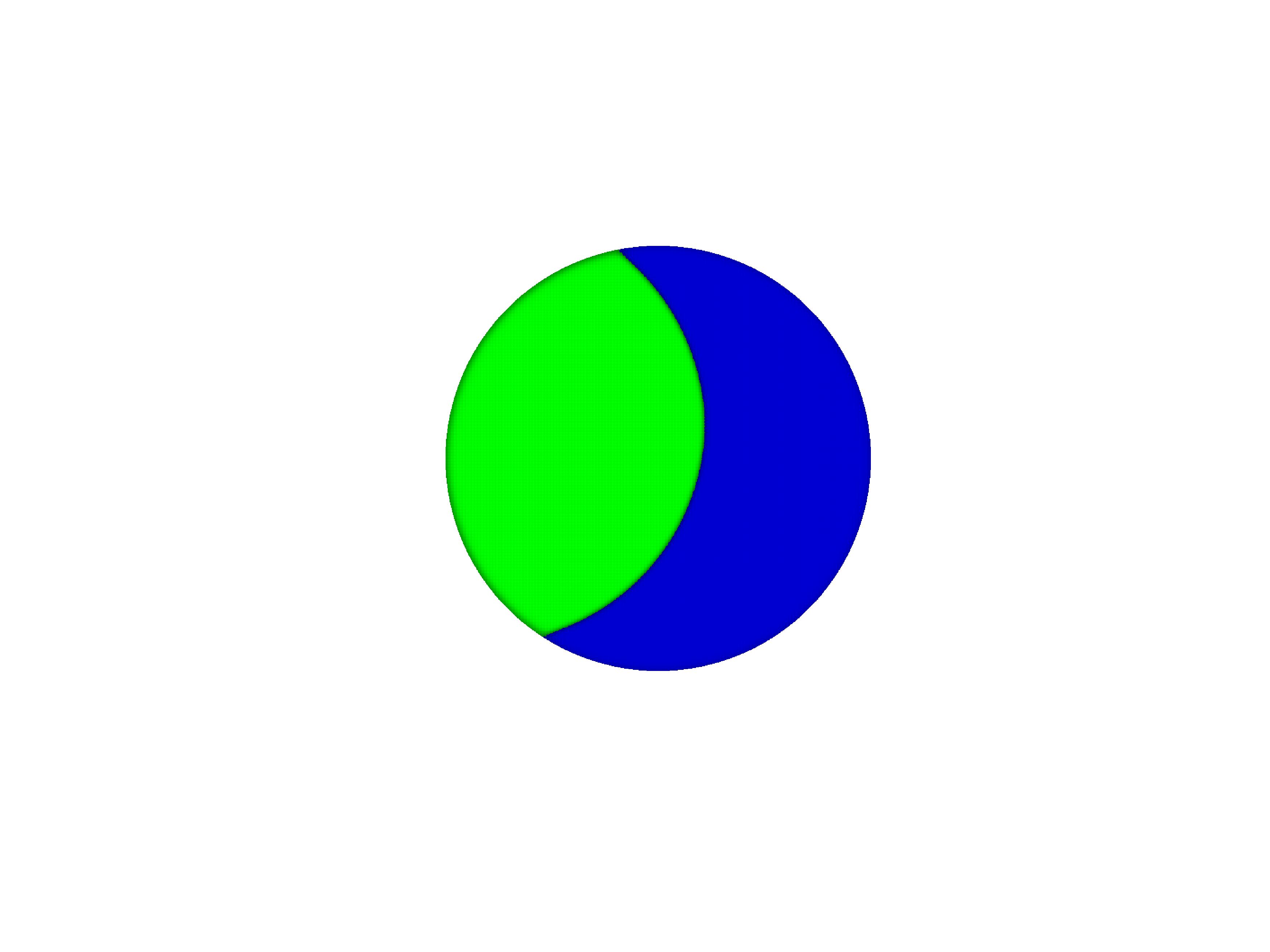}%
   \caption{$t=20000$}
    \label{fig:3e}
\end{subfigure}%

\begin{subfigure}{0.085\textwidth}
\centering%
    \includegraphics[scale=0.03,trim={39cm 7cm 39cm 7cm},clip]{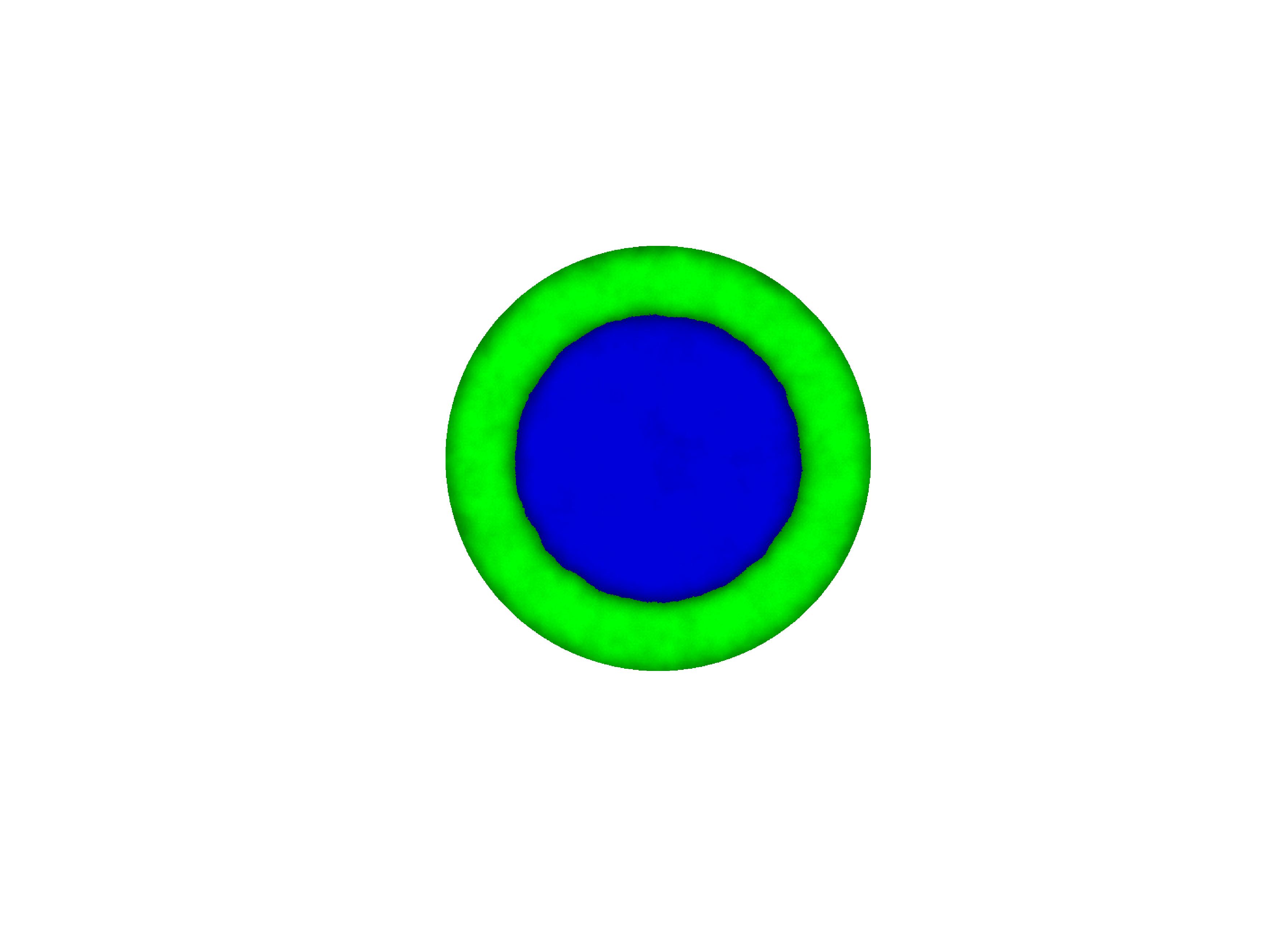}%
   \caption{$t=19000$}
    \label{fig:3f}
\end{subfigure}%
\begin{subfigure}{0.085\textwidth}
\centering%
    \includegraphics[scale=0.03,trim={39cm 7cm 39cm 7cm},clip]{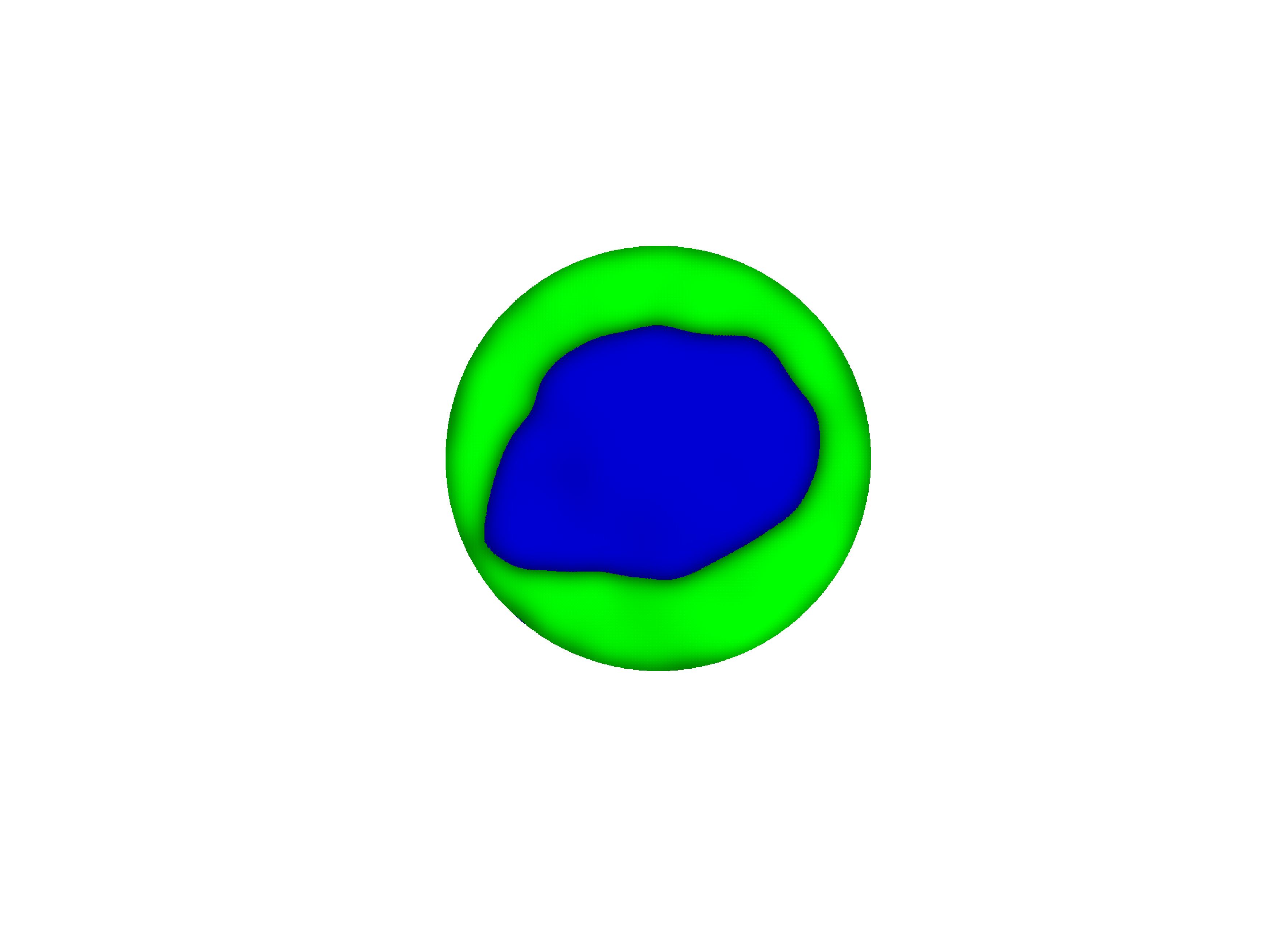}%
   \caption{$t=20600$}
    \label{fig:3g}
\end{subfigure}%
\begin{subfigure}{0.085\textwidth}
\centering%
    \includegraphics[scale=0.03,trim={39cm 7cm 39cm 7cm},clip]{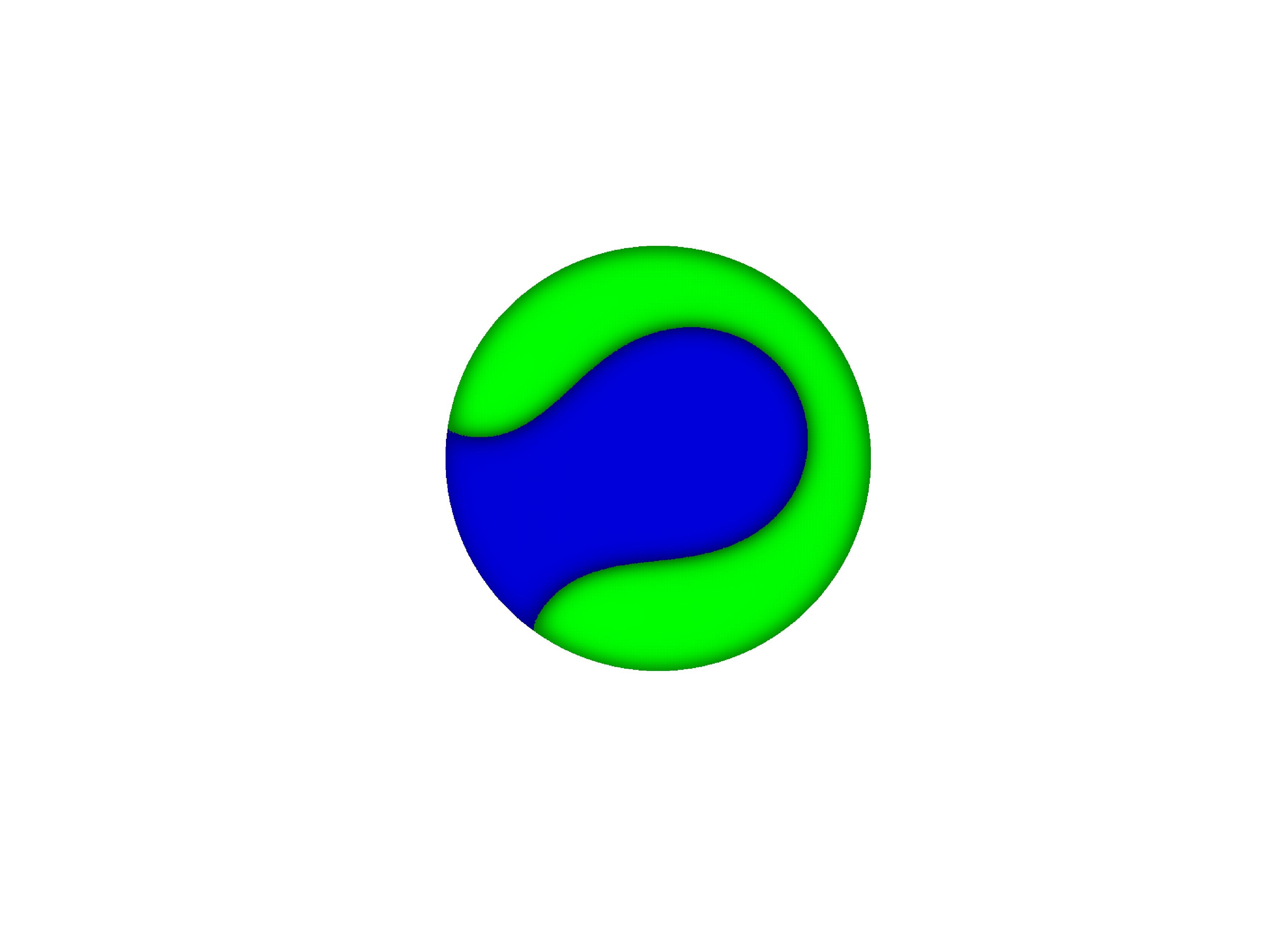}%
   \caption{$t=25000$}
    \label{fig:3h}
\end{subfigure}%
\begin{subfigure}{0.085\textwidth}
\centering%
    \includegraphics[scale=0.03,trim={39cm 7cm 39cm 7cm},clip]{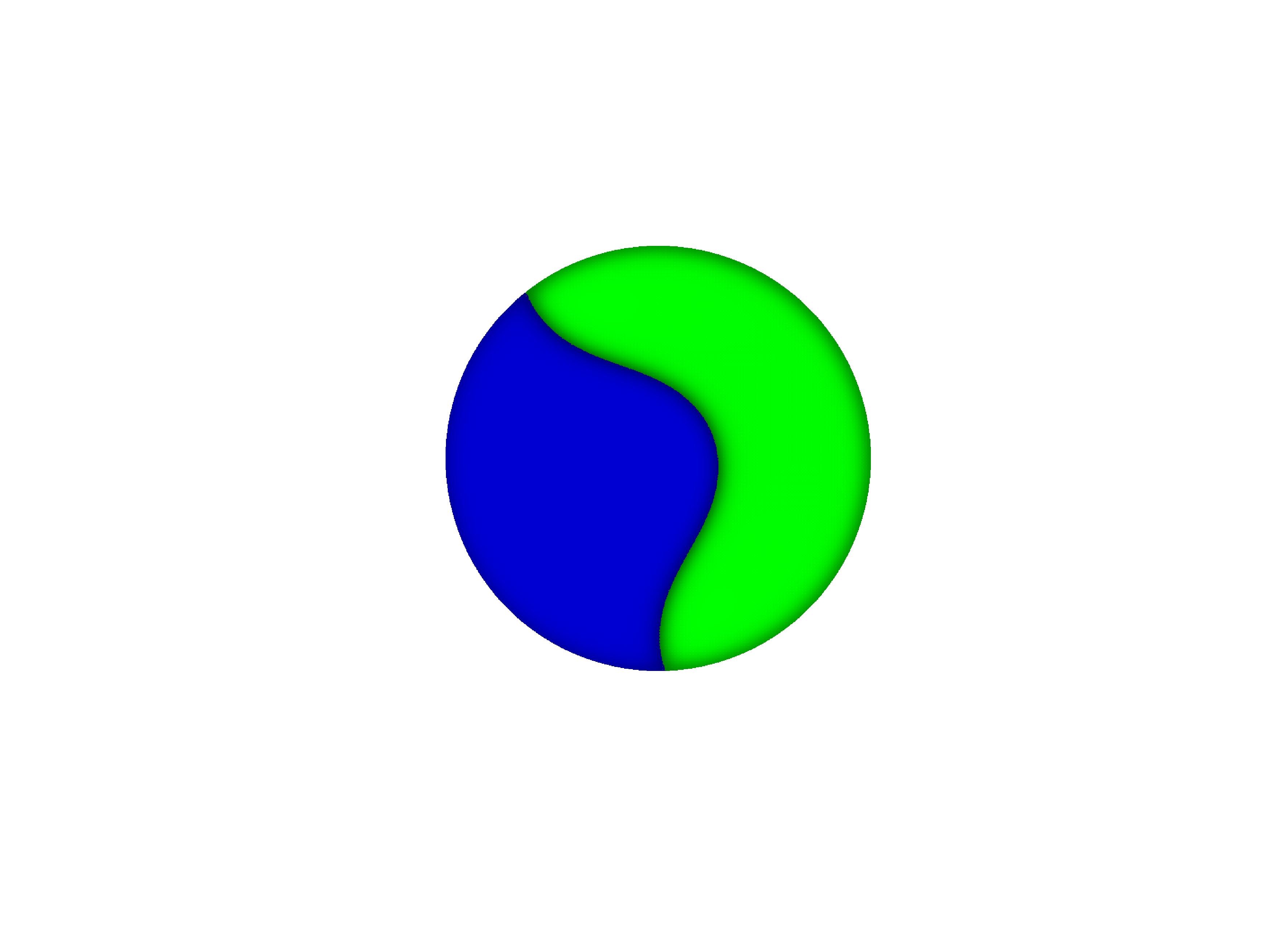}%
   \caption{$t=40000$}
    \label{fig:3i}
\end{subfigure}%
\begin{subfigure}{0.085\textwidth}
\centering%
    \includegraphics[scale=0.03,trim={39cm 7cm 39cm 7cm},clip]{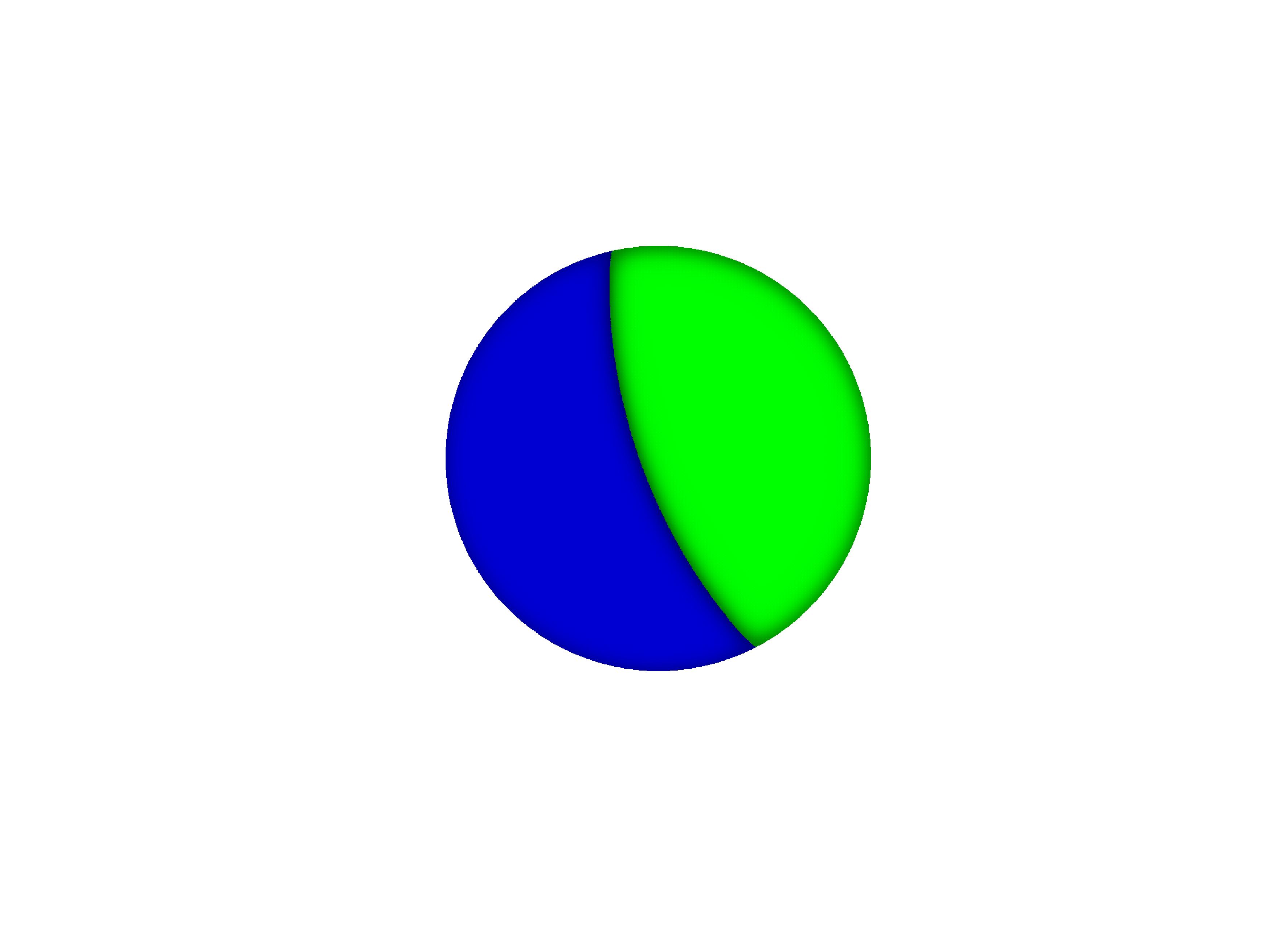}%
   \caption{$t=111600$}
    \label{fig:3j}
\end{subfigure}%

\begin{subfigure}{0.085\textwidth}
\centering%
    \includegraphics[scale=0.03,trim={39cm 7cm 39cm 7cm},clip]{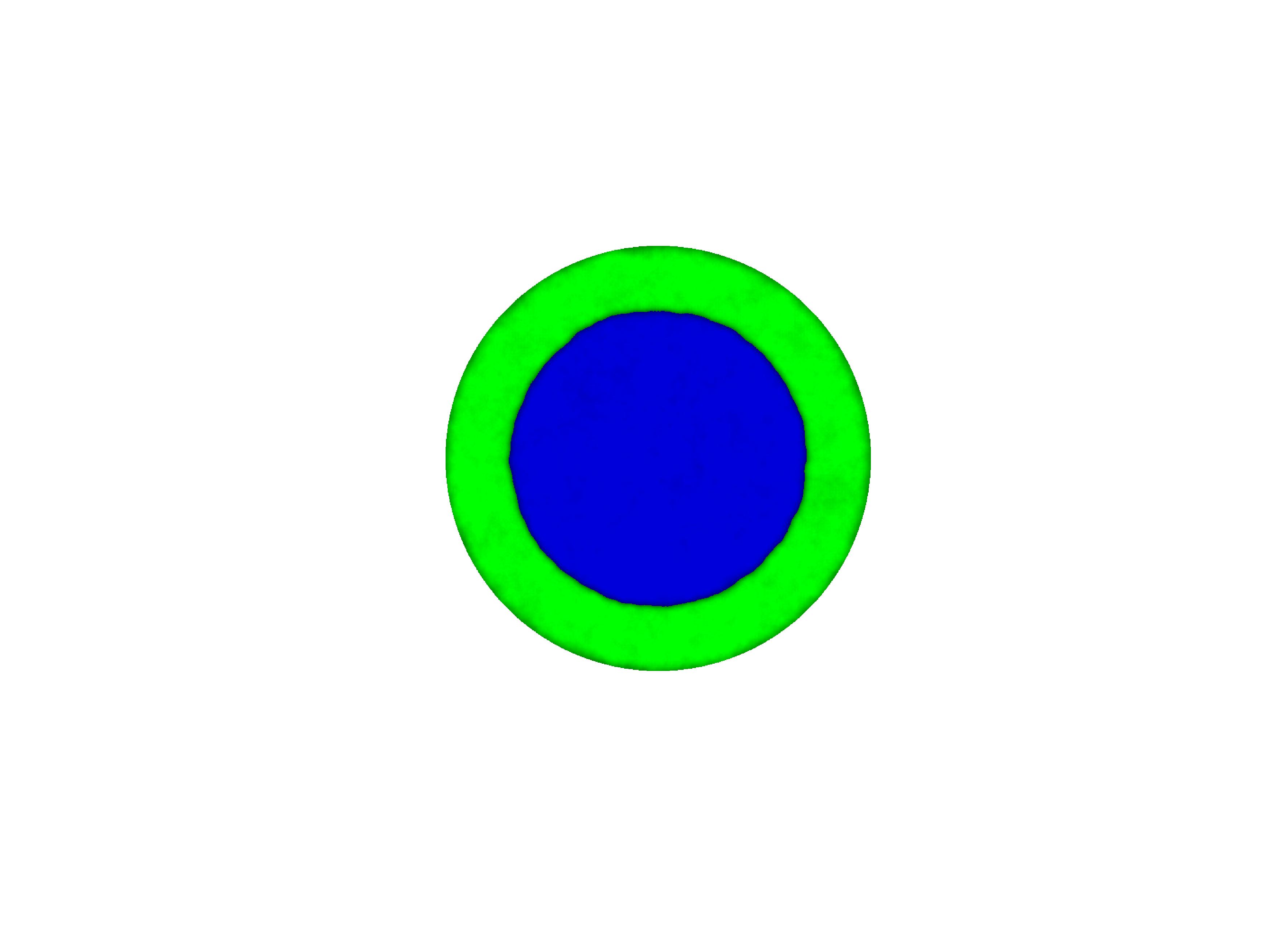}%
    \caption{$t=55000$}
    \label{fig:3k}
\end{subfigure}%
\begin{subfigure}{0.085\textwidth}
\centering%
    \includegraphics[scale=0.03,trim={39cm 7cm 39cm 7cm},clip]{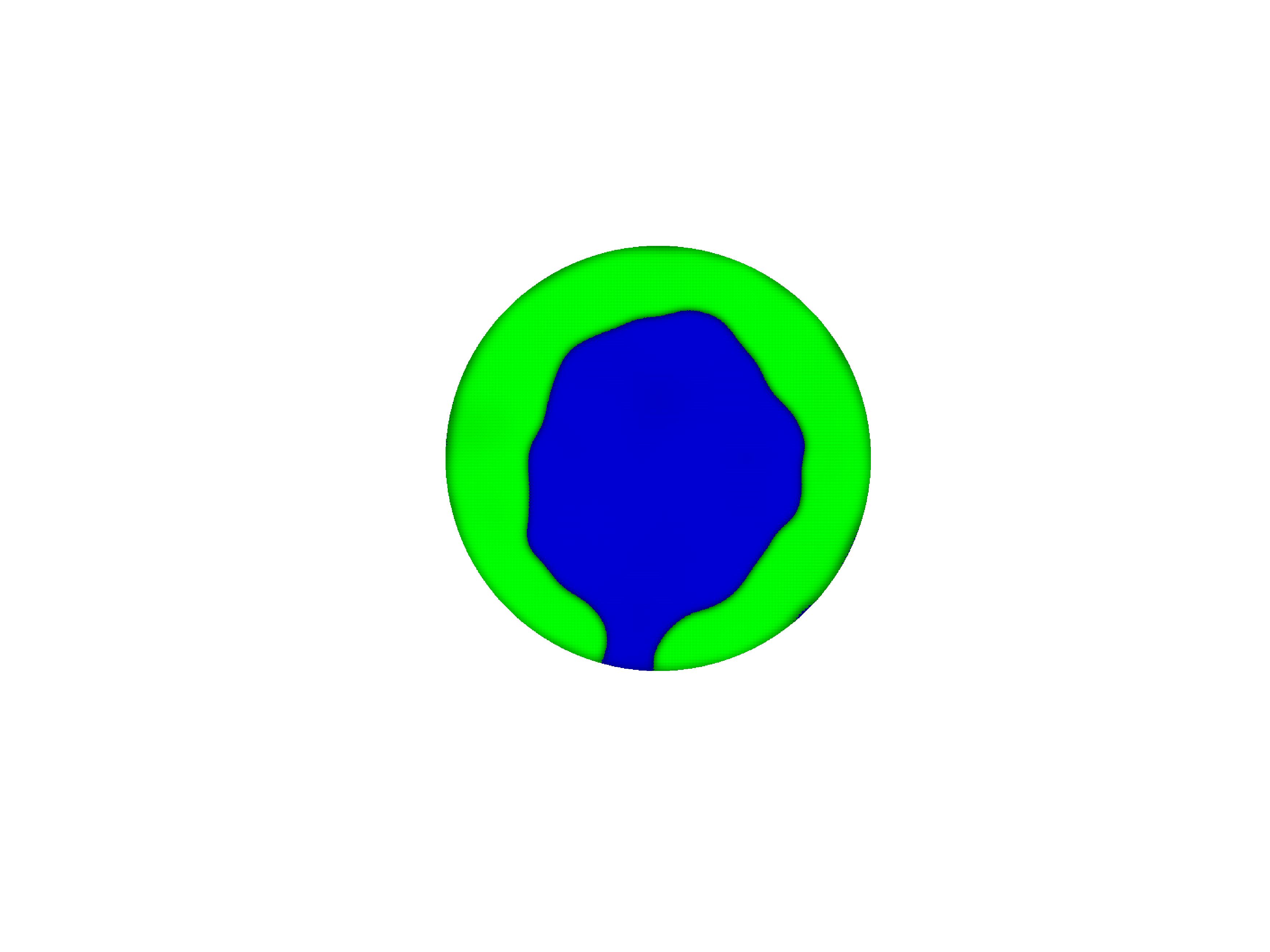}%
   \caption{$t=57000$}
    \label{fig:3l}
\end{subfigure}%
\begin{subfigure}{0.085\textwidth}
\centering%
    \includegraphics[scale=0.03,trim={39cm 7cm 39cm 7cm},clip]{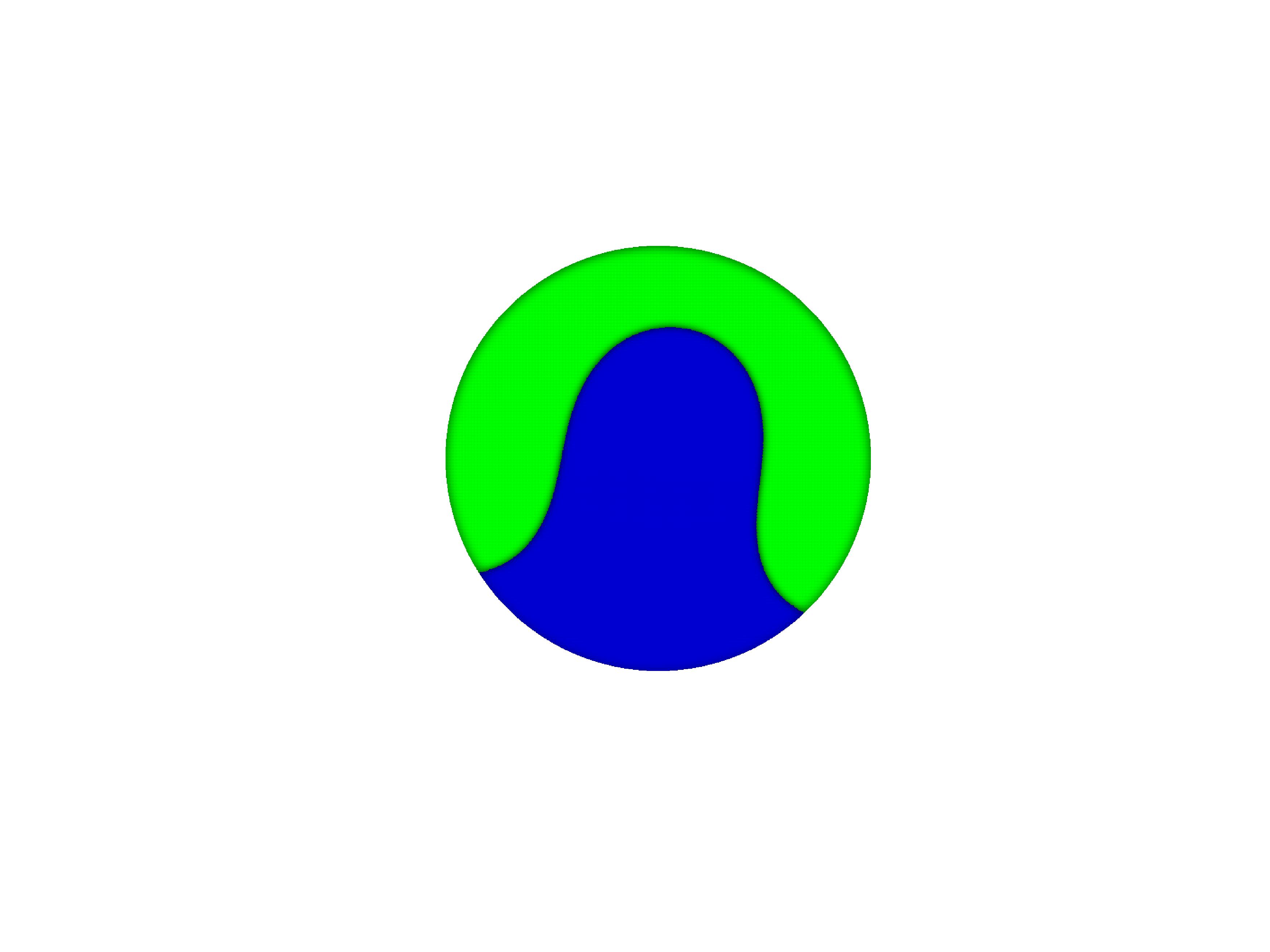}%
   \caption{$t=70000$}
    \label{fig:3m}
\end{subfigure}%
\begin{subfigure}{0.085\textwidth}
\centering%
    \includegraphics[scale=0.03,trim={39cm 7cm 39cm 7cm},clip]{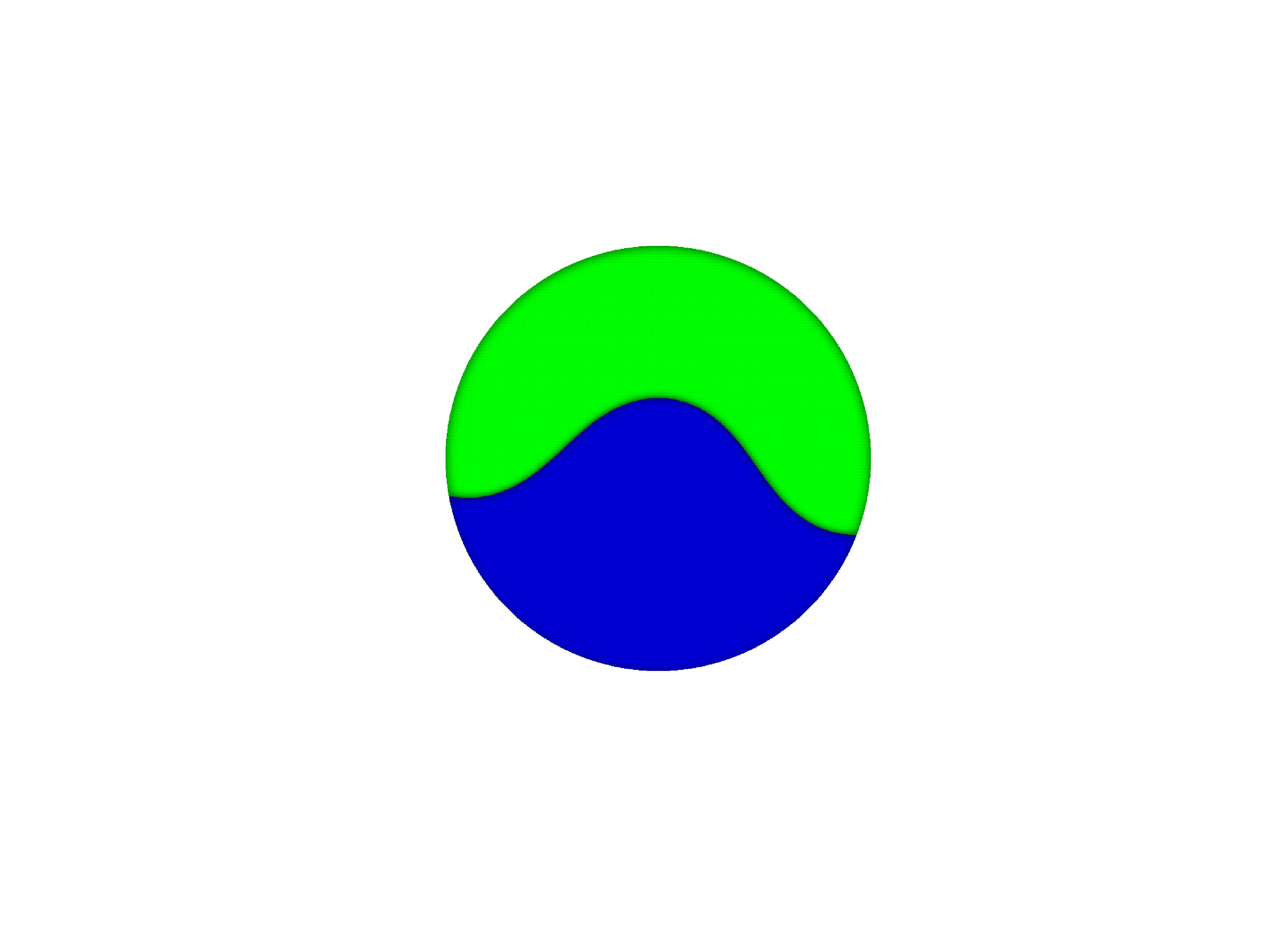}%
   \caption{$t=82000$}
    \label{fig:3n}
\end{subfigure}%
\begin{subfigure}{0.085\textwidth}
\centering%
    \includegraphics[scale=0.03,trim={39cm 7cm 39cm 7cm},clip]{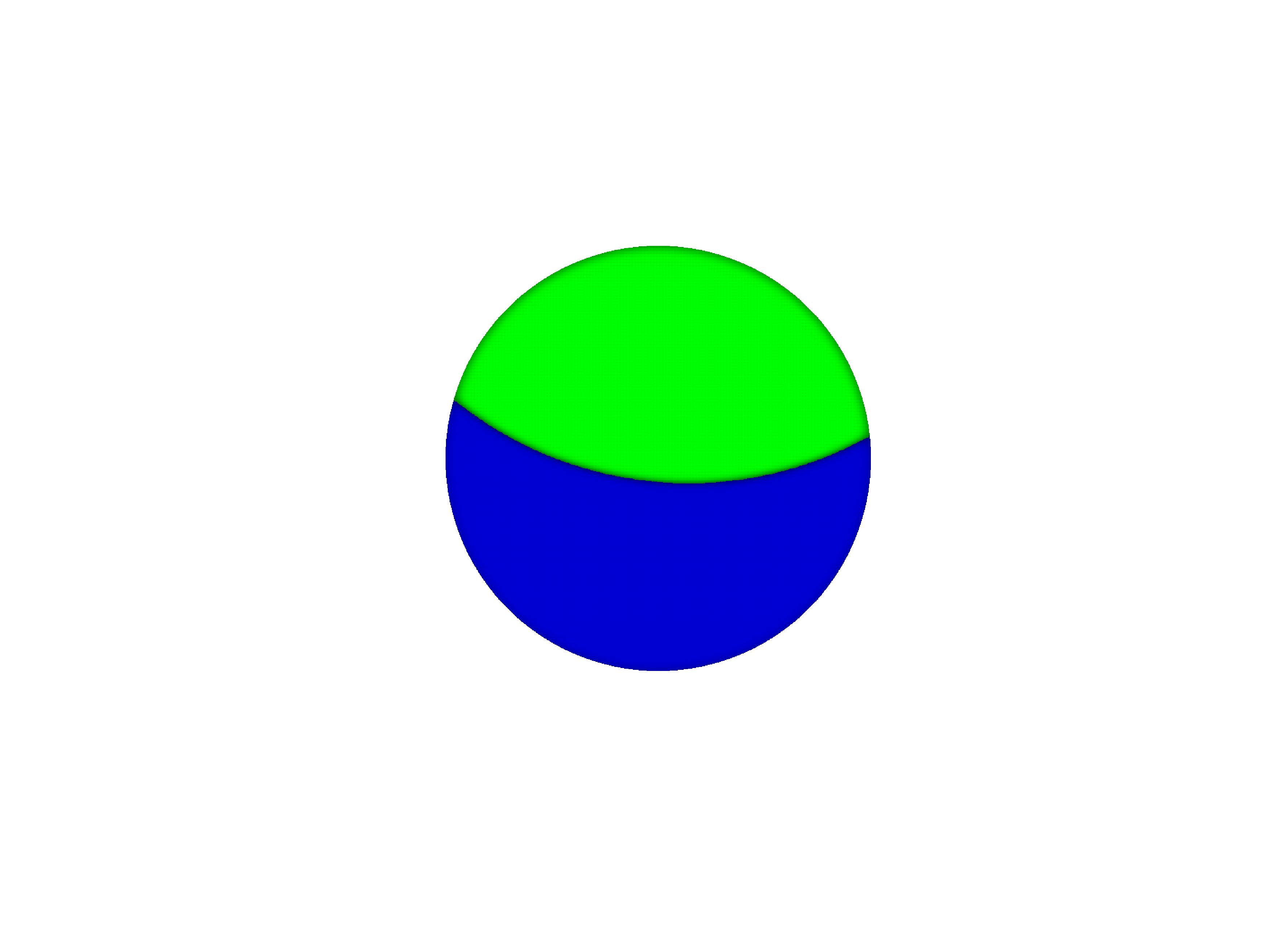}%
   \caption{$t=127000$}
    \label{fig:3o}
\end{subfigure}%
\caption{Time snapshots of transition from metastable CS/ICS to stable CS configurations brought
about by sustained noise. Non-dimensional times are indicated below for each snapshot.}
\label{fig:fig3}
\end{figure}

As shown in Fig.~\ref{fig:fig3}, however, the CS at non-zero $\theta$ and ICS configurations
are metastable, as they relax to more stable structures when subjected to a sustained white
noise~\cite{Cook1970}. Here the initial configurations for the simulations are the final metastable
ones shown in Fig.~\ref{fig:fig2}. With increasing time, the concentric structures break down and
evolve to form Janus. The time taken for the metastable-to-stable morphological transition is
very large because the difference of the energies between the metastable (CS/ICS) and stable 
(Janus) configuration is very small, resulting in very sluggish diffusion. Therefore, these can
be termed as kinetically trapped configurations~\cite{Li2019Nanoscale,Grammati2016,pankaj}.

To understand the formation of different morphologies as a function of the model parameter sets,
one needs to relate the competing processes of bulk SD and SDSD. For bulk SD, $f_0^p$ sets the
driving force $\Delta{f}$, defined as the difference between the free energy of non-equilibrium
initial $\beta$ state and that of the final state made up of equal parts of equilibrium $\beta_1$
and $\beta_2$ phases:

\begin{align}
    \Delta{f}&=f^\beta(c=0.5)-\frac{1}{2}\left[f^\beta(c=c_{\beta_1}^{e})
    +f^\beta(c=c_{\beta_2}^{e})\right]\nonumber\\
    &=f_0^p/16 \,\,(\textrm{using Eq.~\eqref{eq:bulk-free-energy-dens}}).
    \label{eqn:DeltaF}
\end{align}
$\Delta{f}$, together with $\kappa_c$, also determines the interfacial energy 
$\sigma_{12}$ of the system as~\cite{Abinandanan}:
\begin{align}
    \sigma_{12}=\frac{1}{3V_m}\left[c_{\beta_2}^{e}(T)-c_{\beta_1}^{e}(T)\right]^3
    \sqrt{\kappa_c\Delta{f(T)}}.
    \label{eqn:sigma12}
\end{align}
Both $\Delta{f}$ and $\sigma_{12}$ increase with decreasing $T$ (or increasing undercooling
$\Delta{T}$).

The surface energies $\sigma_1$ and $\sigma_2$, and their difference $\Delta\sigma$, on the 
other hand, scale with the height of the free energy barrier $\omega_0$. While $\Delta f$ 
controls bulk SD within the particle, $\omega_0$ influences SDSD by defining $\Delta\sigma$ 
-- the larger the $\Delta\sigma$, the easier it is to initiate SDSD. We now define a
normalized driving force for bulk phase separation, $\Delta\tilde{f}$, by taking the ratio
of $\Delta f$ to $\Delta\sigma$.

\begin{figure}[htbp]
\centering
 \begin{subfigure}{0.45\textwidth}
   \includegraphics[scale=0.2]{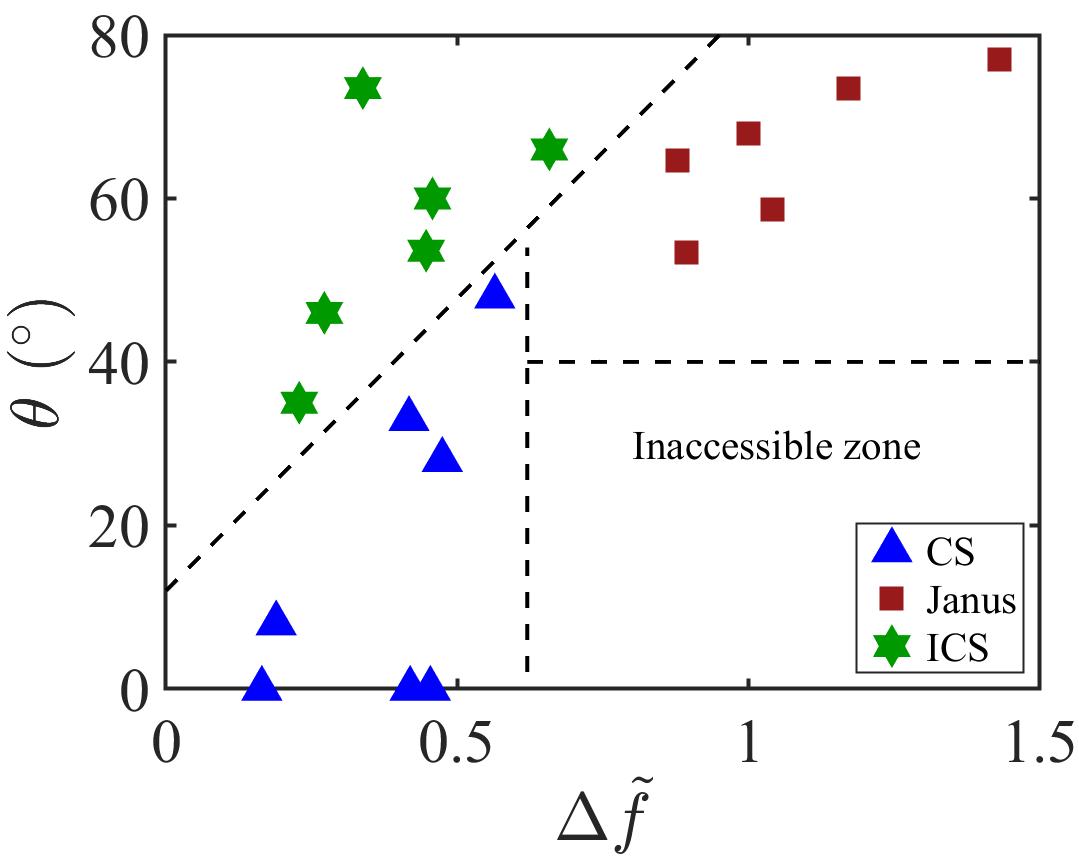}%
   \caption{}
 \end{subfigure}%
 \caption{Simulated stable and metastable BNP configurations in the 
 $\Delta\tilde{f}$--$\theta$ space. Dashed lines are drawn to delineate 
 regions of where Janus, CS and ICS morphologies are observed.} 
 \label{fig:morphologymap}
 \end{figure}

Fig.~\ref{fig:morphologymap} presents the essence of all simulations by assigning the location
of BNP configuration in the space of model parameters $\Delta\tilde{f}$ and $\theta$. The map 
clearly demarcates three distinct regions in this space: high $\Delta\tilde{f}$-high $\theta$ for 
Janus, low $\Delta\tilde{f}$-high $\theta$ for ICS, and the intervening region for CS. The region 
of high $\Delta\tilde{f}$-low $\theta$ is physically inaccessible since high $\Delta\tilde{f}$
implies either high $\Delta{f}$ or low $\omega_0$, which increases $\sigma_{12}$ or decreases
$\Delta\sigma$, respectively, and thereby lead to high $\theta$.

When the spontaneous wetting condition is satisfied ($\theta=\ang{0}$ or undefined), CS is the
final configuration irrespective of $\Delta\tilde{f}$. Also, for a given $\theta$, a metastable
CS configuration is preferred over ICS at larger $\Delta\tilde{f}$. At lower $\Delta\tilde{f}$,
SDSD creates alternate concentric rings extending towards the center, with segregation giving
rise to an outermost $\beta_1$ ring~\cite{pankaj}. Subsequently, ring-coarsening leads to the
ICS structure. In contrast, bulk SD dominates at higher $\Delta\tilde{f}$, resulting in 
interactions between inner interconnected domains and the outermost $\beta_1$ ring during
coarsening. Subsequently, the latter is replaced by $\beta_2$, leading to metastable CS. Note
that these interactions do not yield the stable Janus configuration, because, in this case,
$\Delta\sigma$ remains sufficiently large to cause $\beta_2$ to preferentially spread along the
surface and $\beta_1$ to recede.

 
\subsection{Computation of chemical and capillary forces for Ag-Cu}

Fig.~\ref{fig:morphologymap} demonstrated how different morphologies clustered around
in distinct regions of the driving force -- contact angle space. This space
essentially represents the interplay of chemical and capillary forces, and
temperature is one of the key physical parameters that directly or indirectly controls
them. The chemical driving force for phase separation, $\Delta{f}$, for a given
system can be obtained as a function of temperature in a relatively straightforward
way from its CALPAHD data. Capillarity, on the other hand, is manifested through the
contact angle $\theta$, which itself is defined in terms of interfacial and surface
energies ($\sigma_1, \sigma_2, \sigma_{12}$). Reliable values of the latter as a
function of temperature are often difficult to measure experimentally. Nevertheless,
here we attempt to arrive at fair estimates of these energies using the available data
and thermodynamic correlations. We consider the Ag-Cu system as an example and proceed
to compute the temperature dependence of these quantities.

\subsubsection{Driving force for phase separation}

The molar bulk free energy $F_m$ of the Cu-Ag solid solution can be expressed 
through a Redlich-Kister polynomial as:
\begin{align}
    F_m =&\left[X_\mathrm{Ag}F_\mathrm{Ag}^0 + X_\mathrm{Cu}F_\mathrm{Cu}^0\right] +
            RT\left[X_\mathrm{Cu}\ln{X_\mathrm{Cu}}
            + X_\mathrm{Ag}\ln{X_\mathrm{Ag}}\right]\nonumber\\
            &+ X_\mathrm{Cu}X_\mathrm{Ag}\left[L_0 + L_1(X_\mathrm{Ag}-X_\mathrm{Cu})\right]
\end{align}
where $X_i$'s ($i=$ Cu, Ag) are the mole fractions component $i$ in the solution. The first
bracketed group in the right hand side is the mixture free energy contribution with $F_{i}^0$'s 
being the standard free energies of the pure components which can be found in the SGTE
database~\cite{DINSDALE1991317}. The second group is the contribution from the ideal solution part
of free energy ($F_m^\mathrm{id,bulk}$) and the last term represents the excess contribution 
($F_m^\mathrm{ex,bulk}$). The temperature-dependent interaction parameters for the Cu-Ag solid
solution are given as $L_0=34532-9.178T$ and $L_1=-5996+1.725T$~\cite{subramanian1993ag}.

\begin{figure}[htbp]
\centering
 \begin{subfigure}{0.45\textwidth}
 \centering
   \includegraphics[scale=0.2]{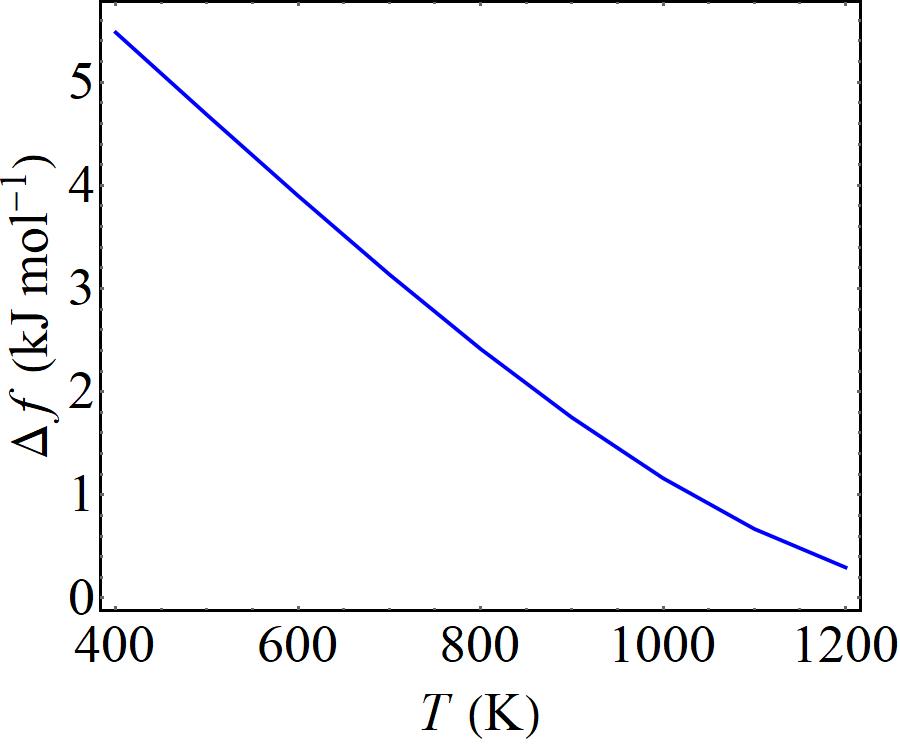}
   \caption{}
   \label{fig:4b}
 \end{subfigure}\\%
 \begin{subfigure}{0.45\textwidth}
   \centering
   \includegraphics[scale=0.22]{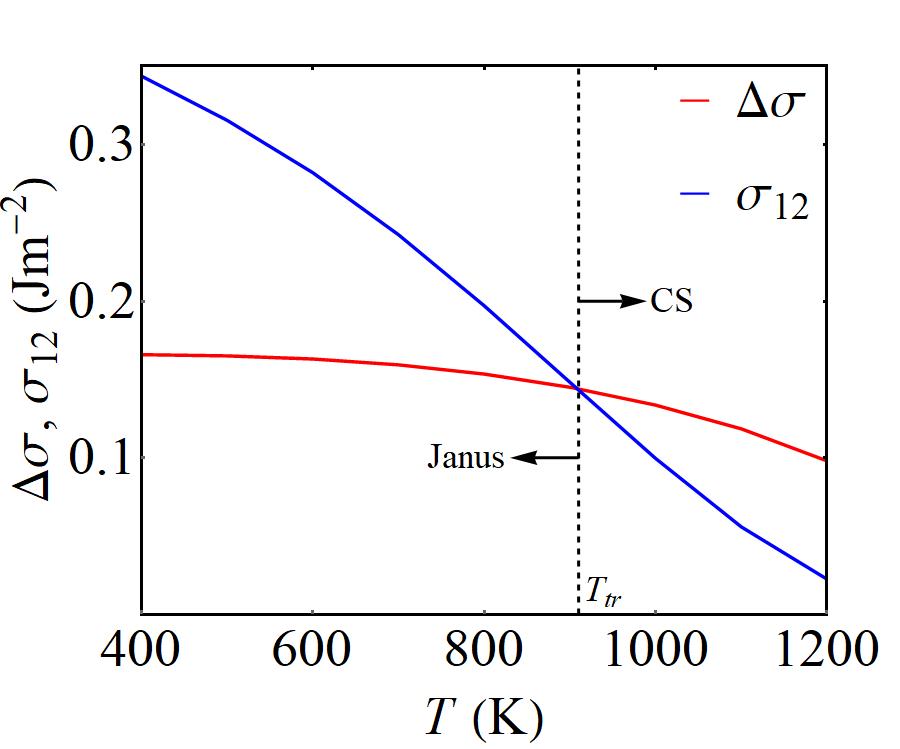}%
   \caption{}
   \label{fig:4c}
 \end{subfigure}
\label{fig:CALPAHD}
 \caption{(a) Variation of $\Delta{f}$ with temperature for the Ag-Cu system computed using
 its CALPHAD data. (b) Temperature-dependence of interfacial and surface energies of Cu-poor
 and Cu-rich solid solution phases. Temperature of transition ($T_{tr}$) from stable CS to
 Janus morphology is indicated.}
\end{figure}

Equilibrium solute mole fractions $X_i^e$ in Cu-rich and Ag-rich solid solutions below the critical
temperature $T_c$ are computed using the conditions for chemical equilibrium:
\begin{align}
    \mu_\mathrm{Ag}^\mathrm{Cu_{ss}} = \mu_\mathrm{Ag}^\mathrm{Ag_{ss}}, \quad
    \mu_\mathrm{Cu}^\mathrm{Cu_{ss}} = \mu_\mathrm{Cu}^\mathrm{Ag_{ss}},
    \label{eq:chempotequality}
\end{align}
where the chemical potentials $\mu_i\,(i=\mathrm{Cu}, \mathrm{Ag})$ evaluated at equilibrium
bulk compositions $X_i^e$ are given by
\begin{align}
    \mu_\mathrm{Cu} = F_m - 
        X_\mathrm{Ag}\frac{\partial F_m}{\partial X_\mathrm{Ag}}; \quad
    \mu_{Ag} = F_m + (1-X_\mathrm{Ag})\frac{\partial F_m}{\partial X_\mathrm{Ag}}.
    \label{eq:chempotdef}
\end{align}
The free energy of mixing for an alloy composition $X_\mathrm{Ag}$ is expressed as 
\begin{align}
\Delta{F}_\mathrm{mix} = F_m - (X_\mathrm{Cu}\mu_\mathrm{Cu}^e+X_\mathrm{Ag}\mu_\mathrm{Ag}^e).
\end{align}
The driving force for bulk spinodal, $\Delta{f}$, is the maximum value of 
$\Delta{F}_\mathrm{mix}$ over the entire composition range (\emph{i.e.}, maximum of the $\Delta{F}_\mathrm{mix}-X$ curve). It is evaluated and plotted in Fig.~\ref{fig:4b} for
the temperature range of 400-800 K. The plot shows that $\Delta{f}$ decreases monotonically
with increasing temperature. Using the same thermodynamic data, we also compute
the equilibrium mole fractions $X_i^e$ ($i=\;\mathrm{Cu, Ag}$) in both Cu- and Ag-rich
solutions which are required for estimating the interfacial energy.

\subsubsection{Interfacial energy}

The temperature dependence of the interfacial energy between the Cu- and Ag-rich solid
solutions ($\sigma_\mathrm{Cu-Ag}$) can be computed using Eq.~\eqref{eqn:sigma12}. To
be able to do so, however, one requires the value of $\kappa_c$. We utilize the value of 
$\sigma_\mathrm{Cu-Ag}=0.197$ \si{\joule\meter^{-2}} at 800 K obtained from molecular
dynamics simulations~\cite{Chandross_2014}, along with the values of $\Delta f$ and
$X_i^e$ at this temperature calculated in the previous step, and a constant molar volume
of $10^{-5}$ \si{\meter^3\mol^{-1}}, to compute $\kappa_c = 1.94\times 10^{-14}$ 
\si{\joule\meter^2\mol^{-1}}. Since $\kappa_c$ is generally considered to be 
temperature-independent for a given system, we use this constant value in 
Eq.~\eqref{eqn:sigma12} to compute $\sigma_\mathrm{Cu-Ag}$ in the temperature range of
400-1200 K.

\subsubsection{Surface energies}

We calculate surface energies $\sigma_{i,\mathrm{ss}}$ of the Cu-rich and Ag-rich solid 
solutions as a function of temperature $T$ using a modified Butler model~\cite{Tanaka2001}:
\begin{align}
    \sigma(T) = &\sigma_i^0 + 
    \frac{RT}{A_i}\log{\left[\frac{X_i^\mathrm{surf}}{X_i^e}\right]}\nonumber\\
    +&\frac{1}{A_i}[F_i^\mathrm{ex,surf}(T,X_i^\mathrm{surf})-F_i^\mathrm{ex,bulk}(T,X_i^e)].
         \label{eq:Butler}
\end{align}
Here, $i=\mathrm{A,\;B}$ denote the components Cu and Ag, respectively, $\sigma_i^0$ are the
surface energies of the pure components, $A_i$ are the molar surface areas, $X_i^\mathrm{surf}$
are the surface compositions at $T$, $F_i^\mathrm{ex,surf}$ and $F_i^\mathrm{ex,bulk}$ are the
excess partial molar free energies of $i$ associated with surface and bulk, respectively.

Surface energies of the pure components Eq.~\eqref{eq:Butler} are obtained from the correlation
\begin{align}
    \sigma_i^0 = 1.25 \sigma_i^\mathrm{liq} + \frac{d\sigma}{dT}(T-T_m^i)
\end{align}
where $\sigma_i^\mathrm{liq}$ is the surface tension of liquid component at its melting point
$T_m^i$ and the temperature coefficient of surface energy is taken as $-10^{-4}$
\si{\joule\meter^{-2}\kelvin}. The partial molar excess free energies for bulk, 
$F_i^\mathrm{ex,bulk}$, can be expressed in terms of $X_i^e$ using Eq.~\eqref{eq:chempotdef},
with excess free energy $F^{ex,bulk}$ replacing the total molar free energy $F_m$. Following
Tanaka and Hara~\cite{Tanaka2001}, we take $F_i^\mathrm{ex,surf} = \beta_{mix}F_i^\mathrm{ex,bulk}$.
The parameter $\beta_{mix}$ is the ratio of coordination number in the surface to that in the bulk; 
it is taken as $0.75$ for face centered cubic solid solutions. Molar surface areas $A_i$ of the pure
components are obtained using the relation $A_i=1.091 N_0^{1/3} V_m^{2/3}$ where $N_0$ is the
Avogadro's number. Eq.~\eqref{eq:Butler} constitutes a set of two simultaneous non-linear algebraic
equations which can now be solved numerically to obtain the unknown surface composition 
$X_i^\mathrm{surf}$ for both the terminal phases. Plugging its values in Eq.~\eqref{eq:chempotdef}, 
one can compute the surface energies of the Cu- and Ag-rich solution phases.

Fig.~\ref{fig:4c} plots the difference of the surface energies, $\Delta\sigma=\sigma_\mathrm{Cu_{ss}}-\sigma_\mathrm{Ag_{ss}}$, and the interfacial
energy, $\sigma_\mathrm{Cu-Ag}$, with temperature. It shows that although the surface
energies themselves vary with temperature, their difference is relatively insensitive to it,
changing only slightly at higher temperatures. The interfacial energy, on the other hand, 
depends strongly with temperature, decreasing steeply at higher temperatures. This is expected,
as it must vanish at the critical temperature for the miscibility gap. In terms of the 
spontaneous wetting criterion (Eq.~\eqref{eq-contact}), temperature of intersection of surface
and interfacial energy lines, $T_{tr}$ ($\sim$ 910 K) marks the transition from \emph{stable} CS
to Janus morphologies.

\subsection{Connecting phase-field results with thermodynamic computations}

\begin{figure*}[htbp]
\centering
 \begin{subfigure}{0.45\textwidth}
 \centering
   \includegraphics[scale=0.2]{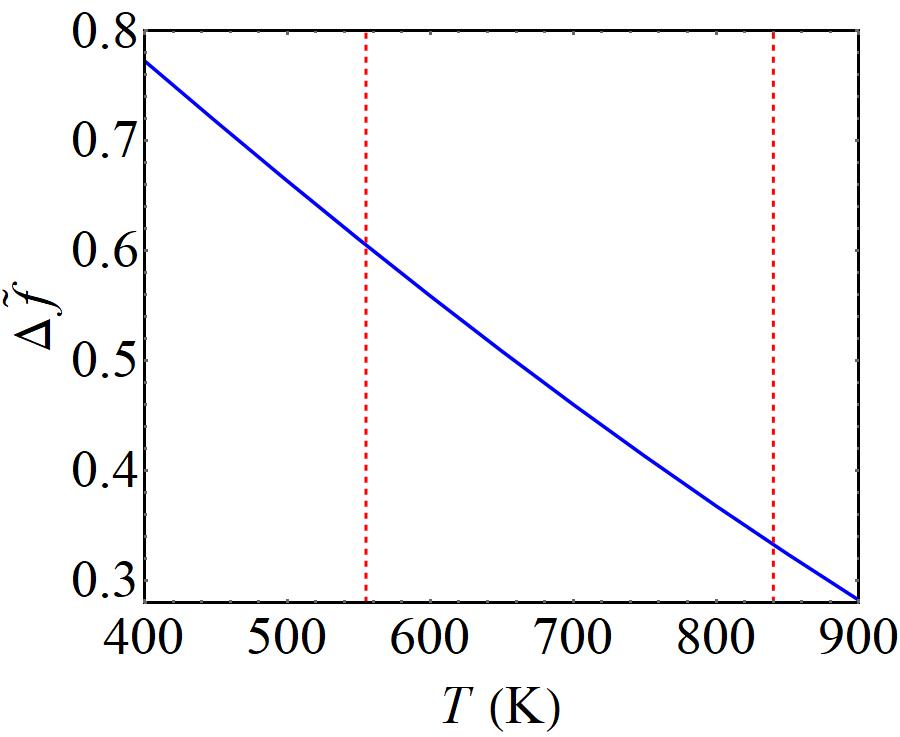}
   \caption{}
   \label{fig:8a}
 \end{subfigure}
 \begin{subfigure}{0.45\textwidth}
 \centering
   \includegraphics[scale=0.2]{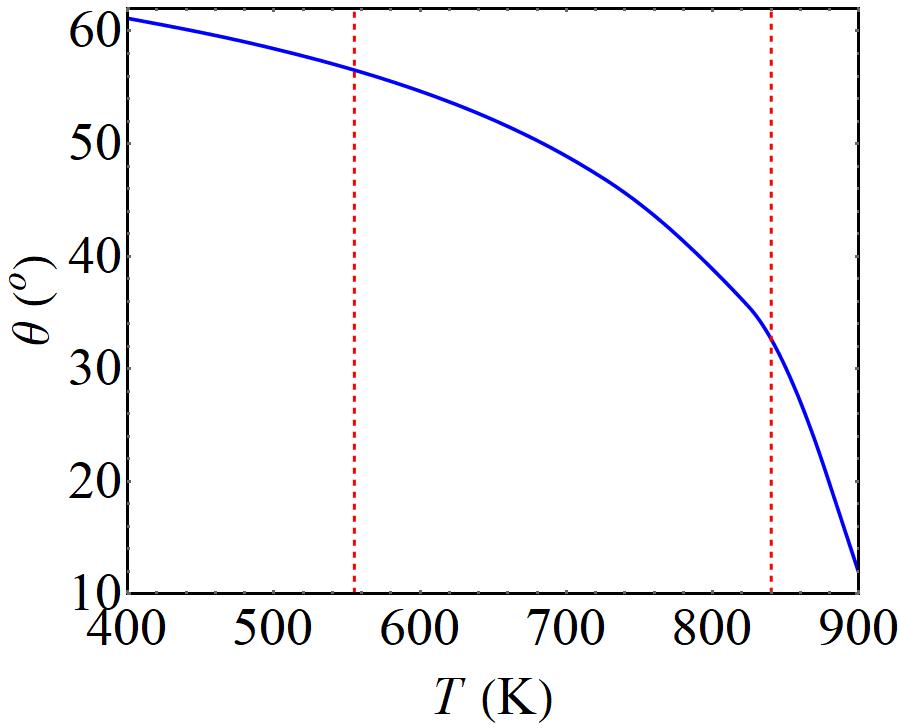}
   \caption{}
   \label{fig:8b}
 \end{subfigure}\\%
 
 \begin{subfigure}{0.45\textwidth}
   \centering
   \includegraphics[scale=0.2]{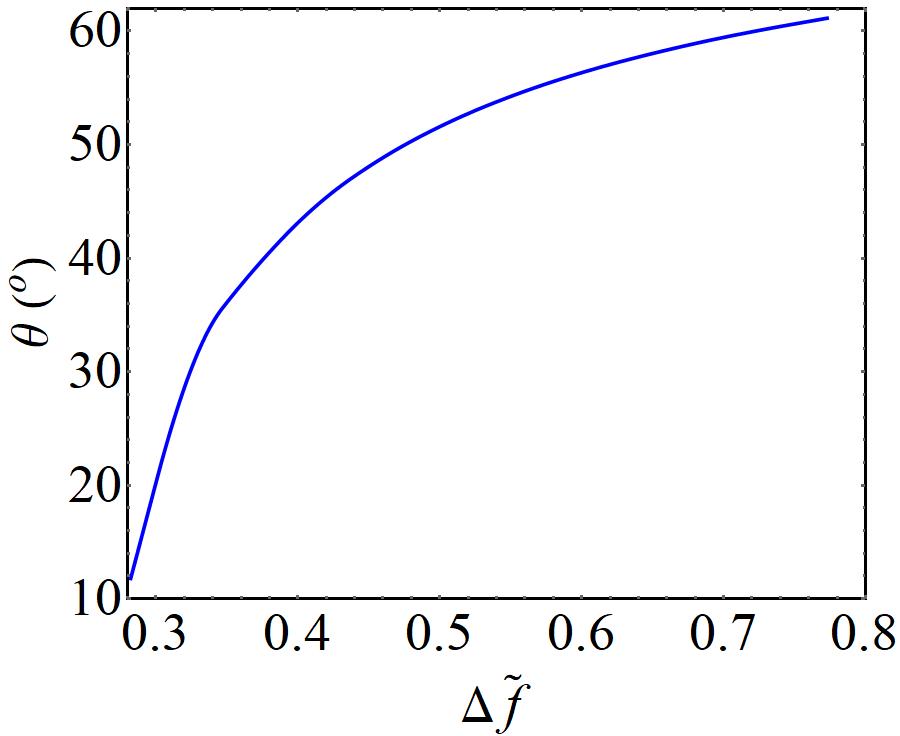}%
   \caption{}
   \label{fig:8c}
 \end{subfigure}
 \begin{subfigure}{0.45\textwidth}
   \centering
   \includegraphics[scale=0.17]{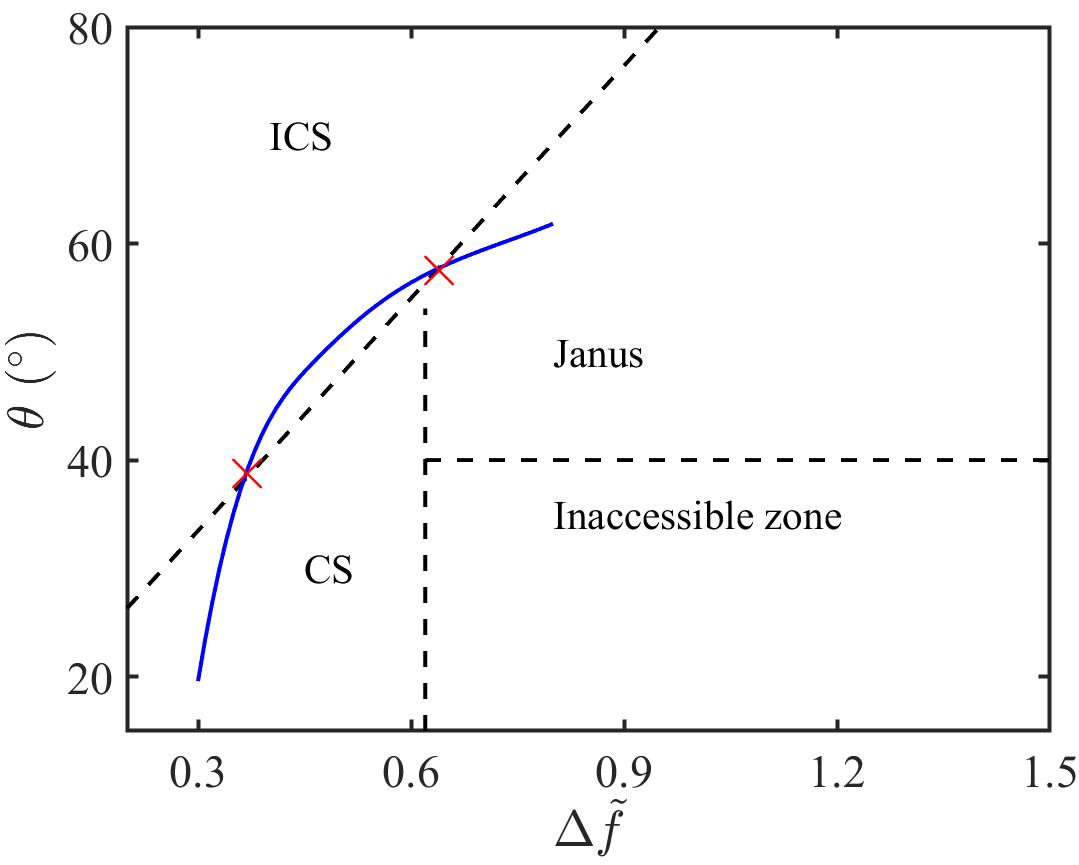}%
   \caption{}
   \label{fig:8d}
 \end{subfigure} 
\label{fig:misc-plots}
 \caption{(a) Temperature-dependence of $\Delta{\tilde{f}}$. (b) Variation of $\theta$
 with temperature for the Ag-Cu system computed using its CALPHAD data. (c) Correlation
 of $\theta$ with $\Delta{\tilde{f}}$. (d) The $\theta-\Delta\tilde{f}$ trajectory
 for Ag-Cu is superimposed on the morphology map obtained from simulations. CS$\to${ICS} 
 and ICS$\to${Janus} transitions are indicated by the crosses, and vertical dashed lines are
 drawn on (a) and (b) to indicate the temperatures for these transitions (the former takes
 place at the higher $T$).}
\end{figure*}

We proceed further to obtain a correlation between the effective driving force 
$\Delta\tilde{f}$ and $\theta$, noting that the latter is defined only for 
$T\le T_{tr}$. First, the bulk driving force $\Delta{f}$ (in J/mol) for phase 
separation in Ag-Cu is normalized by the product of surface energy difference 
$\Delta\sigma$ (in J/m$^2$) and molar surface area (in m$^2$/mol).
Fig.~\ref{fig:8a} shows the temperature dependence of $\Delta\tilde{f}$ estimated this
way, it is very similar to the variation of $\Delta{f}$ with $T$ shown
in Fig.~\ref{fig:4b}. Next, the interfacial and surface energies computed
earlier are used to determine how contact angle varies as a function of temperature.
This is presented in Fig.~\ref{fig:8b}, which shows that at  high temperatures, $\theta$
decreases steeply with decrease in temperature, but
the rate of this decrease reduces at lower temperatures. 
Finally, these variations are combined into a single $\Delta\tilde{f}$-$\theta$ plot
in Fig.~\ref{fig:8c}. It shows a monotonic increase of $\theta$ with $\Delta\tilde{f}$;
however, the slope of the curve is steeper at low-$\Delta\tilde{f}$, gradually becoming
gentler with increase in $\Delta\tilde{f}$ (\emph{i.e.}, decrease in $T$).
This line is the trajectory that the system follows in the 
$\Delta\tilde{f}$-$\theta$ space as temperature is reduced, and therefore captures its
response to a change in the state variable $T$. This trajectory is now superimposed
on the morphology map which is redrawn in Fig.~\ref{fig:8d}. It predicts the 
morphological transitions in Ag-Cu for $T<T_{tr}$: both $\Delta\tilde{f}$ and
$\theta$ increase with decreasing temperature, and the system moves from a metastable
CS to metastable ICS configuration, before finally transitioning into the Janus regime.
Thus, depending on the processing conditions, all the three configurations can form
in Ag-Cu BNPs, as confirmed by the experimentally observed 
configurations~\cite{nakamura_ics,Malviya2014}.

Since the temperature dependence of bulk chemical and capillary forces are system
specific, it should be noted that the trajectory may not always pass through the
metastable ICS region of the map for all alloy systems. However, the insights gained
from the study remains valid and provide crucial guidelines and understanding for
conducting further experiments aimed at tailoring the BNP morphology.

\section{Conclusions}

Results of the present study show how different BNP morphologies can emerge from a
competition between two alternative and concomitant mechanisms, namely, bulk and 
surface-directed spinodal decomposition. Their interplay with capillarity sets up 
the eventual coarsening pathway toward a steady-state configuration. When we express
the results in terms of three physical parameters, namely, driving force, difference
in surface energies and contact angle, different morphologies automatically cluster into
three distinct regions in the $\Delta\tilde{f}-\theta$ space. This identification of relevant
physical parameters appears remarkable, as the computed thermodynamic trajectory for Ag-Cu
involving the same variables traverses all the three distinct regions revealing the
morphological transitions. We note that morphological transitions for a particular alloy
system are sensitive to the nature of temperature-dependence of the relevant variables.
Therefore, the exact transition points for different alloys systems will be different.

The following specific conclusions can be drawn from the study:

\begin{enumerate}
    \item Irrespective of the driving force for bulk spinodal, stable CS forms when the
    spontaneous wetting condition is satisfied ($\theta=\ang{0}$). This happens at 
    $T\geq{T_{tr}}$.
    \item A combination of low driving force and high contact angle gives rise to metastable ICS,
    while metastable CS forms at moderate driving force and lower non-zero $\theta$. Thus, the 
    \emph{spontaneous} wetting condition given by Cahn is found to be a \emph{sufficient}, but
    not a \emph{necessary} condition for the formation of CS.
    \item Janus forms when both bulk driving force and contact angle are large.
    \item Trajectory of Ag-Cu in the driving force-contact angle space shows transitions involving
    CS, ICS and Janus morphologies as a function of temperature.

\end{enumerate}

\section*{Conflict of Interest}
The authors have no conflicts to disclose.

\section*{Acknowledgement}
Authors gratefully acknowledge the computational support from DST-NSM Grant DST/NSM/R\&D-HPC-Applications/2021/03.

\nocite{*}
\bibliography{arxiv}

\section*{Appendix: Non-dimensionalization procedure}

We make the governing equations (Eqs.~\eqref{eq:totalFcal} and~\eqref{eqn:CH}) dimensionless
by using dividing the length, energy and time variables by their characteristic values ($L_c$, 
$E_c$, and $\tau_c$, respectively):
\begin{align}
    x = x^\prime/L_c, f = f^\prime/E_c,t = t^\prime/\tau_c,
\end{align}
where the primed quantities represent the dimensional values of the variables. Now we write
the dimensional form of Eq.~\eqref{eq:totalFcal} and use the above relations to make it
non-dimensional (noting that the dimensional gradient operator $\nabla^\prime$ has the
dimension of inverse of length):
\begin{align}
    \mathcal{F^\prime} & = \frac{1}{V_m^\prime} \int_{\Gamma} \Big[f^\prime(c,\phi)
                  + \kappa_\phi^\prime|\nabla^\prime\phi|^2 +
                  \kappa_c^\prime|\nabla c|^2 \Big] d\Gamma^\prime\nonumber\\
    \Rightarrow E_c\mathcal{F} & = \frac{1}{L_c^3V_m} \int_{\Gamma} \Big[E_c f(c,\phi)
        + (1/L_c^2)\kappa_\phi^\prime|\nabla\phi|^2 \nonumber\\
        &+(1/L_c^2)\kappa_c^\prime|\nabla c|^2 \Big] L_c^3 d\Gamma,\nonumber\\
    \Rightarrow\mathcal{F} & =\frac{1}{V_m} \int_{\Gamma} \Big[f(c,\phi)
        + \frac{\kappa_\phi^\prime}{E_cL_c^2}|\nabla\phi|^2
        + \frac{\kappa_c^\prime}{E_cL_c^2}|\nabla c|^2 \Big] d\Gamma
\end{align}
Thus we see that we get back the original form of Eq.~\eqref{eq:totalFcal} in terms of non-
dimensional variables when $\kappa_c^\prime$ and $\kappa_\phi^\prime$ are scaled by choosing
a reference value of $\kappa=E_cL_c^2$.

The Cahn-Hilliard equation (Eq.~\ref{eqn:CH}) in dimensional form is converted to its
non-dimensional form as follows:

\begin{align}
\label{eq:CHdim}
\frac{\partial c}{\partial t^\prime} &=  
    \nabla^\prime \cdot M^\prime\nabla^\prime\Big(\frac{\delta\mathcal{F}^\prime}{\delta c}\Big)\nonumber\\
     \Rightarrow\frac{1}{\tau_c}\frac{\partial c}{\partial{t}} &= \frac{1}{L_c}\nabla\cdot\frac{1}{L_c}M^\prime\nabla
     \Big[\frac{\delta({E_c}\mathcal{F})}{\delta c}\Big]\nonumber\\
     \Rightarrow\frac{\partial c}{\partial{t}} &= \nabla\cdot\frac{E_c\tau_c}{L_c^2}M^\prime\nabla
     \Big(\frac{\delta\mathcal{F}}{\delta c}\Big)
\end{align}
Thus we get back the non-dimensional Eq.~\eqref{eqn:CH} by scaling the dimensional mobility $M^\prime$
with its characteristic value $L_c^2/E_c\tau_c$. Using these conversion expressions and choosing
appropriate values for the reference variables $L_c$, $E_c$ and $\tau_c$, the dimensionless model 
parameters used in the simulations can now be connected to their dimensional counterparts~\cite{pankaj}.

Choosing $L_c = (V_m/N_0)^{1/3}$ where $N_0$ is the Avogadro's number gives a reference length of
$\sim$ \SI{0.25}{\nano\meter}. If reference energy is taken as $k_B T_c$, it yields a value of
$E_c=1.863\times10^{-20}~\si{\joule} = \SI{0.116}{\electronvolt}$ for $T_c=$\SI{1350}{\kelvin}. 
Finally, a reference time can be obtained, for example, by using a typical mobility 
value~\cite{Wang2012AgCu} of $M^{\prime}\approx\dfrac{1.8\times10^{-18}}{k_BT}$
\SI{}{\metre^2\per\joule\second}, which for $T =$ \SI{800}{\kelvin} yields 
$\tau_c=L_c^2/E_c M^\prime \approx 21$ ms.
Dimensional values of relevant model parameters, along with their conversion factors, are listed in 
Table~\ref{SimulationParameters}, while the corresponding values of surface and interfacial energies
resulting from different parameter sets are provided in Table~\ref{tab:IEdimensional}.

\begin{table*}[ht]
    \captionsetup{font=small}
    \centering
    \caption{Simulation parameters (all energies are in \emph{per atom} basis). Conversion factors
    from non-dimensional to dimensional form are based on characteristic length 
    $L_c=\SI{0.25}{\nano\meter}$, characteristic energy $E_c=\SI{0.116}
    {\electronvolt}$ and characteristic time $\tau_c=\SI{21}{\milli\second}$.}
\begin{adjustbox}{max width=0.95\textwidth}
\begin{tabular}{lllll}
    \toprule
    \multirow{2}{*}{Parameter name} & \multirow{2}{*}{Symbol} & \multicolumn{2}{c}{Value} &~\multirow{2}{*}{Conversion factor}\\
   \arrayrulecolor{Gainsboro} \cmidrule[\heavyrulewidth](l){3-4}
   & & Non-dimensional & Dimensional & \\
   \arrayrulecolor{Black} \midrule
    \small{Grid size} & $\Delta x$ & $0.5$ & $\SI{0.125}{\nano\meter}$ & $L_c$\\ 
    \small{Time step} & $\Delta t$ & $0.001$ & $\SI{21}{\micro\second}$ & $L_c^2/E_cM^\prime$\\
    \small{Particle diameter} & $d$ & 140 & $\SI{35}{\nano\meter}$ & $L_c$ \\ 
    \small{Matrix free energy coefficient} & $f_0^m$ & 2 & \SI{0.232}{\electronvolt}
    & $E_c$\\ 
    \small{Particle free energy coefficient} & $f_0^p$ & 2, 4, 6, 8 & 0.23, 0.46, 0.69, 0.93 eV
    & $E_c$\\ 
    \small{Barrier height} & $\omega_0$ & 3.75, 5, 6, 12 & 0.43, 0.58, 0.67, 1.39 eV
    & $E_c$\\ 
    \small{Gradient energy coefficient} & $\kappa_c$ & 1, 2, 8 
    & 0.007, 0.014, 0.06 \SI{}{\electronvolt\nano\meter^2} & $E_c L_c^2$\\ 
    & $\kappa_\phi$ & 1 &  \SI{0.007}{\electronvolt\nano\meter^2} & $E_c L_c^2$\\ %
    \bottomrule
\end{tabular}
\end{adjustbox}
\label{SimulationParameters}
\end{table*}

\begin{table*}[htbp]
 \small\caption{Non-dimensional (first sub-row in a row) and dimensional (second sub-row in a row) values of surface and interfacial energies. Unit for the dimensional values is \si{\milli\joule \meter^{-2}}.}
 \label{tab:IEdimensional}
 \begin{tabular*}{0.5\textwidth}{@{\extracolsep{\fill}}*{4}l}
 \hline
  ($f_0^p$, $\kappa_c$, $\omega_0$) & $\sigma_1$ & $\sigma_2$ & $\sigma_{12}$  \\
 \hline
   
  ($8,1,12$) & $4.19$ & $3.09$ & $0.94$ \\
  & $1248$ & $921$ & $280$   \\
  \hline
  ($6,2,3.75$) & $1.77$ & $1.44$ & $1.15$ \\
  & $527$ & $429$ & $343$   \\
  \hline
  ($4,1,6$) & $2.21$ & $1.76$ & $0.67$ \\
  & $658$ & $524$ & $200$   \\
  \hline
  ($2,8,6$) & $2.24$ & $1.86$ & $1.33$ \\
  & $667$ & $554$ & $396$   \\
  \hline
  ($4,2,5$) & $1.98$ & $1.6$ & $0.94$ \\
  & $590$ & $477$ & $280$   \\
 \hline

 \end{tabular*}
\end{table*}

\end{document}